%% file: paper1.tex
\pdfoutput=1
\documentclass[iop, apj, twocolappendix, numberedappendix]{emulateapj}

\newif\iflocal
\localfalse

\usepackage{xspace}
\usepackage{amsmath}
\usepackage{framed} 
\usepackage{txfonts}
\usepackage{epstopdf}
\usepackage[dvipsnames]{xcolor}
\usepackage{rotating}
\usepackage{natbib}
\usepackage{ulem}
\usepackage{xspace}
\usepackage[colorlinks=true,urlcolor=black,linkcolor=blue,citecolor=blue]{hyperref}
\usepackage{sistyle}
\SIthousandsep{,}

\iflocal
\input{../../../Docs/_LatexInclude/commands}

\input{macros}

\input{../../../Docs/_LatexInclude/citation_fix}
\else

\input{commands}
\input{macros}
\input{citation_fix}
\fi

\shorttitle{Hydrogen in IllustrisTNG}
\shortauthors{Diemer et al.}

\journalinfo{The Astrophysical Journal Supplement Series, {\rm 238:33 (22pp), 2018 October}}
\submitted{Received 2018 June 6; revised 2018 September 17; accepted 2018 September 20; published 2018 October 23}

\begin{document}


\newcommand\cas[1]{{\color{Cyan} AS: #1}\xspace}
\newcommand\capp[1]{{\color{ForestGreen} AP: #1}\xspace}
\newcommand\ccl[1]{{\color{RubineRed} CL: #1}\xspace}
\newcommand\cdn[1]{{\color{Orange} DN: #1}\xspace}
\newcommand\cfv[1]{{\color{Plum} FVN: #1}\xspace}
\newcommand\cgp[1]{{\color{DarkOrchid} GP: #1}\xspace}
\newcommand\cjf[1]{{\color{ForestGreen} JF: #1}\xspace}
\newcommand\cfm[1]{{\color{Cyan} FM: #1}\xspace}


\iflocal
\def\figdir{figs1}
\else
\def\figdir{.}
\fi


\defcitealias{leroy_08}{L08}
\defcitealias{gnedin_11}{GK11}
\defcitealias{krumholz_09_kmt1}{KMT}
\defcitealias{krumholz_13}{K13}
\defcitealias{gnedin_14}{GD14}
\defcitealias{sternberg_14}{S14}

\defcitealias{springel_03}{SH03}


\title{Modeling the atomic-to-molecular transition in cosmological simulations of galaxy formation}
\author{
Benedikt Diemer\altaffilmark{1},
Adam R. H. Stevens\altaffilmark{2},
John C. Forbes\altaffilmark{1},
Federico Marinacci\altaffilmark{1},
Lars Hernquist\altaffilmark{1},
Claudia del P. Lagos\altaffilmark{2,3},
Amiel Sternberg\altaffilmark{4,5},
Annalisa Pillepich\altaffilmark{6},
Dylan Nelson\altaffilmark{7},
Gerg{\"o} Popping\altaffilmark{6}, 
Francisco Villaescusa-Navarro\altaffilmark{5},
Paul Torrey\altaffilmark{8},
Mark Vogelsberger\altaffilmark{9}
}

\affil{
$^1$ Institute for Theory and Computation, Harvard-Smithsonian Center for Astrophysics, 60 Garden St., Cambridge, MA 02138, USA; \href{mailto:benedikt.diemer@cfa.harvard.edu}{benedikt.diemer@cfa.harvard.edu}\\
$^2$ International Centre for Radio Astronomy Research, University of Western Australia, 35 Stirling Highway, Crawley, WA 6009, Australia\\
$^3$ ARC Centre of Excellence for All Sky Astrophysics in 3 Dimensions (ASTRO 3D)\\
$^4$ Raymond and Beverly Sackler School of Physics and Astronomy, Tel Aviv University, Ramat Aviv 69978, Israel\\
$^5$ Center for Computational Astrophysics, Flatiron Institute, 162 5th Avenue, 10010, New York, NY, USA\\
$^6$ Max-Planck-Institut f{\"u}r Astronomie, K{\"o}nigstuhl 17, D-69117 Heidelberg, Germany\\
$^7$ Max-Planck-Institut f{\"u}r Astrophysik, Karl-Schwarzschild-Str 1, D-85741 Garching, Germany\\
$^8$ Hubble Fellow, MIT Kavli Institute for Astrophysics and Space Research, Cambridge, MA 02139, USA\\
$^9$ Department of Physics, Massachusetts Institute of Technology, Cambridge, MA 02139, USA
}


\begin{abstract}
Large-scale cosmological simulations of galaxy formation currently do not resolve the densities at which molecular hydrogen forms, implying that the atomic-to-molecular transition must be modeled either on the fly or in postprocessing. We present an improved postprocessing framework to estimate the abundance of atomic and molecular hydrogen and apply it to the IllustrisTNG simulations. We compare five different models for the atomic-to-molecular transition, including empirical, simulation-based, and theoretical prescriptions. Most of these models rely on the surface density of neutral hydrogen and the ultraviolet (UV) flux in the Lyman--Werner band as input parameters. Computing these quantities on the kiloparsec scales resolved by the simulations emerges as the main challenge. We show that the commonly used Jeans length approximation to the column density of a system can be biased and exhibits large cell-to-cell scatter. Instead, we propose to compute all surface quantities in face-on projections and perform the modeling in two dimensions. In general, the two methods agree on average, but their predictions diverge for individual galaxies and for models based on the observed midplane pressure of galaxies. We model the UV radiation from young stars by assuming a constant escape fraction and optically thin propagation throughout the galaxy. With these improvements, we find that the five models for the atomic-to-molecular transition roughly agree on average but that the details of the modeling matter for individual galaxies and the spatial distribution of molecular hydrogen. We emphasize that the estimated molecular fractions are approximate due to the significant systematic uncertainties.
\end{abstract}



\section{Introduction}
\label{sec:intro}

Cosmological simulations that follow volumes of up to a few hundred megaparsecs on a side are integral to the study of galaxy formation \citep{schaye_10, schaye_15, vogelsberger_13, vogelsberger_14_illustris, vogelsberger_14_nature, dubois_14, somerville_15, dave_16, pillepich_18_tng}. These calculations achieve statistically representative samples of galaxies, but at the cost of relatively low resolution, typically about a million solar masses and roughly kiloparsec spatial resolution. Thus, many important physical processes cannot be followed explicitly. For example, while some higher-resolution, zoom-in simulations have been performed with chemical networks and radiative transfer \citep{pelupessy_06, robertson_08, gnedin_09, christensen_12, kuhlen_12, rosdahl_15, hu_16, katz_17, nickerson_18}, such calculations would not necessarily be applicable at lower resolution and would also be too costly to perform in cosmological volumes and over a Hubble time. Thus, a number of physical quantities are not directly predicted by cosmological simulations.

One particularly desirable quantity is the phase structure of galactic gas: while most simulations now approximate the ionization balance of hydrogen due to ultraviolet (UV) radiation and self-shielding in low-density regions, they do not compute the abundances of atomic and molecular hydrogen (hereafter abbreviated \hi and \htwo, respectively) at high densities \citep[see, however,][]{thompson_14, dave_16}. The \hi and \htwo masses and distributions are critical for comparisons to observations because cold gas is often observed in a specific phase. For example, \hi can be detected via its 21 cm emission or absorption in quasar spectra, while \htwo can be inferred via spectroscopic observations of CO and other molecular tracers or via the dust continuum \citep[see][for a review]{bolatto_13}. 

To facilitate meaningful comparisons with observations of the gas content of galaxies, we must post-process cosmological simulations. This step is nontrivial because the transition from atomic to molecular hydrogen is physically complicated: \htwo molecules form predominantly on the surface of dust grains (which are not modeled in the simulations) and are mostly destroyed by UV radiation in the Lyman--Werner (LW) band, which is shielded by dust, \hi, and \htwo \citep[e.g.,][]{spitzer_74, jura_75_thick, jura_75_thin, black_76, shull_78, vandishoeck_86, sternberg_89, elmegreen_93, draine_11}. Thus, we expect molecules to form most efficiently in dense regions, at high metallicity, and to be surrounded by a layer of shielding atomic gas \citep{sternberg_88, elmegreen_89, browning_03, krumholz_09_kmt1, krumholz_13, sternberg_14, bialy_17}. However, the geometry of molecular clouds can be highly irregular, and the properties of dust are poorly understood \citep[in particular, its distribution may not follow the gas;][]{gnedin_08, bekki_15, hopkins_16_dust, mckinnon_16, mckinnon_17, mckinnon_18}. Moreover, the efficiency of dissociation is not even across the LW band but instead is due to a series of lines, introducing complications such as line overlap \citep{draine_96, gnedin_14}. In cosmological simulations, there is no way to model any of these effects in detail because the gas cells are large enough to contain a significant number of unresolved molecular clouds and sources of UV radiation. 

Instead, we rely on prescriptions for the dependence of the molecular fraction, $\fmol$, on several averaged quantities, such as density, surface density, neutral gas fraction, the UV field, and metallicity. Providing physically motivated estimates of these quantities emerges as the main challenge, but once they have been computed, there are, broadly speaking, three commonly used classes of models that can be used to estimate the molecular fraction. First, it is observationally well established that, in the local universe, $\fmol$ is correlated with the midplane pressure of galaxies \citep{blitz_04, blitz_06, leroy_08, robertson_08}. We can estimate the midplane pressure from surface densities and velocity dispersions, with no dependence on the UV field. Second, high-resolution simulations including advanced chemical networks have been used to calibrate $\fmol$ as a function of surface density, metallicity, and the UV field \citep{gnedin_11, gnedin_14}. Third, analytical equilibrium models of molecular clouds provide the most accurate modeling of \htwo creation and destruction, but such models need to be modified when applying them in the context of cosmological simulations, where they represent a collection of molecular clouds rather than an individual cloud \citep{sternberg_88, sternberg_14, krumholz_09_kmt1, mckee_10, krumholz_13, bialy_17}. We use at least one representative from each of these classes of models to estimate the molecular fraction.

Similar postprocessing exercises were first applied to semianalytical models of galaxy formation and 1D simulations \citep{obreschkow_09b, fu_10, lagos_11_hih2, lagos_11_sflaw, forbes_12, forbes_14, krumholz_12, popping_14, popping_15} and were later extended to the OWLS and EAGLE simulations \citep{schaye_10, schaye_15}. Many aspects of our work are based on the effort of \citet{lagos_15} who computed $\fmol$ for EAGLE galaxies on a particle-by-particle basis. A number of follow-up studies used similar methodology and compared EAGLE to observations of the neutral and molecular gas fraction, the respective mass functions, the structure of \hi disks, environmental dependencies, and the mass--size relation, among others \citep[][see also \citealt{duffy_12}]{rahmati_15, lagos_16, bahe_16, marasco_16, crain_17}. \citet{marinacci_17} applied the method of \citet{lagos_15} to the AURIGA high-resolution zoom-in simulations \citep{grand_17} and \citet{villaescusanavarro_18} investigated the overall abundance and clustering of \hi in the IllustrisTNG simulations. Notably, some studies found good agreement between the \hiht models tested \citep[e.g.,][]{marinacci_17} while others reported significant differences between the $\fmol$ predictions, especially in detailed, spatially resolved analyses \citep[e.g.,][]{bahe_16}. Such differences are surprising because the \hiht models are supposed to agree reasonably well when they are based on the same underlying UV field, metallicity, and density \citep[e.g.,][]{krumholz_11_comparison, sternberg_14}.

In this work, we attempt to understand and to remedy such disagreements, and to provide a comprehensive investigation of the modeling methods in general. Based on the IllustrisTNG simulations, we find that the conversion between volume and surface density can introduce systematic errors and propose a new method based on two-dimensional projections of all relevant quantities. We introduce a new way to crudely estimate the LW-band UV flux by following the propagation of radiation from star-forming populations in the optically thin limit. We investigate five different models for the \hiht transition and show that they are in good agreement on average, though their predictions diverge for individual galaxies and the spatial distribution of molecules. We will not attempt to evaluate which models are most accurate but accept their range of predictions as a systematic uncertainty.

In this paper, we focus on our methodology and reserve the quantitative comparison between IllustrisTNG and the observed gas content of galaxies for future work (Diemer et al. 2018, Stevens et al. 2018, Popping et al. 2018, all in preparation). The paper is organized as follows. In Section~\ref{sec:methods}, we lay out our methodology, though some of the details are to be found in Appendix~\ref{sec:app:uv}. We compare the predictions of the different \hiht models in Section~\ref{sec:results}. We discuss the most important uncertainties in Section~\ref{sec:discussion} and summarize our conclusions in Section~\ref{sec:conclusion}. We give detailed mathematical expressions and algorithmic information about the \hiht models in Appendix~\ref{sec:app:models}.


\section{Methods}
\label{sec:methods}

In this section, we describe how we compute the molecular fraction in gas cells and galaxies in our simulations. We briefly introduce the IllustrisTNG simulation suite, focusing on those parts of the setup that are particularly relevant for our study. We discuss the two fundamental types of modeling used (volumetric and projected) and describe how we compute the neutral fraction, the UV field, and eventually the \hiht transition.

\subsection{Notation}
\label{sec:intro:notation}

As the notation of gas densities can easily get confusing due to the large number of subspecies, we use a consistent set of subscripts. In particular, we denote the surface densities of all gas as $\sigmag$, of all hydrogen as $\sigmah$, of neutral hydrogen as $\sigman$, of atomic hydrogen as $\sigmahi$, and of molecular hydrogen as $\sigmaht$. The same subscripts are used for volumetric mass densities ($\rhog$, $\rhoh$, $\rhon$, $\rhohi$, and $\rhoht$), as well as for number densities (e.g., $n_{\rm H}$), column densities (e.g., $N_{\rm H}$), and galaxy-integrated masses (e.g., $M_{\rm H}$). The molecular fraction can be expressed either as the ratio of atomic to molecular gas,
\begin{equation}
\rmol \equiv \frac{\sigmaht}{\sigmahi} \,,
\end{equation}
or as the fraction of neutral hydrogen that is molecular:
\begin{equation}
\fmol \equiv \frac{\sigmaht}{\sigmahi + \sigmaht} = \frac{\rmol}{\rmol + 1}\,.
\end{equation}
When quoting the fraction of neutral hydrogen, $\fneutral$, we note that it refers to the fraction of all gas (including helium and metals) that is in neutral hydrogen, not the fraction of hydrogen that is neutral. Furthermore, we denote the sound speed in gas as $c_{\rm s}$, the ratio of heat capacities as $\gamma = 5/3$, the internal energy per unit mass as $u$, and the metallicity as $Z$.

\subsection{The IllustrisTNG Simulations}
\label{sec:methods:illustris}

IllustrisTNG is a suite of cosmological hydrodynamical simulations run using the moving-mesh code \textsc{Arepo} \citep{springel_10}. We predominantly use the two highest-resolution runs for each box size, hereafter referred to as TNG100 and TNG300, which model box sizes of about $100$ and $300$ Mpc, respectively \citep{marinacci_18, naiman_18, nelson_18, pillepich_18, springel_18}. The simulations assume the \citet{planck_16} cosmology: $\Omega_{\rm m} = 0.3089$, $\Omega_{\rm b} = 0.0486$, $h = 0.6774$, and $\sigma_8 = 0.8159$. Physical models in IllustrisTNG include prescriptions for gas cooling, star formation, metal enrichment, black hole growth, stellar winds, supernovae, and active galactic nuclei (AGNs); we refer the reader to the respective papers for details \citep{weinberger_17, pillepich_18_tng}. The TNG galaxy formation model is built on the original Illustris model \citep{vogelsberger_13, vogelsberger_14_illustris, vogelsberger_14_nature, genel_14, torrey_14, sijacki_15}. 

Galaxies and halos are identified using the \textsc{Subfind} algorithm \citep{davis_85, springel_01_subfind, dolag_09}. In this work, virtually all quantities are computed directly from the particle distributions, but we do use the total gravitationally bound mass in both gas and stars to select galaxies. Table~\ref{table:sims} shows a list of the particular simulations used in this paper, as well as the lower mass limits in gas and stellar mass used to select galaxies. We emphasize that we include galaxies in our sample if they pass either the gas {\it or} stellar mass threshold to ensure a well-defined sample completeness regardless of which selection is applied. 

\begin{deluxetable}{lcccrrr}
\tablecaption{Simulations and galaxy sample selection criteria
\label{table:sims}}
\tablewidth{0pt}
\tablehead{
\colhead{Simulation} &
\colhead{$m_{\rm baryon}$} &
\colhead{$M_{\rm gas,min}$} &
\colhead{$M_{\rm *,min}$} &
\colhead{$N_{\rm gas}$} &
\colhead{$N_{\rm gal}$}
}
\startdata
TNG100   & $1.4 \times 10^{6}$ & $2.0 \times 10^{8}$ & $2.0 \times 10^{8}$  & 142 & \num{152122} \\
TNG300   & $0.9 \times 10^{7}$ & $5.0 \times 10^{9}$ & $5.0 \times 10^{10}$ & 454 & \num{452664} \\
TNG100-2 & $1.1 \times 10^{7}$ & $1.1 \times 10^{9}$ & $1.1 \times 10^{9}$  & 100 &  \num{78308} \\
TNG100-3 & $8.9 \times 10^{7}$ & $9.0 \times 10^{9}$ & $9.0 \times 10^{9}$  & 100 &  \num{20635}
\enddata
\tablecomments{All masses are given in $\msun$. To be included in our sample, a galaxy needs to have either $M_{\rm gas,min}$ gas mass or $M_{\rm *,min}$ stellar mass, counting all particles bound to the galaxy. The baryonic mass units in IllustrisTNG (i.e., the stellar population particles and gas cells) can slightly evolve in mass; the mass quoted as $m_{\rm baryon}$ is the target mass \citep{pillepich_18}. On average, the minimum gas and stellar masses correspond to the listed particle numbers $N_{\rm gas}$ and result in sample sizes of $N_{\rm gal}$ galaxies.}
\end{deluxetable}

\subsubsection{Interstellar Medium Model and Equation of State}
\label{sec:methods:illustris:ism}

\begin{figure*}
\centering
\includegraphics[trim = 10mm 10mm 2mm 2mm, clip, scale=0.49]{\figdir/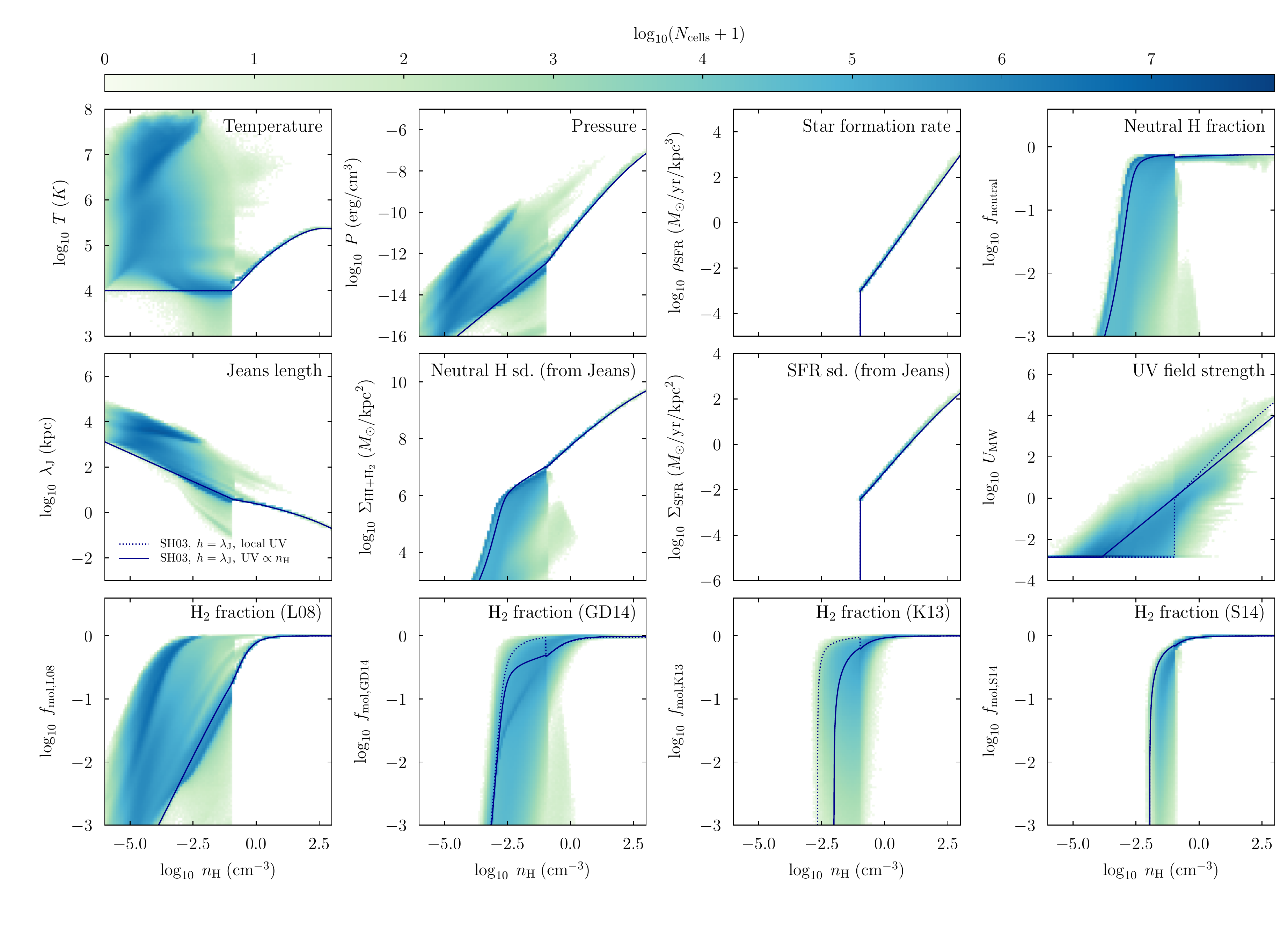}
\caption{Physical conditions and the resulting molecular fractions as a function of density. In each panel, the histograms show the distribution of all gas cells in the analyzed galaxies in TNG100, about $1.2$ billion cells. The solid blue lines represent the ISM model of \citetalias{springel_03} (Section~\ref{sec:methods:illustris:ism}). {\it Top row:} At densities below the star formation threshold of $n_{\rm H} = 0.106\ {\rm cm}^{-3}$, hydrodynamics governs the state of the gas, leading to a range of temperatures and pressures. To guide the eye, the blue lines correspond to an ideal-gas equation of state with a temperature of \num{10000} K. Above the threshold, the \citetalias{springel_03} effective equation of state determines the temperature, pressure, and SFR. The top right panel shows the fraction of gas in neutral hydrogen, which is computed from the model of \citet{rahmati_13} below the star formation threshold (assuming $Z = Z_{\odot}$) and as the cold cloud fraction in the \citetalias{springel_03} model above the threshold (Section~\ref{sec:methods:illustris:neutral}), introducing a slight discontinuity. {\it Middle row:} the most important ingredients used in the cell-by-cell modeling of the \hiht transition (Section~\ref{sec:methods:volmap}), namely the Jeans length, the surface densities of neutral hydrogen and star formation, and the UV field strength. The surface densities are computed by multiplying the respective volume densities by the Jeans length. The UV field is estimated from optically thin propagation (Section~\ref{sec:methods:uv}) and approximated as $\umw \propto n_{\rm H}$ in the solid blue lines. {\it Bottom row:} the predicted molecular fraction according to the various \hiht models (the \modelgk predictions are very similar to that of \modelgd). For comparison, the dotted lines show an alternative UV model where $\umw$ is equal to the UV background below the star formation threshold and proportional to the SFR surface density above the threshold. This assumption tends to introduce an unphysical dip in $\fmol$ at the star formation threshold (though it has no impact on the \modell model, which does not depend on the UV field). However, even with the new modeling of the UV field, the star formation threshold introduces a slight discontinuity into many of the computed quantities, including the molecular fractions.}
\label{fig:ismconditions}
\end{figure*}

The results of our \hiht modeling will critically depend on the hydrodynamics and star formation in dense gas in IllustrisTNG galaxies, both of which are regulated by the two-phase interstellar medium (ISM) model of \citet[][hereafter \citetalias{springel_03}]{springel_03}. This model assumes that there is a threshold density above which star formation sets in. Below the threshold, the gas physics is determined by hydrodynamics, assuming an ideal-gas equation of state. Above the star formation threshold, the model presupposes that the ISM consists of cold, star-forming clouds with a temperature of \num{1000} K and hot, ionized gas. The model computes an equilibrium star formation rate (SFR) that is proportional to the cold cloud density over a free-fall time, leading to a scaling of ${\rm SFR} \propto \rho^{1.5}$, which roughly matches the observed Kennicutt--Schmidt (KS) relation \citep{schmidt_59, kennicutt_98}. The SFR and the resulting heating of the ISM via supernovae balance the cooling rate. By averaging over the cold and hot gas phases, the model prescribes an effective equation of state, that is, an effective pressure and temperature as a function of density. These quantities are understood to be averages over many cold clouds and the hot ISM, and thus do not necessarily correspond to values that would be physically realized. The predictions of the \citetalias{springel_03} model are shown as solid lines in the top row of Figure~\ref{fig:ismconditions}. 

The \citetalias{springel_03} ISM model used in IllustrisTNG is a modification of the original version that accounts for a different stellar initial mass function and provides a less-steep equation of state. In particular, the equation of state is softened such that the internal energy is $u_{\rm eff} = 0.3 u_{\rm eff,SH03} + 0.7 u_4$, where $u_4$ corresponds to a temperature of \num{10000} K. Moreover, the parameters were changed to a star formation timescale of $3.28$ Gyr, a supernova temperature of $5.73 \times 10^{7}$ K, and a cloud evaporation factor of $573$. The mass fraction in massive stars changed to $\beta = 0.226$ due to the adoption of the \citet{chabrier_03} stellar initial mass function instead of the \citet{salpeter_55} function used in \citetalias{springel_03}. With these parameters, the threshold density for star formation is $n_{\rm H} = 0.106\ {\rm cm}^{-3}$. We refer the reader to \citetalias{springel_03} for further details.

The histograms in Figure~\ref{fig:ismconditions} show the distribution of all gas cells in the analyzed galaxies in TNG100. As expected, the \citetalias{springel_03} model determines the gas properties above the star formation threshold (with a few exceptions where cells are prevented from forming stars because they are too hot). Below the star formation threshold, the cells exhibit a wider distribution of temperatures and pressures. Here, the blue lines refer to an ideal gas with a constant temperature of \num{10000} K, shown to guide our understanding of how the atomic and molecular fractions evolve as a function of density.

In the following, we will often treat star-forming and quiescent gas cells differently, either because of the effects of star formation or because the pressure and temperature calculations above the threshold do not represent physical values. For this purpose, we check whether the SFR in a cell is greater than zero and treat the cell accordingly as specified in the following sections.

\subsubsection{Hydrogen Fraction and Neutral Fraction}
\label{sec:methods:illustris:neutral}

The primordial hydrogen fraction is assumed to be $0.76$ in IllustrisTNG, but this fraction decreases as the gas becomes enriched with helium and metals. We use the exact hydrogen fraction in each cell for all computations in this paper. For cells within galaxies, the difference is typically a few percent.

For gas below the star-formation threshold density, the neutral fraction is computed self-consistently in the IllustrisTNG simulations. The cooling rate is the sum of primordial cooling (computed from the temperature), metal-line cooling (computed from \textsc{Cloudy} lookup tables, \citealt{ferland_98}), and Compton cooling off cosmic microwave background photons. The photoionization rate due to the UV background is computed using the model of \citet[][in the updated 2011 version]{fauchergiguere_09}, also taking into account nearby AGNs. The self-shielding of gas from the UV background at high densities is estimated based on Equation A1 in \citet{rahmati_13}. The result is used to modify the input UVB photoionization rate to the \textsc{Cloudy} computation of the cooling tables, and to modify the computation of the neutral fraction.

For star-forming cells, the situation is more complicated because the neutral fraction is not computed self-consistently in the star formation model of \citetalias{springel_03}. However, the model splits the gas into a hot and a cold phase, and it is safe to assume that the hot phase would be entirely ionized (Section~\ref{sec:methods:illustris:ism}). Realistically, the cold clouds would be partially ionized due to the UV radiation from young stars, but we neglect this contribution and assume that all cold gas is neutral, giving a fraction of neutral hydrogen of $f_{\rm HI+H_2} = f_{\rm H} \rho_{\rm cold} / \rho$ (where $\rho_{\rm cold}$ is the overall volume density of cold clouds, not the density of individual clouds). The \citetalias{springel_03} model predicts a cold fraction between $0.9$ and $1$, meaning that this estimate is almost equivalent to simply assuming that all star-forming gas is neutral. We further discuss the uncertainty in the neutral fraction in Section~\ref{sec:discussion:neutral}.

\vspace{1cm}

\subsubsection{Dust Abundance}
\label{sec:methods:illustris:dust}

Following numerous previous works \citep[e.g.,][]{gnedin_11, krumholz_13, lagos_15}, we assume that the dust-to-gas ratio is proportional to the metallicity. Thus, the dust-to-gas ratio observed locally is the same as the metallicity in solar units, $D_{\rm MW} = \zmw \equiv Z/Z_{\odot}$, where we use the solar metallicity assumed in IllustrisTNG, $Z_{\odot} = 0.0127$ \citep{asplund_09}. Recently, there has been observational and theoretical evidence that this assumption may break down at low metallicities \citep{remyruyer_14, mckinnon_17, popping_17}.

\subsection{Volumetric and Projected Modeling}
\label{sec:methods:volmap}

\begin{figure*}
\centering
\includegraphics[trim = 23mm 6mm 0mm 5mm, clip, scale=0.44]{\figdir/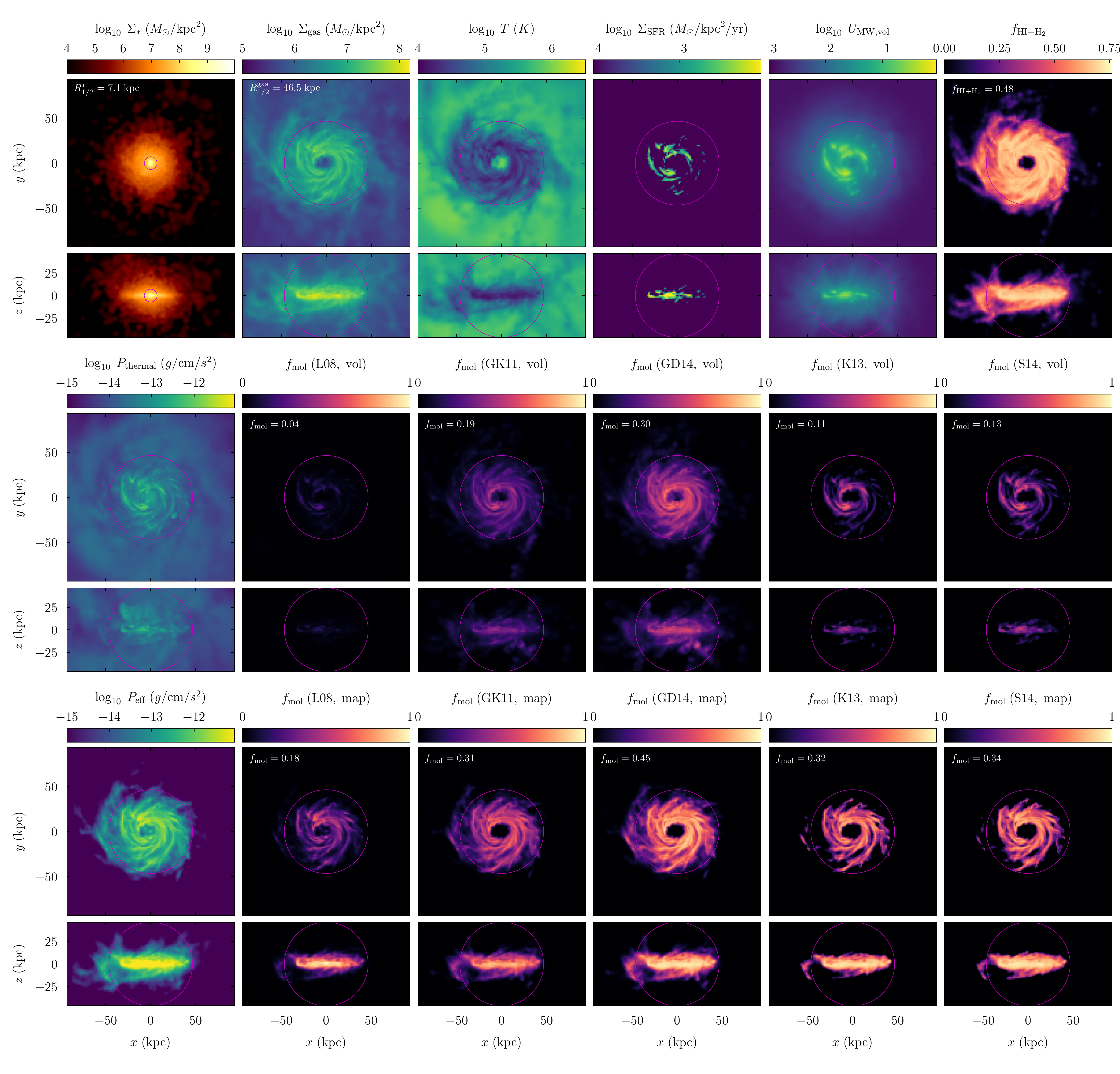}
\caption{Projections of physical properties and the resulting molecular fractions for a gas-rich disk galaxy in TNG100. At $z = 0$, this galaxy has a halo mass of $M_{\rm 200c} = 10^{12} \msun$, a total gas mass of $10^{11} \msun$, a stellar mass of $4 \times 10^{10} \msun$, an SFR of $1.4 \msun/{\rm yr}$, and a relatively uniform of metallicity of $1.2\ Z_{\odot}$ on average, and it hosts a black hole of about $10^{8} \msun$. The purple circles indicate the stellar and gas half-mass radii. Each set of panels shows the face-on (perpendicular to the rotation axis) and edge-on projections using all stellar population particles or gas cells that are gravitationally bound to the galaxy. The maps of temperature, thermal pressure, UV field strength, and neutral fraction are weighted by total gas mass, and the maps of molecular fractions by the neutral hydrogen mass. {\it Top row:} The UV field strength $\umw$ is sourced from star-forming regions and propagated into the surrounding regions (Section~\ref{sec:methods:uv}). The neutral fraction refers to the total gas density and thus cannot exceed the primordial hydrogen fraction, $0.76$. {\it Middle row:} predictions for the molecular fraction according to the volumetric version of the \hiht models. The white labels list the overall molecular fractions of the galaxy. {\it Bottom row:} same as the middle row, but for the projection-based version of the \hiht models. With the exception of the volumetric \modell model, the \hiht models vary in their predictions between 11\% and 45\% molecular fraction, a typical spread for individual galaxies (Section~\ref{sec:results:models}). The large difference between the volumetric and projected versions of the \modell model is clearly caused by the underlying estimates of the pressure: the thermal pressure (middle left panel) is a poor approximation to the midplane pressure (bottom left panel) that the model should be based on (Equation~(\ref{eq:peff})). The volume-weighted thermal pressure would be even lower than the mass-weighted pressure shown here.}
\label{fig:galaxy}
\end{figure*}

In simulations, the natural unit is a particle or gas cell. Many of the models for the \hiht transition, however, work in 2D quantities such as the surface densities of neutral gas or star formation. In most previous \hiht modeling efforts \citep[e.g.,][]{lagos_15, marinacci_17}, the molecular fraction was computed on a particle-by-particle (or cell-by-cell) basis, and the conversion to surface quantities was performed by multiplying with the Jeans length, $\Sigma_{\rm X} = \lambda_{\rm J} \rho_{\rm X}$, where
\begin{equation}
\label{equ:jeans}
\ljeans = \sqrt{\frac{c_{\rm s}^2}{G \rho}} = \sqrt{\frac{\gamma (\gamma - 1) u}{G \rho}} \,.
\end{equation}
Here, $u$ and $\rho$ are the internal energy per unit mass and total density of the gas cell, respectively. The underlying idea is that the Jeans length should approximate the size of a self-gravitating system \citep{schaye_01, schaye_08}. For the typical densities of IllustrisTNG gas cells, the Jeans length varies between about $0.1$ and $10^4$ kpc (Figure~\ref{fig:ismconditions}). 

We will refer to this approach as ``volumetric'' or ``cell-by-cell'' modeling (abbreviated to ``vol'' in some figure captions). In Section~\ref{sec:discussion:jeans}, we test the Jeans approximation explicitly by comparing its predictions to the projected surface densities of gas, neutral gas, and SFR in simulated galaxies. We find that the approximation leads to a significant bias in the recovered quantities and exhibits large cell-to-cell scatter. Thus, we propose an alternative way to model the \hiht transition from the projected quantities directly, a method that we will refer to as ``projected'' or ``map'' modeling. Here, we compute the face-on surface quantities of a galaxy ab initio as follows. We first rotate the galaxy into a face-on projection by aligning its angular momentum vector with the $z$ direction. There is no unique answer as to what constitutes the angular momentum of a galaxy because the rotation of dark matter, stars, and baryons is not necessarily aligned and depends on radius \citep[e.g.,][]{bett_10}. As we are most concerned with the properties of gas, we attempt to measure the angular momentum of all gas cells within half the 3D gas half-mass radius. If there are fewer than $50$ gas cells within this radius, we consider the measurement unreliable and revert to using the angular momentum of the stellar particles within two stellar 3D half-mass radii. We find, by visual inspection, that this algorithm reliably aligns disk-like galaxies in the desired face-on orientation. Some galaxies do not exhibit clear rotational symmetry, meaning that the algorithm produces a more or less random orientation.

Once a galaxy has been aligned, we project the quantities in question (density, stellar density, SFR, and so on) onto a face-on-oriented grid of $128^2$ pixels, representing a physical size of
\begin{equation}
L_{\rm map} = 2 \times {\rm max}(2 R_{1/2}^{\rm gas},\, 6 R_{1/2}^{*},\, 30 {\rm kpc}) \,.
\end{equation}
This expression accounts for both gas-rich and gas-poor galaxies, while the absolute minimum forces the map to contain a reasonable number of force resolution lengths ($\epsilon = 710\ {\rm pc}$ in TNG100). The results shown in this paper correspond to projections of all cells or particles bound to the galaxy according to the \textsc{Subfind} halo finder. When comparing to observations with a limited integration depth, this procedure may need to be adjusted.

Given the complex shapes of gas cells in moving-mesh codes such as \textsc{Arepo}, it is numerically difficult to project the cells onto a pixel grid exactly. Instead, we apply a 2D Gaussian smoothing kernel with a width $\sigma = 1/2\ L_{\rm cell} = 1/2\ (m_{\rm cell} / \rho_{\rm cell})^{1/3}$. This width is comparable to an SPH-like smoothing length computed from $64$ nearest neighbors but leads to less diffuse maps. We have confirmed that increasing or decreasing the smoothing scale by a factor of two changes the median \htwo mass of galaxies by less than 10\%. We have also experimented with a simple particle-in-cell algorithm (i.e., without any smoothing), but we find that such maps systematically overestimate the molecular fraction because they tend to concentrate too much mass in a few pixels. Finally, we have confirmed that our results are converged with map resolution, in other words, that doubling the resolution to $256^2$ leads to no appreciable change in the predicted average molecular masses. The median pixel size for our $128^2$ maps is $0.8\ {\rm kpc}$ in TNG100 and $1.3\ {\rm kpc}$ in TNG300.

Figure~\ref{fig:galaxy} shows an example of maps for a gas-rich, Milky Way-sized disk galaxy from TNG100. The projected galaxy properties allow us to compute more faithful representations of required physical quantities such as surface densities. However, we have lost access to any volumetric quantity. For example, we cannot distinguish star-forming and quiescent cells any more. For the remainder of the paper, we will compute all models in both their volumetric and projected incarnations, implying computations on a cell-by-cell and pixel-by-pixel basis, respectively.

We note that there are characteristic length scales other than the Jeans length that could be used in cell-by-cell modeling. For example, there is the Sobolev length, $\rho / |\nabla \rho|$, the scale over which the density of a gas distribution changes by order unity \citep[e.g.,][]{gnedin_09}. However, we do not experiment with such approximations because it is nontrivial to compute the Sobolev length from the moving-mesh grid cells in \textsc{Arepo}. 

\subsection{The UV Field}
\label{sec:methods:uv}

The \htwo abundance is governed by the balance between recombination on dust grains and dissociation by UV photons in the LW band. Thus, the majority of the \hiht models we consider rely on an estimate of the UV radiation field. Since cosmological simulations such as IllustrisTNG currently do not include radiative transfer, we need to estimate it. This computation turns out to be one of the most uncertain and challenging aspects of modeling the \hiht transition. In this section, we briefly summarize our algorithm and refer the reader to Appendix~\ref{sec:app:uv} for a more detailed description. We parameterize the UV field as $\umw$, the field at $1000$\AA\ in units of the observationally measured field in the local neighborhood according to \citet{draine_78}.

We begin by recognizing that the vast majority of LW photons originate from very young stars whose UV flux decreases by orders of magnitude within $100$ Myr \citep[e.g.,][]{leitherer_99}. In IllustrisTNG, the SFR evolves on timescales of more than half a gigayear at $z \lsim 1$ and a few hundred megayears at higher redshift \citep{torrey_18}. Thus, we can safely estimate the UV field from the SFR, though this approximation becomes less accurate at high redshift. In contrast, the stellar population particles in the simulation are stochastic tracers of the SFR, and only the very most recently formed ones would contribute significant UV flux, rendering the calculation spatially inhomogeneous.

In previous works, the UV field in star-forming gas cells has been scaled to the observed local star-formation surface density, $\umw = \sigmasfr / \Sigma_{\rm SFR,local}$ \citep{lagos_15}. With this method, however, it is not clear how to treat quiescent cells below the star formation threshold. These cells can contain a significant fraction of neutral hydrogen (Figure~\ref{fig:ismconditions}). In the past, either their \htwo contribution was ignored \citep{marinacci_17} or their UV field was set to the cosmic UV background \citep{lagos_15}, a small fraction of the local field at $z = 0$ (Appendix~\ref{sec:app:uv:bg}). The latter choice creates a sharp break in the UV field at the star formation threshold, which is itself an arbitrary byproduct of the ISM model (Figure~\ref{fig:ismconditions}). 

Instead, we assume that some fraction of the UV emitted by star-forming cells is absorbed in situ, and that the remaining UV propagates through an optically thin medium. This approximation is likely flawed for dust-rich, high-metallicity systems, but has the advantage that it relies on only one free parameter: the escape fraction. We calibrate this parameter to $10\%$ by comparing the SFR--UV relation to solar neighborhood values (Section~\ref{sec:discussion:fesc}). The UV background serves as a lower limit for cells far away from any star formation. An example of the resulting UV field is shown in Figure~\ref{fig:galaxy}. In the projection-based models, we proceed equivalently for each pixel in the projected maps. We describe our procedure in detail in Appendix~\ref{sec:app:uv}. In Section~\ref{sec:app:uv:convergence}, we show that even drastic changes in the escape fraction have a relatively modest effect on the molecular fraction. 

The center-right panel of Figure~\ref{fig:ismconditions} shows a histogram of $\umw$ for the gas cells in TNG100 galaxies. The UV field follows a roughly linear relation with density on average, $\umw \approx 10\ n_{\rm H} / (1 {\rm cm}^{-3})$, but with large scatter. The scatter is expected from higher-resolution simulations that resolve the ISM structure better \citep[see, e.g., Figure 11 in][]{hu_17}. Compared to the model based only on local star formation and the UV background (dotted line in Figure~\ref{fig:ismconditions}), our optically thin propagation transports UV radiation to the lower-density parts of galaxies. Comparing to the dotted lines in the bottom panels of Figure~\ref{fig:ismconditions}, it is clear that this radiation suppresses \htwo formation in those regions. Furthermore, our modeling avoids the large unphysical jump in $\fmol$ that results from the discontinuity in the UV field at the star formation threshold. However, a smaller discontinuity remains, due to the discontinuous nature of other quantities such as the estimated neutral surface density. This discontinuity is removed in the projection-based models because a threshold in the volumetric density does not translate into a threshold in surface density. We discuss the uncertainties of our optically thin UV modeling and the escape fraction in Section~\ref{sec:discussion:fesc}.

\subsection{Models for the ${\rm H}\,${\sc i}\ /\ ${\rm H}_2$ Transition}
\label{sec:methods:models}

Of the numerous models for the molecular fraction that have been proposed, we consider those of \citet[][\modell]{leroy_08}, \citet[][\modelgk]{gnedin_11}, \citet[][\modelgd]{gnedin_14}, \citet[][\modelk]{krumholz_13}, and \citet[][\models]{sternberg_14}. As described in Section~\ref{sec:methods:volmap}, we compute each model on a cell-by-cell basis and in projection.

\subsubsection{Observed Correlations (L08)}
\label{sec:methods:models:l08}

The ratio of molecular to atomic gas has been observed to correlate strongly with the midplane pressure \citep{wong_02, blitz_04, blitz_06, leroy_08}:
\begin{equation}
\label{eq:l08}
\rmol = \left( \frac{\peff}{P_0} \right)^\alpha \,,
\end{equation}
where we use the \citet{leroy_08} values of $P_0/k_{\rm B} = 1.7 \times 10^4\ {\rm K}/{\rm cm}^3$ and $\alpha = 0.8$ (their Equation 34). This model depends only on hydrodynamical and stellar quantities but not on the UV field or metallicity.

For the cell-by-cell computation, we follow \citet{marinacci_17} in setting $\peff \equiv P_{\rm thermal} = u \rho (\gamma - 1)$. For star-forming cells, \citet{marinacci_17} used the partial pressure of the cold phase; instead, we use the total pressure. However, Equation (\ref{eq:l08}) actually refers to the midplane pressure of a disk in hydrostatic equilibrium \citep{elmegreen_89},
\begin{equation}
\label{eq:peff}
\peff \approx \frac{\pi}{2}G \sigmag^2 \left( 1 + \frac{\sigma_{\rm gas,z}}{\sigma_{*,{\rm z}}} \frac{\Sigma_*}{\sigmag}  \right) \,,
\end{equation}
where $\sigma_{\rm gas,z}$ and $\sigma_{*,{\rm z}}$ are the vertical velocity dispersions of gas and stars, respectively. Thus, for the projection-based version of the \modell model, we compute $\peff$ from surface maps of the relevant quantities, as described in Appendix~\ref{sec:app:models:l08}. 

We find that the mass-weighted average thermal pressure maps significantly differ from $\peff$. Because the thermal pressure tends to be lower in high-density regions, the volumetric \modell model predicts systematically lower molecular fractions than the projected version (Figure~\ref{fig:galaxy} and Section~\ref{sec:results:models}). This mismatch is not surprising because the resolution of IllustrisTNG is not sufficient to resolve the disk scale height, meaning that the pressure cannot be expected to match the true midplane pressure. In high-resolution zoom-in simulations, the pressure should represent the actual hydrostatic equilibrium, which explains why \citet{marinacci_17} found that the \modell and \modelgk models agree reasonably well when applied to the AURIGA simulations.

In the context of this work, we conclude that the $\peff = p_{\rm thermal}$ assumption does not hold, meaning that the volumetric \modell model is not physical and that its predictions should not be taken at face value. We thus omit the volumetric \modell model from most figures and conclusions.

\subsubsection{Calibration from Simulations (GK11 and GD14)}
\label{sec:methods:models:gk11}

\begin{figure*}
\centering
\includegraphics[trim = 3mm 1mm 23mm 1mm, clip, scale=0.7]{\figdir/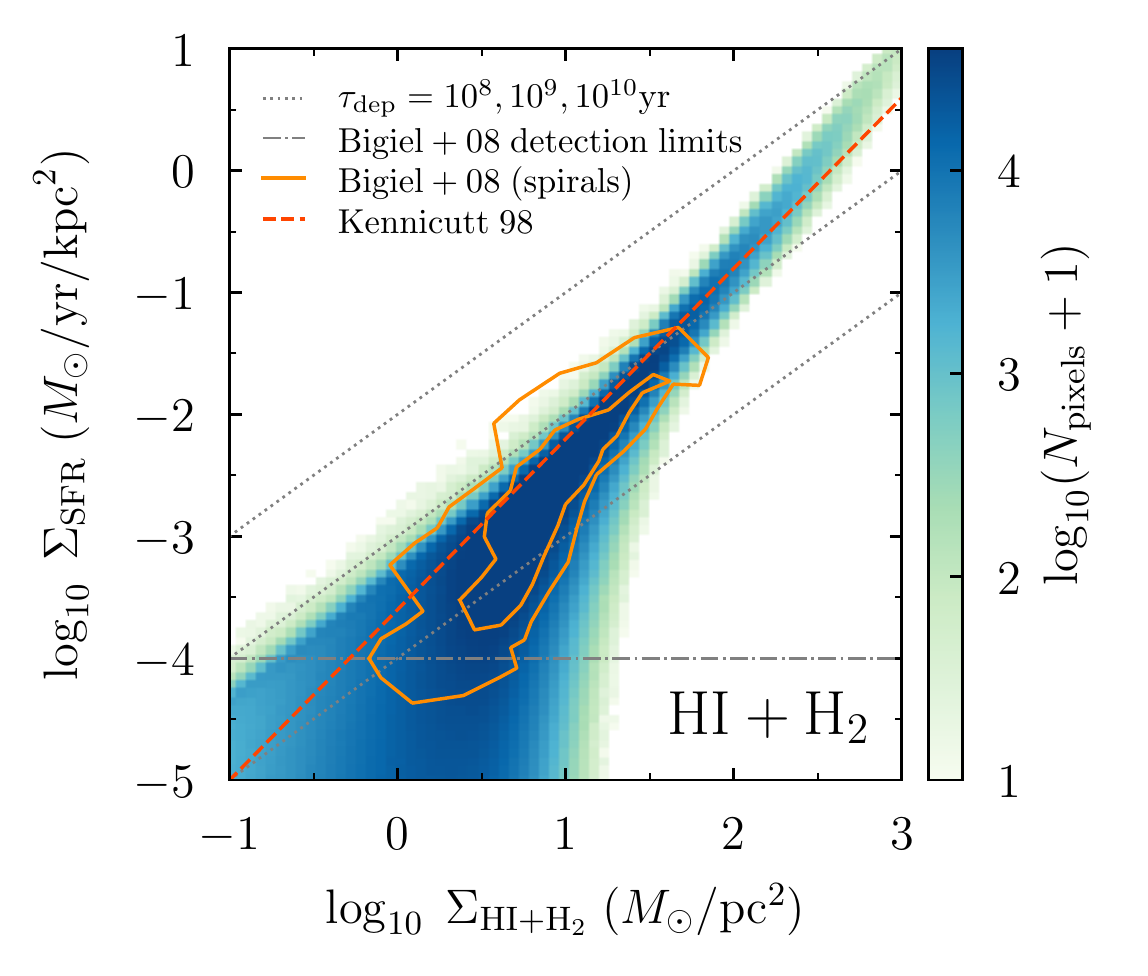}
\includegraphics[trim = 21mm 1mm 23mm 1mm, clip, scale=0.7]{\figdir/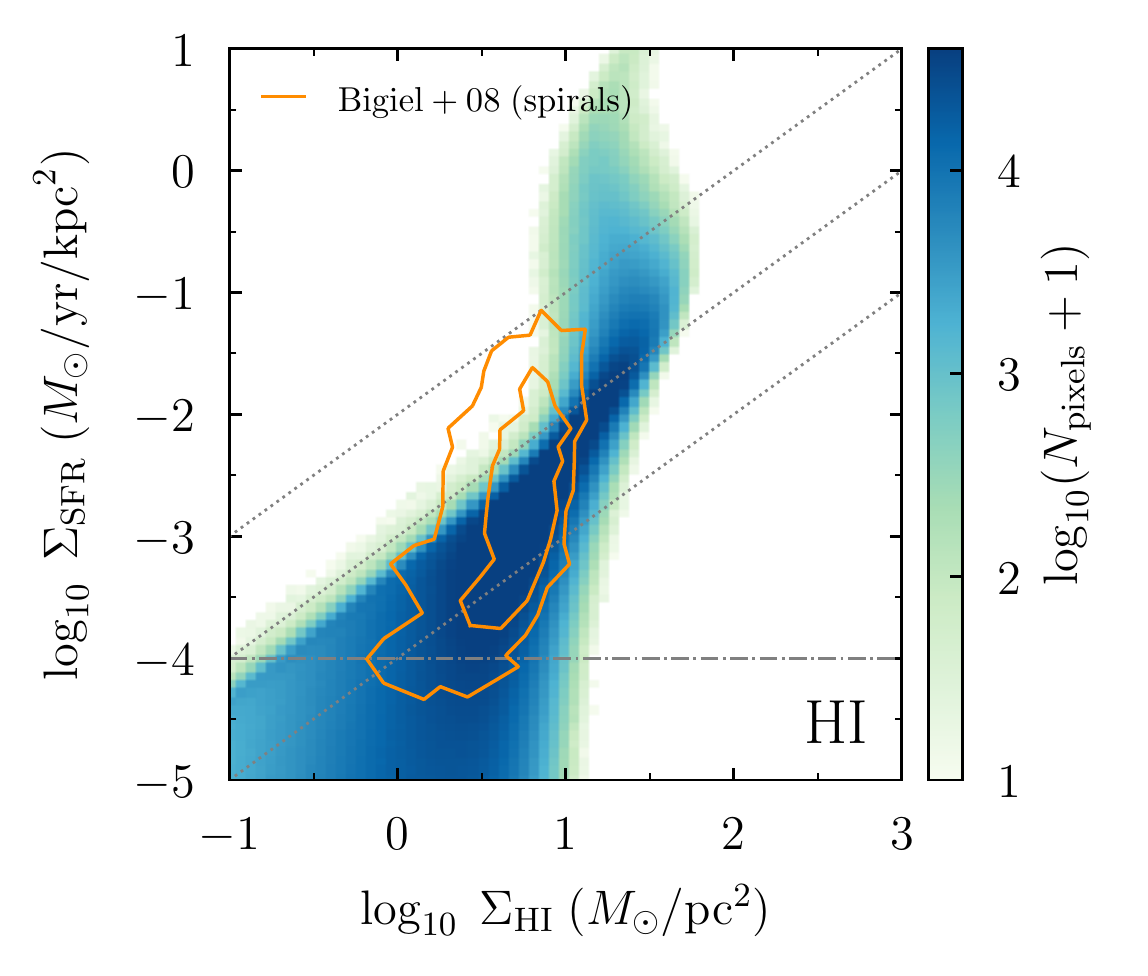}
\includegraphics[trim = 21mm 1mm 3mm 1mm, clip, scale=0.7]{\figdir/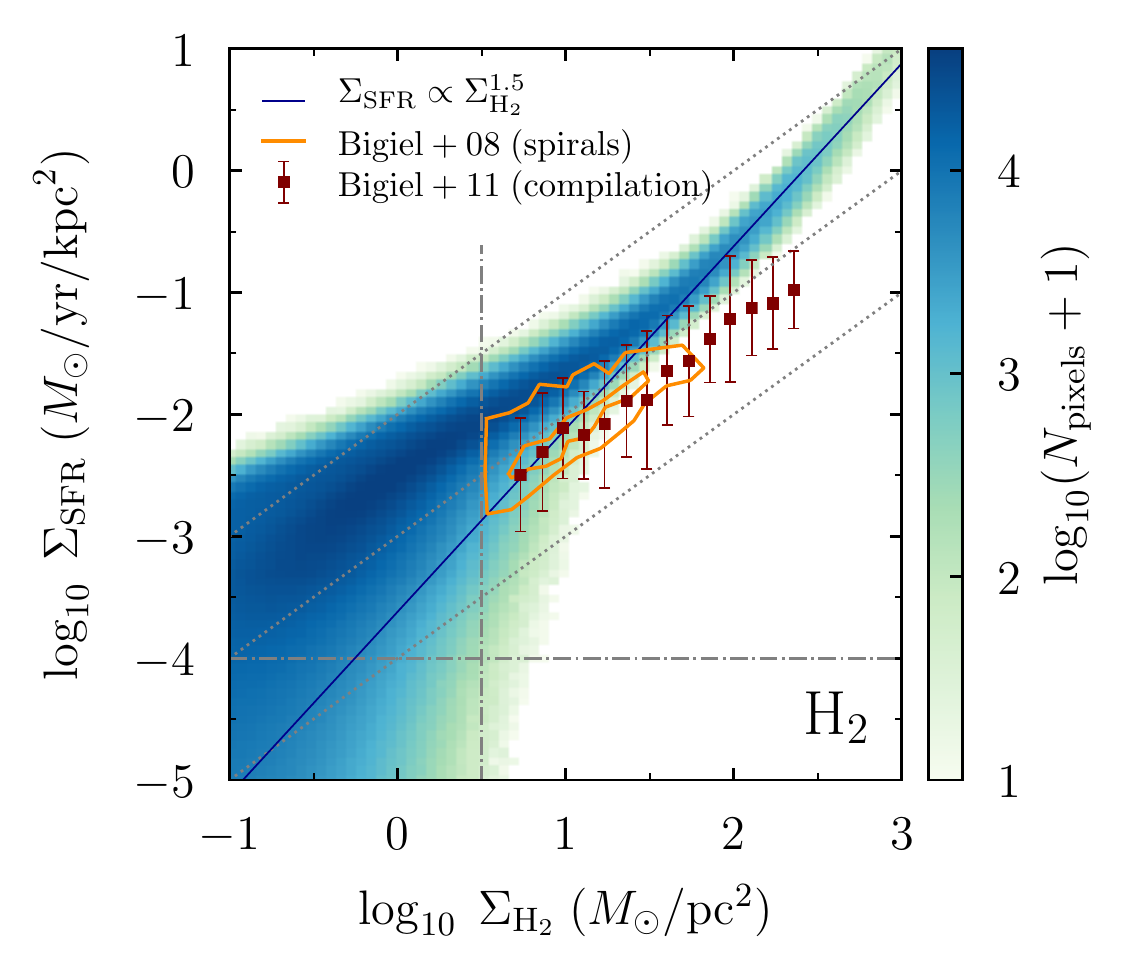}
\caption{Kennicutt--Schmidt relation for neutral (left), atomic (center), and molecular (right) gas. The histograms show the distribution of pixels from the projected maps of TNG100 galaxies. In each panel, the dotted lines correspond to constant depletion times of $\tau_{\rm dep} = \sigmag / \sigmasfr$. The orange contours roughly indicate the location of the spatially resolved observations of seven spiral galaxies of \citet{bigiel_08}, and the horizontal and vertical dot-dashed lines mark their sensitivity limits for star formation and \htwo surface density. {\it Left panel:} The total gas KS relation in IllustrisTNG is independent of the \hiht models and roughly follows the \citet{kennicutt_98} relation (red dashed line) at high densities, largely by construction. Around a surface density of $1\ \msun/{\rm pc}^2$, the relation steepens and falls below the \citet{kennicutt_98} relation, in agreement with observations. {\it Center panel:} the distribution of $\sigmahi$ according to the projected \modelgd model. The other models give very similar trends. The relation between $\sigmahi$ and $\sigmasfr$ is less pronounced than for $\sigman$ and $\sigmaht$, in agreement with \citet{bigiel_08}. {\it Right panel}: the molecular KS relation is not quite obeyed by the simulation. The \citet{bigiel_08, bigiel_11} data exhibit a roughly constant depletion time, whereas the simulation results are better described by a superlinear relation with a slope of $1.5$ (dark blue line). See Section~\ref{sec:results:ks} for a detailed discussion.}
\label{fig:ks}
\end{figure*}

The \modelgk model is based on 35 high-resolution simulations of isolated disk galaxies that follow the detailed chemical evolution of the gas \citep[see also][]{gnedin_09}. Each simulation is initialized with a particular value of the dust-to-gas ratio and the interstellar radiation field. \modelgk present a fitting function for the resulting molecular fraction averaged over a scale of $500$ pc. This function takes $\umw$, metallicity, and the surface density of neutral gas as input parameters (Appendix~\ref{sec:app:models:gk11}). 

Here, $\sigman$ is computed based on the Jeans approximation in the volumetric version of the model, and directly in the projected version. Our pixel sizes of roughly a kiloparsec are similar to the $500$ pc AMR cells that \modelgk used to calibrate their formulae. 

The \modelgd model builds on the \modelgk model by considering the effects of line overlap. The model is based on the same quantities as \modelgk, but its mathematical expressions are somewhat different and explicitly take resolution into account (Appendix~\ref{sec:app:models:gd14}). We expect the \modelgk and \modelgd models to predict similar molecular fractions.

\subsubsection{Analytic Models (K13 and S14)}
\label{sec:methods:models:k13}

The \modelk model represents an extension of the model of \citealt{krumholz_09_kmt1} (hereafter \citetalias{krumholz_09_kmt1}; see also \citealt{krumholz_09_kmt2} and \citealt{mckee_10}), who developed a formula for $\fmol$ as a function of the neutral gas surface density and metallicity. Their model, however, predicts that $\fmol$ rapidly approaches zero below some critical, $Z$-dependent surface density. This prediction is inconsistent with observations and with the requirement that the ISM maintain vertical hydrostatic equilibrium. Following \citet{ostriker_10}, \modelk added the latter condition, providing an iterative model that computes $\fmol$ and the surface density of star formation as a function of $\sigman$ and metallicity. Instead of computing $\sigmasfr$, we impose the UV field externally and iterate to find $\fmol$ as a function of $\sigman$, $\umw$, and $Z$. 

The mathematical expressions and details of our algorithm are given in Appendix~\ref{sec:app:models:k13}. In the projected version of this model, we proceed similarly to the \modelgk model, meaning that we compute all surface densities directly and apply the model equations to the 2D maps. We expect that the \modelk and \modelgk models should agree relatively well \citep{krumholz_11_comparison}.

The model of \models is similar in spirit to the \citetalias{krumholz_09_kmt1} prescription but improves on some aspects of the physical modeling. Instead of approximating $\fmol$ directly, the \models model computes the column density of \hi as a function of $\sigman$, $\umw$, and $Z$. This \hi column can then be subtracted from the total column density to find the molecular column density. We further discuss this model in Appendix~\ref{sec:app:models:s14}.


\section{Results}
\label{sec:results}

In this section, we evaluate the predictions of the various \hiht models. We begin by checking whether the well-known relation between surface density and star formation is obeyed on a pixel-by-pixel basis. We compare the model results on a galaxy-by-galaxy basis and in summary statistics such as gas fractions and mass functions. Eventually, we consider the spatial distribution of molecular gas within galaxies and discuss potential changes at high redshift.

\subsection{The KS Relation}
\label{sec:results:ks}

\begin{figure*}
\centering
\includegraphics[trim = 8mm 6mm 6mm 0mm, clip, scale=0.62]{\figdir/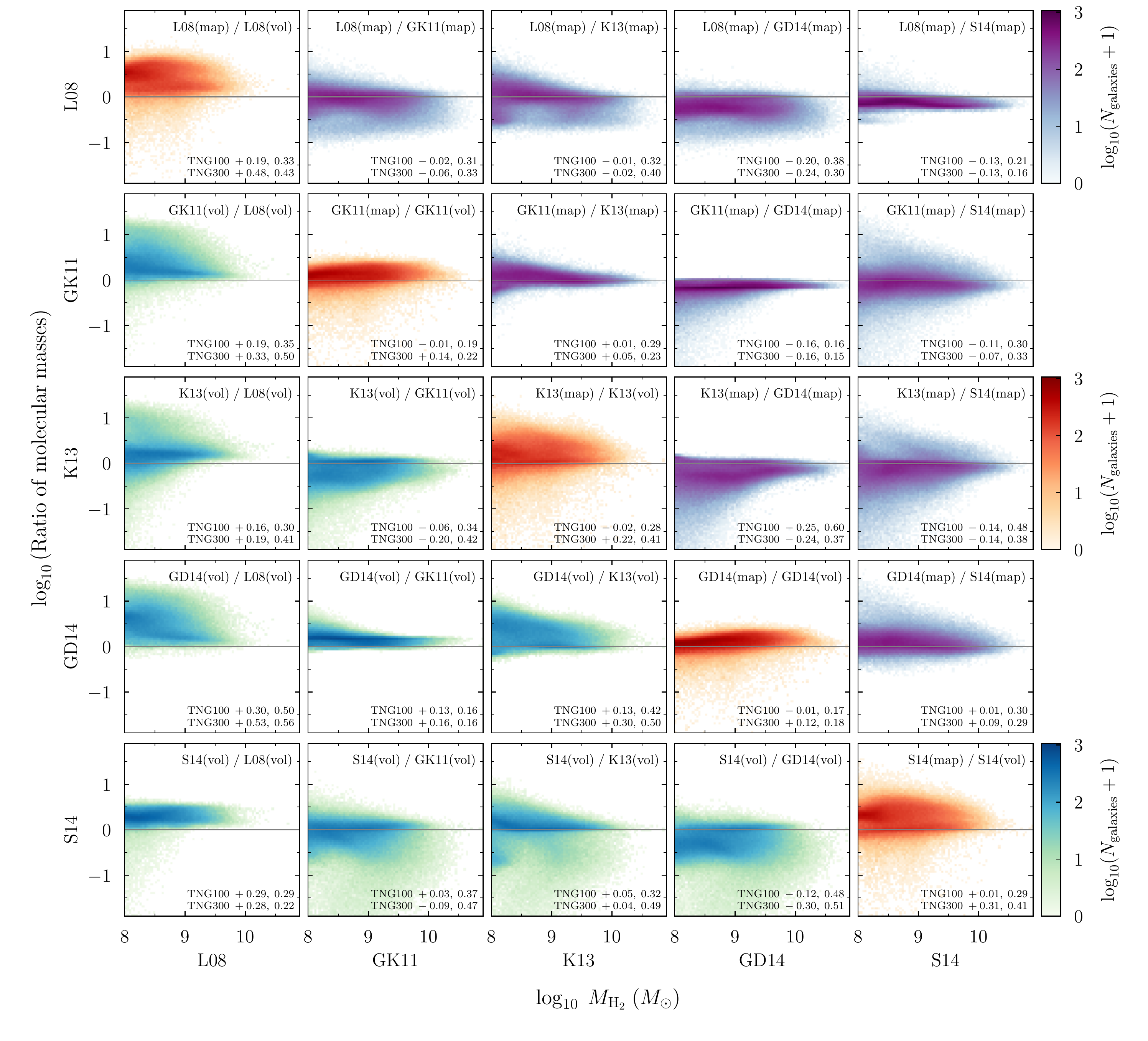}
\caption{Galaxy-by-galaxy comparison of the \htwo masses predicted by the models studied in this paper. Each panel shows a histogram of TNG100 and TNG300 galaxies, with equal weight on each of the two simulations. The bottom left (blue) panels compare the \htwo masses predicted by the volumetric versions of the models, the top right (purple) panels the projection-based versions, and the diagonal (red) panels compare the volumetric and projection results for each model. The numbers at the bottom right of each panel give the median offset in dex as well as the difference between the 84th and 16th percentiles (i.e., the 68\% scatter in dex). Those numbers are given for TNG100 and TNG300 separately and refer only to the galaxies above the threshold of $\mht > 10^8 \msun$ according to the model plotted on the $x$ axis. The volumetric and projection-based models tend to give similar predictions on average (especially in TNG100) but exhibit significant galaxy-to-galaxy scatter, about $0.2$--$0.4$ dex. The exception is the \modell model (left column), where the projection-based model predicts systematically more \htwo because the volumetric version is not physical (Section~\ref{sec:methods:models:l08}). See Section~\ref{sec:results:models} for a detailed discussion.}
\label{fig:modelcomp}
\end{figure*}

\begin{figure*}
\centering
\includegraphics[trim = 2mm 0mm 1mm 0mm, clip, scale=0.7]{\figdir/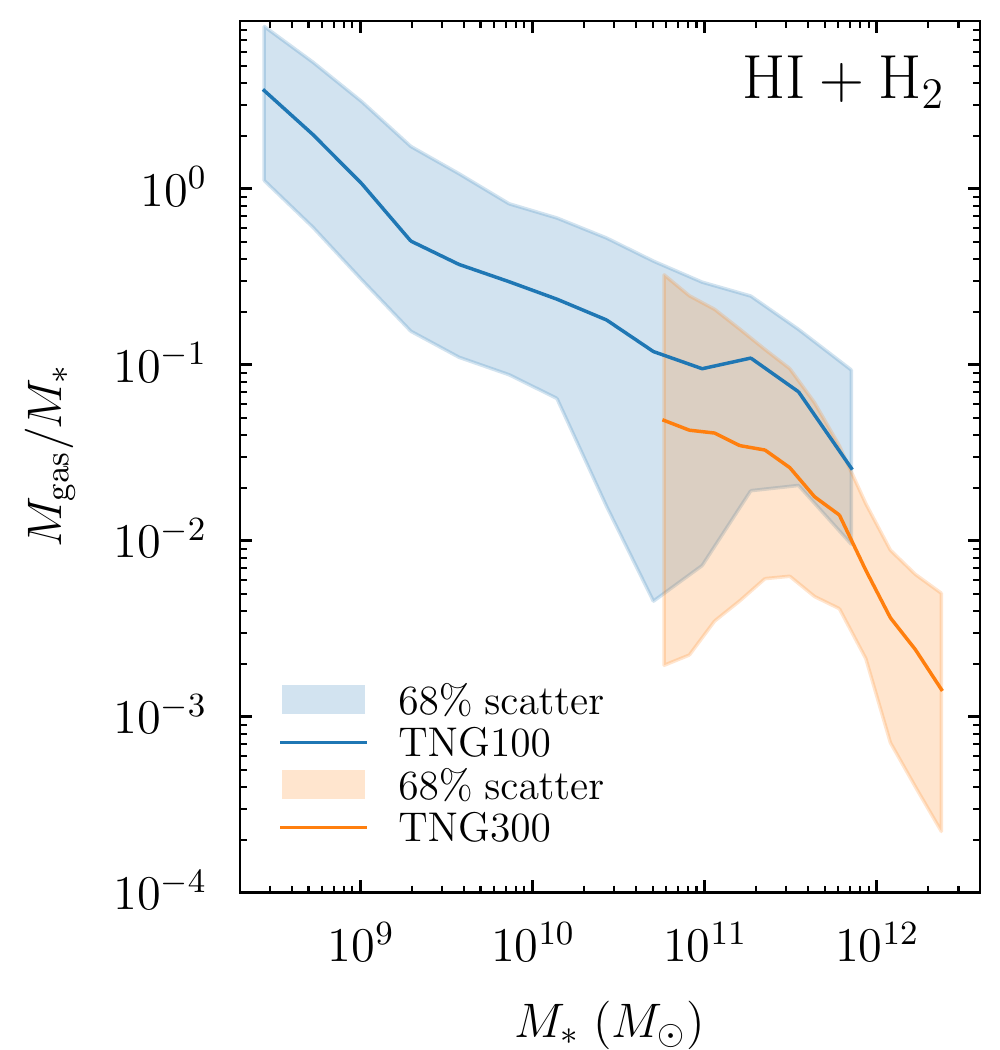}
\includegraphics[trim = 24mm 0mm 1mm 0mm, clip, scale=0.7]{\figdir/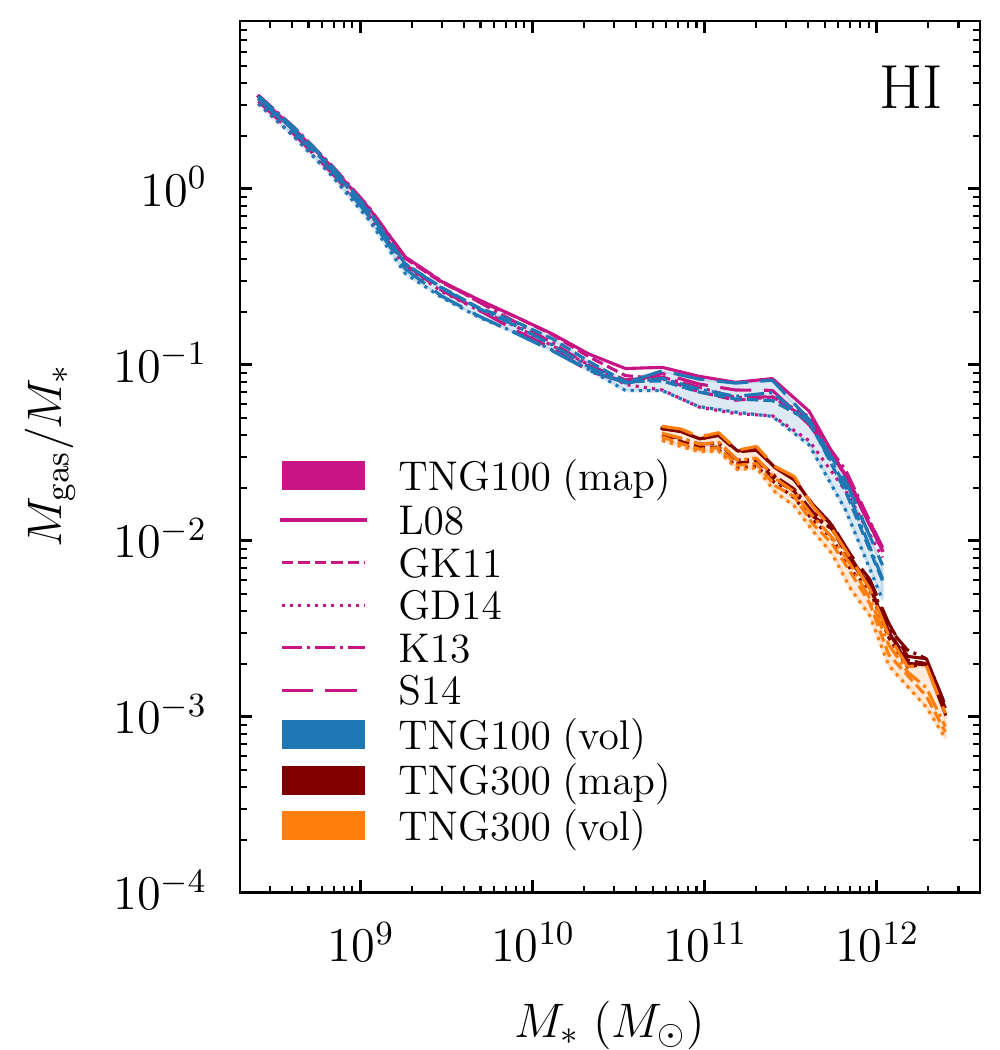}
\includegraphics[trim = 24mm 0mm 1mm 0mm, clip, scale=0.7]{\figdir/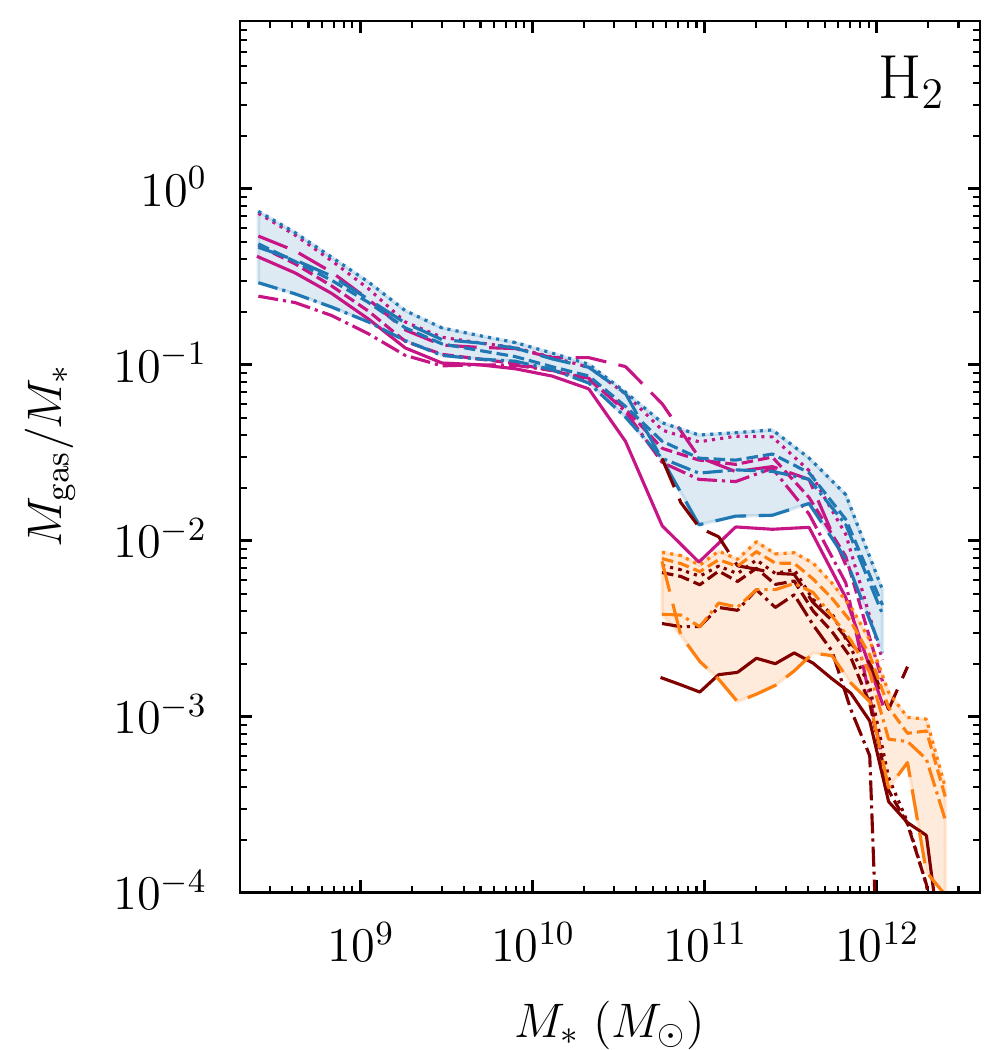}
\caption{Gas-to-stellar mass ratio for neutral (left), atomic (center), and molecular gas (right). The total neutral gas fraction is shown as a median (solid lines) and 68\% scatter (shaded areas). The gas fraction is a factor of two to three lower in TNG300 at fixed stellar mass, indicating that this quantity is not converged with mass resolution. In the center and right panels, the blue and orange lines show the median predictions of the volumetric versions of the \hiht models, and the pink and brown lines show the projected versions. The area between the volumetric models is shaded to indicate the range of their predictions (the shading does not indicate scatter). For TNG100, the molecular masses agree to about $0.4$ dex in most mass bins, with differences up to one dex. The agreement is worse in TNG300, with a typical spread of $0.8$ dex. Because molecular gas is subdominant at $z = 0$, these differences translate into smaller spreads in the \hi mass, typically $0.1$ dex in TNG100 and $0.15$ dex in TNG300, with a maximum of $0.3$ dex. The gas fractions include all gas and stellar mass bound to the galaxy and are thus not suitable for comparisons to observations that correspond to a particular aperture or sensitivity limit.}
\label{fig:fracs}
\end{figure*}

The KS relation \citep{schmidt_59, kennicutt_98} is a correlation between the surface densities of gas and star formation. As our \hiht models do not explicitly tie $\fmol$ to the SFR, this relation provides a basic test of how our modeling assigns molecular and atomic gas in relation to star formation. We perform this comparison in a spatially resolved manner, i.e., by considering the pixels in the projected maps of our galaxies. We could also average the surface densities over entire galaxies, but those quantities would be subject to the definition of the size of a galaxy.

Figure~\ref{fig:ks} shows histograms of $\sigman$, $\sigmahi$, and $\sigmaht$ versus $\sigmasfr$ for all pixels in the maps of all galaxies in TNG100 (a large fraction of pixels lie outside the plotted range). This stack is dominated by relatively small galaxies, but we have confirmed that the plots do not change systematically if only galaxies with Milky Way-like stellar masses are considered. We are showing $\sigmahi$ and $\sigmaht$ according to the projected \modelgd model, but the other models produce similar results.

The left panel of Figure~\ref{fig:ks} shows the relation between $\sigman$ and $\sigmasfr$. The red line shows the \citet{kennicutt_98} relation with an exponent of $N = 1.4$, and orange contours roughly indicate the spatially resolved observations of \citet{bigiel_08}. Their contours refer to resolution elements of a size of about $0.75$ kpc, well matched with the median pixel size for TNG100 (Section~\ref{sec:methods:volmap}). \citet{bigiel_08} cite lower limits on the detection of star formation surface density (roughly $10^{-4} \msun/{\rm yr}/{\rm kpc}^2$) and molecular surface density ($10^{0.5} \msun/{\rm pc}^2$), which are shown as dot-dashed gray lines in Figure~\ref{fig:ks}.

Though not explicitly enforced in the simulation, we expect IllustrisTNG to obey a relation similar to that of \citet{kennicutt_98} because the star formation law was designed to match such observations. In particular, the SFR is proportional to $\rho^{1.5}$ in the \citetalias{springel_03} model (Section~\ref{sec:methods:illustris:ism}). While the projection onto surface densities could change this slope, roughly constant scale heights will naturally lead to a similar relation between the respective surface densities. Nevertheless, the simulation also matches the observed turnoff from the relation around $\sigman \approx 1 \msun/{\rm pc}^2$. 

The center panel of Figure~\ref{fig:ks} shows the relation between $\sigmahi$ and $\sigmasfr$. Our simulation results qualitatively match the \citet{bigiel_08} contours. In particular, the observed ``saturation'' of \hi at $\sigmahi \approx 1 \msun/{\rm pc}^2$ is obeyed, presumably because the \hiht models were designed to match the observationally known transition to molecular gas at this surface density. 

Finally, the right panel of Figure~\ref{fig:ks} shows the KS relation for molecular gas. The IllustrisTNG data match the observed contours well at intermediate densities around $\sigmaht \approx 10 \msun/{\rm pc}^2$. At higher densities, however, the simulation results are well described by a steep relation, $\sigmasfr \propto \sigmaht^{1.5}$ (dark blue line). In contrast, observations favor a shallower relation with a more or less constant depletion time, as shown by the contours and literature compilation \citep{bigiel_11, schruba_11, leroy_13, leroy_17, bolatto_17}. We note that the \hiht modeling cannot be blamed for this difference: at surface densities above $\sigmaht \approx 1 \msun/{\rm pc}^2$, virtually all hydrogen is molecular, meaning that no change in the models could shift the IllustrisTNG data to the higher surface densities required to match the observations. At lower $\sigmaht$, the scatter increases, but it is clear that there is a population of pixels with a detectable $\sigmaht \approx 10^{0.5} \msun/{\rm pc}^2$ and very little or no star formation. High-resolution simulations with \htwo physics have also found that \htwo is not always correlated with star formation \citep[][see also \citealt{krumholz_11_h2sfr} and \citealt{glover_12}]{pelupessy_06, hu_16}. 

Of course, the comparisons in Figure~\ref{fig:ks} are crude because we did not take observational systematics into account. For example, the observational star-formation indicators used by \citet{bigiel_08}, FUV and $24\ \mu$m fluxes, are sensitive to star formation on a $0.1$--$1$ Gyr timescale, whereas we consider instantaneous SFRs. Moreover, the \htwo depletion time has been shown to depend on a number of variables, including galaxy stellar mass and the dynamical state of the star-forming gas \citep{saintonge_11_depletion, leroy_17}. Similarly, \citet{schruba_18} find that the \hi saturation density depends on metallicity. Finally, there are hints of a steeper KS relation at high redshift \citep{hodge_15}. These effects complicate the interpretation of the results shown in Figure~\ref{fig:ks}. Nevertheless, we conclude that our \hiht models give reasonable results for the surface densities of \hi and \htwo and for their relation to the star-formation surface density.

\subsection{Galaxy-by-galaxy Comparison of Model Results}
\label{sec:results:models}

We now assess how well the \hiht models agree with each other by considering the integrated molecular masses of individual galaxies. Figure~\ref{fig:modelcomp} shows histograms of the ratio of \htwo masses predicted by each pair of models. In each panel, the numbers on the bottom right give the median ratio and scatter in dex. These values are shown separately for TNG100 and TNG300, and the two simulations are weighted equally in the histograms to account for the larger number of galaxies in TNG300. The lower left sector compares the volumetric versions of the models, the upper right sector the projected versions, and the panels along the diagonal compare the volumetric and projected version of each model.

As expected from the previous discussion, the volume-based \modell model (left column) predicts molecular masses $0.2$--$0.6$ dex lower than other models, and we will ignore it for the rest of this section. The volumetric and projected versions of each model (red histograms) agree to better than $0.02$ dex in TNG100, though much less well ($0.3$ dex) in TNG300. In both simulations, the scatter around this average is large, between $0.2$ and $0.4$ dex. We conclude that while the volumetric and projected versions agree reasonably well on average, their predictions diverge for many individual galaxies. This scatter is likely a consequence of the fact that the Jeans approximation to the scale height works on average but exhibits significant scatter (Section~\ref{sec:discussion:jeans}).

Comparing the different volumetric models, we find that they agree to $0.13$ dex in TNG100 and $0.3$ dex in TNG300. Given that the \modelgk and \modelgd models are based on the same underlying physics, it is not surprising that they exhibit a smaller scatter than the other pairs of models. Comparing the projected models, we find a large range of offsets between zero and $0.25$ dex. Interestingly, the agreement is roughly equally good in TNG100 and TNG300, indicating that the projected modeling is less sensitive to resolution effects.

We note that the scatter visibly increases toward lower \htwo masses, meaning that the values quoted in Figure~\ref{fig:modelcomp} are sensitive to the lower mass cutoff. For this reason, we will consider only galaxies with $\mht > 10^8 \msun$ when comparing mass functions in Section~\ref{sec:results:mfunc}. We have confirmed that the trends shown in Figure~\ref{fig:modelcomp} hold even when only galaxies with high stellar mass or high SFR are selected. Such cuts do not necessarily reduce the scatter between the models, indicating that the differences are not unique to low-mass galaxies.

\subsection{Gas Fractions}
\label{sec:results:fracs}

\begin{figure*}
\centering
\includegraphics[trim = 4mm 11mm 0mm 0mm, clip, scale=0.77]{\figdir/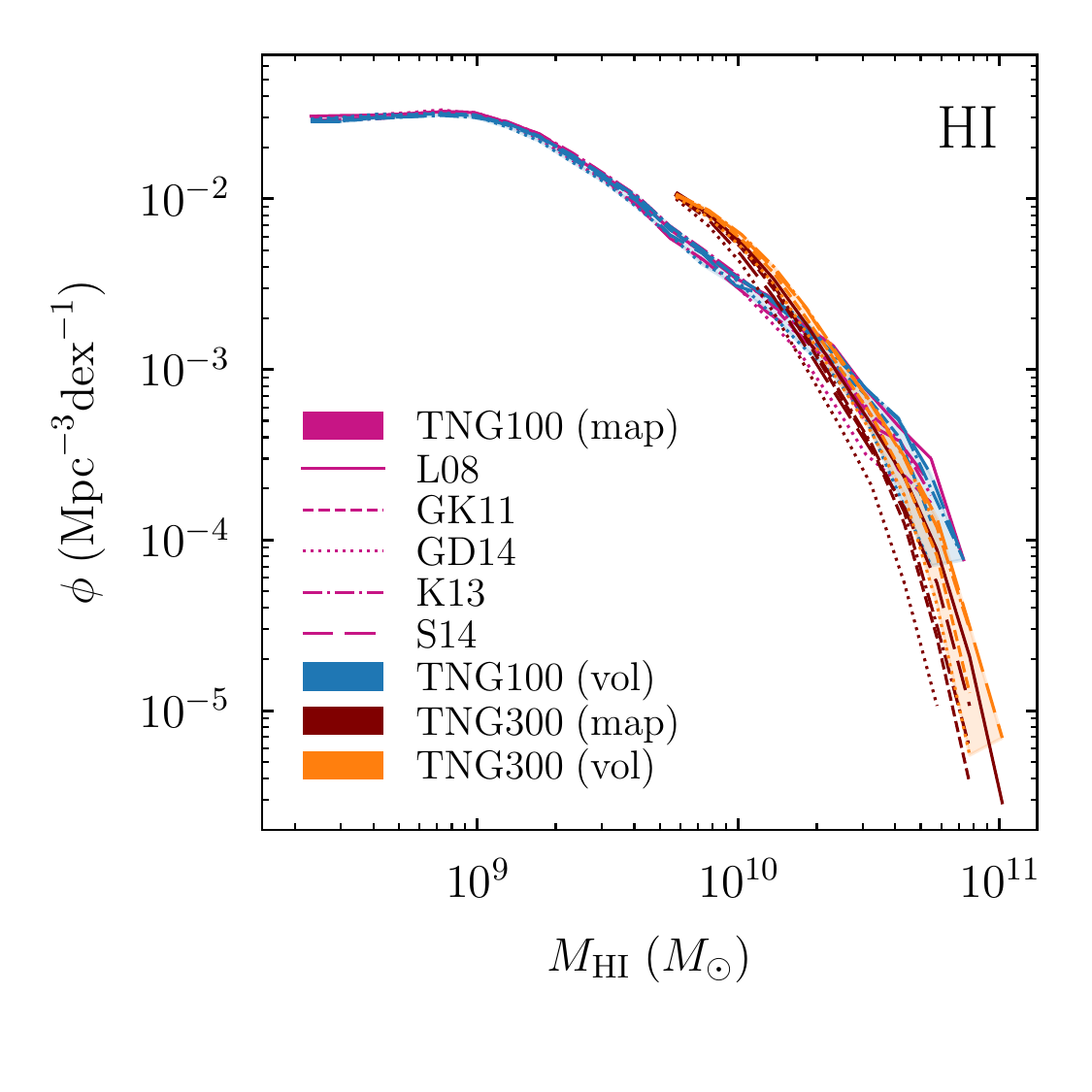}
\includegraphics[trim = 26mm 11mm 2mm 3mm, clip, scale=0.77]{\figdir/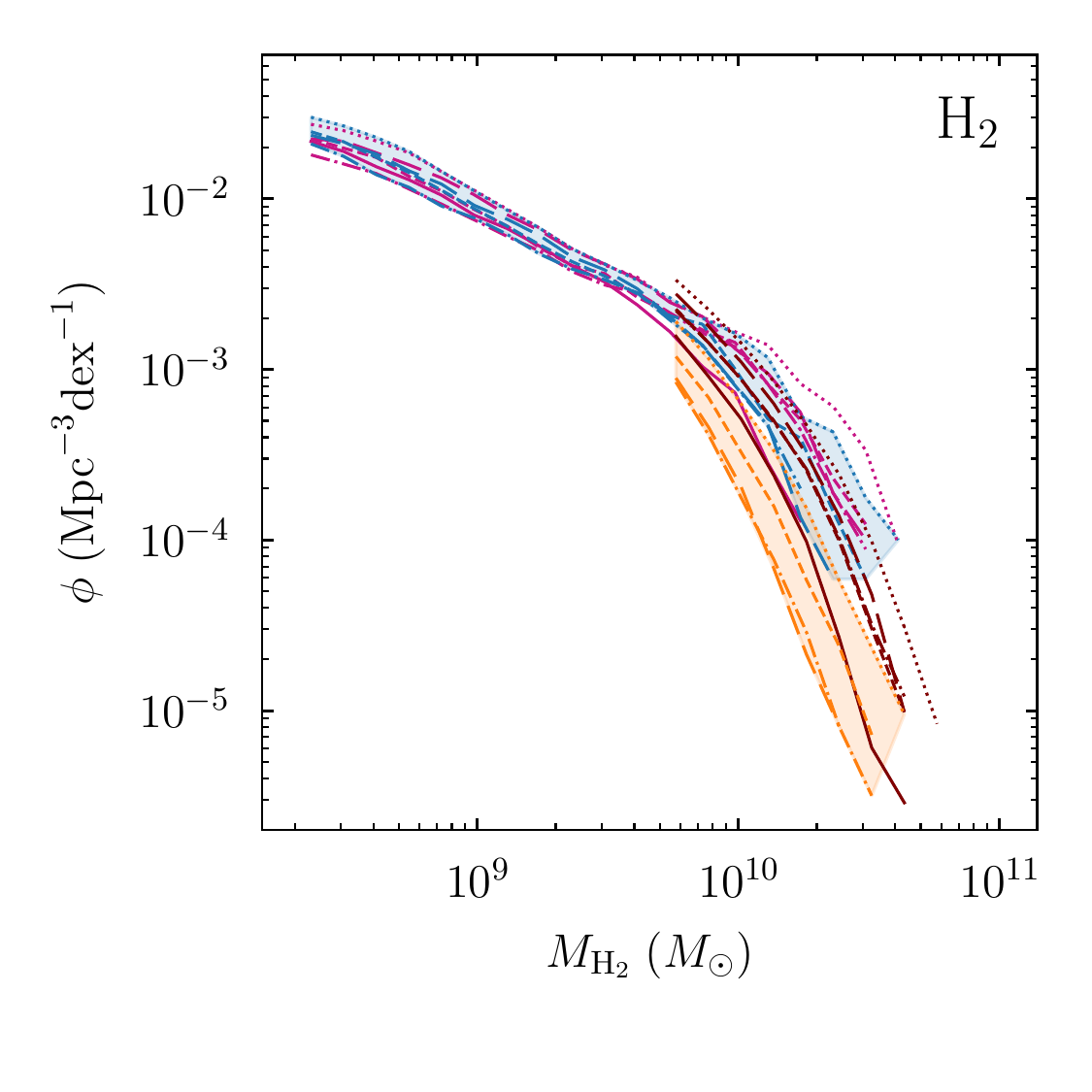}
\caption{Mass functions of \hi (left) and \htwo (right). The lines and shaded regions have the same meaning as in Figure~\ref{fig:fracs}. The different \hiht models agree very well with each other, with the exception that the volumetric versions predict a lower \htwo mass function in TNG300. Generally, the agreement between TNG100 and TNG300 appears to be better than in the gas mass fractions. As in Figure~\ref{fig:fracs}, the gas masses contain all cells bound to a galaxy and should thus not be compared to observations.}
\label{fig:mfunc}
\end{figure*}

Having assessed the differences between \hiht models for individual galaxies, we now turn to averaged quantities such as the gas fractions shown in Figure~\ref{fig:fracs}. We refrain from observational comparisons at this point because all masses correspond to the total gas or stellar mass bound to a galaxy, which is generally not what is measured observationally. Moreover, we have not made any attempt to mimic observational sample selections or account for nondetections. Our sample is complete down to the stellar masses shown because we selected galaxies by either their stellar or gas mass (Table~\ref{table:sims}).

Before analyzing the trends in \hi and \htwo, it is instructive to consider the total neutral gas fraction because it is independent of the \hiht modeling and sets a baseline for the abundance of the atomic and molecular phases. At fixed stellar mass, TNG300 galaxies contain about three times less neutral gas than TNG100 galaxies  (left panel of Figure~\ref{fig:fracs}), a resolution effect that also affects the \hi and \htwo fractions because they depend on the surface density of neutral gas. The stellar mass in TNG100 is about $1.4$ times greater than in TNG300 at fixed halo mass \citep{pillepich_18}, but we find the same disagreement in neutral gas mass when comparing at fixed halo mass.

The center and right panels of Figure~\ref{fig:fracs} shows the \hi and \htwo fractions, respectively. As expected from the average agreement between \hiht models in Figure~\ref{fig:modelcomp}, we find that the median gas fractions according to the different models agree reasonably well. In particular, we cannot discern significant overall offsets between the volumetric or projected models. The \htwo fraction in TNG100 exhibits a typical (median) range of $0.4$ dex within stellar mass bins, with a maximum of one dex. These differences lead to smaller disagreements in the \hi fraction, typically $0.1$ dex and up to $0.3$ dex. The agreement is slightly worse in TNG300. 

We conclude that the spread in the \hi and \htwo fractions due to differences between the \hiht models is relatively small compared to the systematic uncertainties in our modeling, especially for the better resolved TNG100. The \hi gas fraction is particularly well constrained, allowing detailed comparisons to observations in future work.

\subsection{Mass Functions}
\label{sec:results:mfunc}

Given that the gas fractions at fixed stellar mass agree reasonably well between the \hiht models, we might expect that the mass functions would match as well. However, we investigated only the median fractions in Figure~\ref{fig:fracs}. Large scatter in any one model could change the predictions for the mass functions significantly because they fall off steeply toward high masses.

However, Figure~\ref{fig:mfunc} shows that the \hi and \htwo mass functions agree very well between the different \hiht models, particularly for TNG100. In TNG300, we witness a surprising reversal of the trend seen in Figure~\ref{fig:fracs}: while the volumetric models predict a slightly higher median $\fmol$, they predict a slightly lower \htwo mass function at the high-mass end. This seeming contradiction highlights that the mass function depends on the full distribution of gas masses and thus represents a separate check on the modeling. Overall, the mass functions agree better between TNG100 and TNG300 than the gas fractions, particularly for the projected models.

\subsection{Spatial Distribution of \htwo}
\label{sec:results:spatial}

\begin{figure}
\centering
\includegraphics[trim = 0mm 4mm 4mm 3mm, clip, scale=0.78]{\figdir/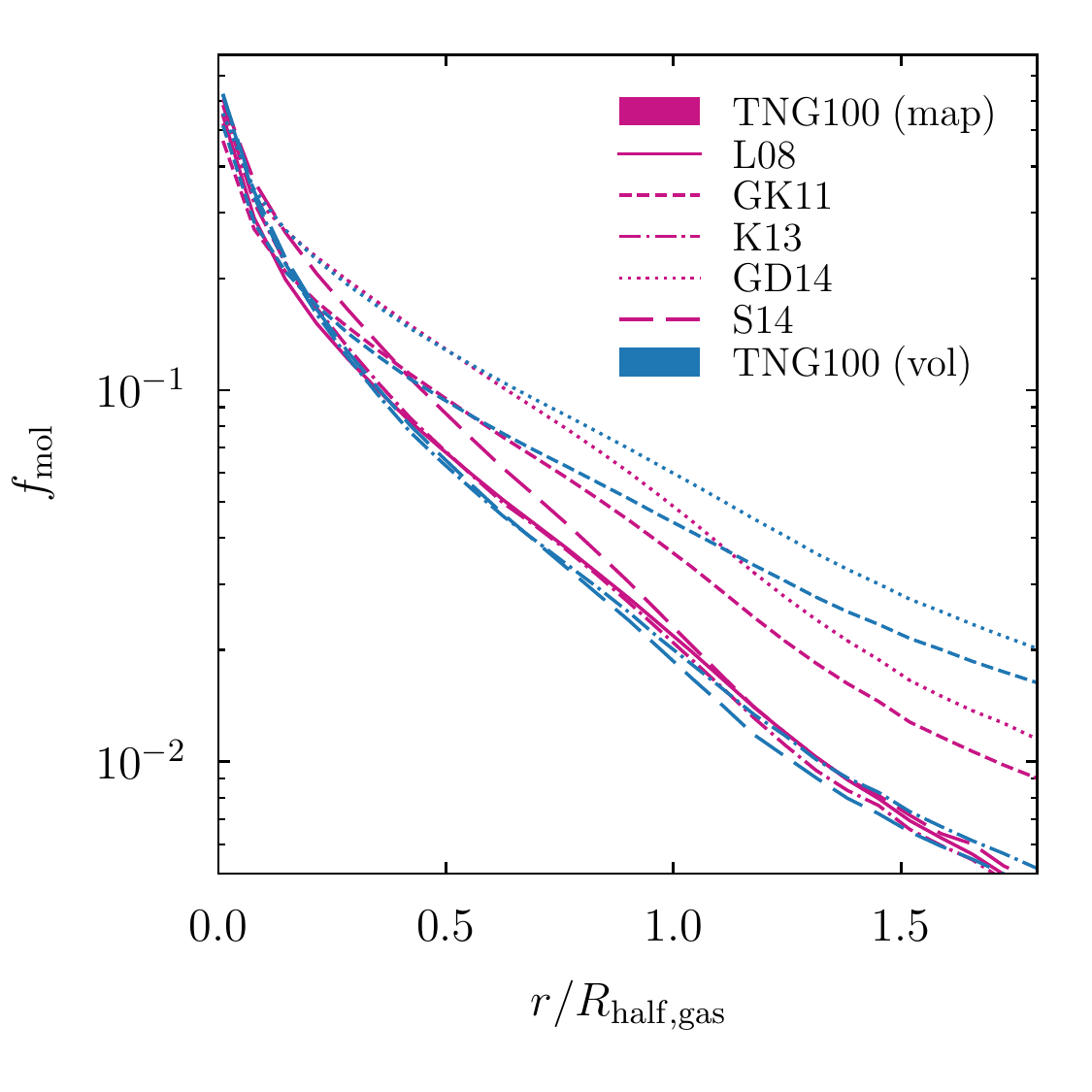}
\caption{Projected mean radial profiles of the molecular fraction for TNG100 galaxies with $M_{{\rm H}_2} > 10^8 \msun$. The line styles and colors have the same meaning as in Figures \ref{fig:fracs} and \ref{fig:mfunc}. The \modelgk and \modelgd models (short-dashed and dotted lines) predict higher molecular fractions at large radii than the other models because they allow for more molecular gas at low densities (Appendix~\ref{sec:app:models}). While the mean profiles are dominated by low-mass galaxies, the profiles for higher-mass samples lead to the same qualitative conclusions. The corresponding median profiles fall to zero between $0.3$ and $1.5$ half-mass radii.}
\label{fig:spatial}
\end{figure}

While the \hiht models generally agree on the average gas fractions and mass functions, they differ in the spatial distribution of \htwo. Figure~\ref{fig:spatial} shows mean radial profiles of $\fmol$ (without any mass weighting). We include only galaxies with $M_{{\rm H}_2} > 10^8 \msun$ according to the respective \hiht model because the profiles of molecule-poor galaxies tend to be noisy. While the stacked sample is dominated by low-mass galaxies, we have confirmed that the profiles look qualitatively the same for galaxies with high stellar mass. We scale the radii by the gas half-mass radius, excluding galaxies with a half-mass radius below 5 kpc. Both the volumetric and projected model profiles refer to projected radii in the face-on orientation.

Figure~\ref{fig:spatial} demonstrates that the mean radial profiles of the molecular fraction are remarkably similar when the \modell, \modelk, or \models models are used. In contrast, the \modelgk and \modelgd profiles fall off less steeply with radius. We show in Appendix~\ref{sec:app:models} that the latter models allow for slightly more molecular gas at lower densities, which explains the offset in the profiles.

\subsection{Results at Higher Redshifts}
\label{sec:results:redshift}

All of the results shown to this point have referred to $z = 0$. In this section, we check whether our conclusions are changed at higher redshifts, in particular $z = 2$ and $z = 4$ in TNG100. We expect some systematic trends because high-redshift galaxies tend to be richer in gas, denser, and subject to a higher UV background field.

Comparing to Figure~\ref{fig:modelcomp}, we find that the galaxy-to-galaxy scatter between the volumetric and projected versions of the models decreases significantly at high $z$, from about $0.2$--$0.3$ dex at $z = 0$ to roughly $0.14$ dex at $z = 4$, while the good average agreement remains intact. The scatter also tends to decrease between pairs of models, although there are some exceptions. The agreement between different pairs of models evolves differently. For example, the agreement between the projected \modelgk and \modelk models gets better ($0.25$ dex at $z = 0$ and $0.06$ at $z = 4$), whereas the volumetric \modelgk and \models models agree much less ($0.03$ dex at $z = 0$ and $0.45$ at $z = 4$). We defer a detailed investigation of the reasons for this behavior to future work.

Compared to Figure~\ref{fig:fracs}, both the average neutral gas fraction ($\mn / \mstar$) and molecular fraction ($\mht / \mstar$) increase by about an order of magnitude by $z = 4$, with unchanged dependencies on stellar mass. The differences between the \hiht models remain roughly the same as at $z = 0$, which translates into larger differences in the \hi fractions because molecular gas makes up for a larger fraction of the total neutral masses. These conclusions are independent of whether the projected or volumetric model versions are used.

The \htwo mass function at $z = 2$ is essentially unchanged from $z = 0$ and increases by a factor of two to five by $z = 4$. Again, the spread between the \hiht models and the agreement between volumetric and projected models remain roughly constant. The \hi mass function falls slightly with redshift, and the spread of the model increases for the same reasons as explained above.

We conclude that the scatter between the \hiht models decreases at high $z$, though some models disagree more systematically. Overall, these details have little effect on the averaged gas fractions and mass functions.


\section{Discussion}
\label{sec:discussion}

\begin{figure*}
\centering
\includegraphics[trim = 10mm 11mm 53mm 0mm, clip, scale=0.56]{\figdir/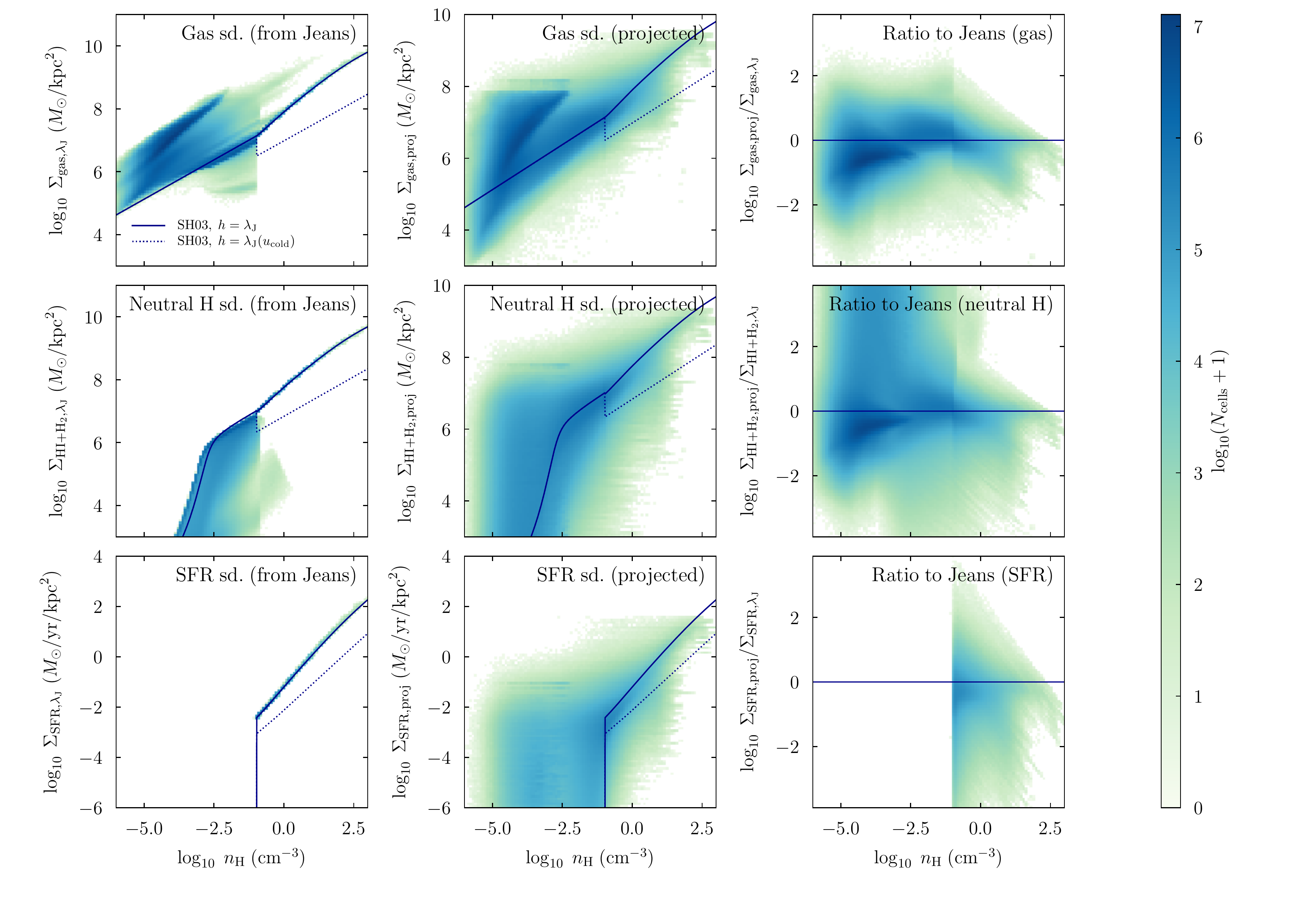}
\includegraphics[trim = 304mm 11mm 15mm 0mm, clip, scale=0.56]{\figdir/ism_conditions_file_scale_height.pdf}
\caption{Test of the Jeans approximation by comparing it to projected surface densities. The histograms in each panel show the distribution of particles in TNG100 (see Figure~\ref{fig:ismconditions}). The left column shows the surface densities estimated from the Jeans approximation ($\Sigma_{\rm X} = \lambda_{\rm J} \rho_{\rm X}$) for each gas cell, the center column shows the corresponding face-on projected surface density at the 2D position of each gas cell, and the right column shows the ratio between the two surface densities. The three rows refer to total gas density, neutral gas density, and the density of star formation. As in Figure~\ref{fig:ismconditions}, the blue lines show a calculation based on the \citetalias{springel_03} equation of state, and the dotted lines show a version that uses the Jeans length of cold gas. The Jeans approximation leads to both a bias in the average surface density of gas as well as large cell-to-cell scatter. The approximation gets worse for the neutral and SFR surface densities.}
\label{fig:scaleheight}
\end{figure*}

We have presented two avenues for modeling the molecular fraction in simulated galaxies, based on volume elements and projections. We have applied these methods to five models of the \hiht transition and shown that the models agree well in their average predictions but disagree for individual galaxies and in the spatial dependence of the molecular fraction. In this section, we further discuss the reasons for the differences between the volumetric and projected modeling and highlight some of the most important systematic uncertainties that affect our analysis.

\subsection{How Accurate Is the Jeans Approximation?}
\label{sec:discussion:jeans}

The \hiht models proposed in the literature generally depend on column densities or surface densities of various quantities. Whenever such calculations are to be performed for each gas cell, the cell's volume density must be converted to a column (or surface) density, necessitating a characteristic length scale. A common solution has been to multiply the volume density by the Jeans length, that is, to compute the surface density of a quantity $X$ as $\Sigma_{\rm X} = \lambda_{\rm J}\ \rho_{\rm X}$ \citep[e.g.,][]{lagos_15, marinacci_17}. In this section, we investigate the accuracy of this approximation specifically in the context of galaxies in cosmological simulations.

By definition, the Jeans length quantifies the size of a system in hydrostatic equilibrium (Equation~\ref{equ:jeans}). \citet{schaye_01} thus argued that it should be a good proxy for the sizes of the gas clouds responsible for the Lyman-$\alpha$ forest. This assumption was further tested when applied to the \hi column density distribution \citep{teppergarcia_12, rahmati_13}, but those results refer to random lines of sight through the universe. While \citet{schaye_08} argued that the Jeans length should also be a good proxy for the disk scale height, our simulated galaxies contain large amounts of gas outside their disks, and many objects are not disk-dominated in the first place. Moreover, the internal energy is governed by the effective equation of state of the \citetalias{springel_03} ISM model at high densities (Section~\ref{sec:methods:illustris:ism}), rendering the resulting Jeans length somewhat artificial.

To clarify the impact of these caveats, we directly test the Jeans approximation by comparing its predictions to the actual projected surface densities. We caution that we specifically choose the face-on orientation of the galaxies for the projection and include all gas bound to the subhalo (according to \textsc{Subfind}). In the context of our \hiht modeling, this comparison is appropriate because we are interested in the surface densities to which a gas cell contributes. These surface densities are, in turn, used to compute the molecular fraction of the gas cell. We emphasize that this comparison represents a special case that may not apply in other contexts. 

The top row of Figure~\ref{fig:scaleheight} shows histograms of the distribution of the same gas cells in TNG100 as in Figure~\ref{fig:ismconditions}, about $1.2$ billion cells. Mass-weighted histograms would appear very similar because the range of gas cell masses is, by design, narrow in \textsc{Arepo} simulations. The panels show the surface density of gas derived using the Jeans approximation, the projected face-on surface density at the 2D position of the center of the gas cell, and the ratio of the two surface densities. The solid blue lines represent the Jeans length according to the effective equation of state used in IllustrisTNG (as in Figure~\ref{fig:ismconditions}). At high densities, the Jeans length (and thus the surface density) is a function of only density, whereas the projected surface densities exhibit scatter. On average, the Jeans approximation predicts surface densities $0.4$ dex (a factor of $2.6$) lower than the projection, with a scatter of $0.4$ dex. The scatter increases toward low densities.

The second and third rows show the same comparison for the surface densities of neutral hydrogen and star formation, respectively. These quantities are more relevant for this work because they are used in computing the molecular fraction. The Jeans approximation performs worse than for the total gas density, with scatters of $1$ and $0.6$ dex, respectively. This change is not surprising: the Jeans length does not depend on the neutral fraction or SFR of a cell and is thus insensitive to how these quantities evolve with density. The scatter shown in the right column of Figure~\ref{fig:scaleheight} largely explains why the volumetric and projection-based versions of the \hiht models tend to perform similarly on average but disagree for a given galaxy.

A particular problem arises for star-forming cells. Here, the internal energy (and temperature) are not physical because they correspond to an average of the hot ISM and star-forming gas \citepalias{springel_03}. The Jeans length corresponding to this temperature does not correspond to a scale on which gas would actually collapse. One way to attempt to remedy this issue is to use the properties of the cold component, $u_{\rm cold} = 1000$ K and $\rho_{\rm cold}$ \citep[e.g.,][]{marinacci_17}. The change in density is not significant because $\rho_{\rm cold} / \rho \approx 1$ in the \citetalias{springel_03} model, but the lower temperature reduces the Jeans length and thus the estimated surface densities. This assumption appears to provide a slightly worse match to the projected surface densities (dotted lines in the center column of Figure~\ref{fig:scaleheight}). Thus, we use the Jeans length based on the effective temperature.

We emphasize that our evaluation of the Jeans approximation corresponds to a particular scenario where we are interested in projected surface densities. Even for a perfectly spherical ball of gas in hydrostatic equilibrium, the geometry would lead to some scatter around the projected surface density. However, our test does demonstrate that the surface densities predicted using the Jeans approximation are off by many orders of magnitude for a significant fraction of gas cells. This issue is exacerbated at low densities and when estimating quantities other than total surface density.

\subsection{Uncertainties in the UV Flux}
\label{sec:discussion:fesc}

We estimate the LW-band UV flux by assuming a constant escape fraction $f_{\rm esc}$ from the local star-forming region and optically thin propagation through the galaxy (Section~\ref{sec:methods:uv} and Appendix~\ref{sec:app:uv}). Thus, $f_{\rm esc}$ is the only free parameter in our modeling and corresponds to an overall normalization of the UV field that impacts our results significantly (Appendix~\ref{sec:app:uv:convergence}). In this section, we discuss our assumptions and attempt to calculate rough estimates for $f_{\rm esc}$.
 
Unfortunately, the escape fraction is hard to simulate because of the unknown small-scale density structure, the uncertain behavior of dust, and absorption below the resolution level. Let us illustrate these difficulties with a simple calculation. For example, we could assume that the radiation from young stars is attenuated by the dust contained in a certain column density of hydrogen $N_{\rm H}$ such that the optical depth becomes
\begin{equation}
\tau = \frac{1}{2}\ N_{\rm H}\ \sigma_{H,1000}\ \frac{Z}{Z_{\odot}}
\end{equation}
where $\sigma_{H,1000}$ is the cross section for absorption. At solar metallicity, $\sigma_{H,1000} = 1.4 \times 10^{-21} {\rm cm}^2$ per atom \citep{draine_03a, draine_03b}. We could estimate the column density in several ways: for example, we could assume that all stars are born in molecular clouds with a roughly constant column density of $N_{\rm H} \approx 2 \times 10^{22}/{\rm cm}^2$ \citep{larson_81}. However, the corresponding optical depth would be $\tau \approx 14\ Z/Z_{\odot}$, resulting in a negligible escape fraction wherever $Z/Z_{\odot} \gsim 0.1$. Clearly, this scenario is not realized in nature; for example, we would not observe any $H_{\alpha}$ flux either. In reality, star-forming clouds are optically thick initially but can be quickly dispersed by photoionization from young stars \citep[e.g.,][]{dale_12}. 

Alternatively, we could assume that the bulk attenuation happens as photons traverse the lower-density regions of the ISM and derive the optical depth from the disk scale height. This approximation leads to much higher escape fractions but is still almost entirely dictated by the metallicity. While the effect of metallicity is, of course, real, it does not take into account complex geometries where certain lines of sight might be entirely unobscured while clouds block other sight lines \citep[e.g.,][]{narayanan_18}.

Instead, we estimate the escape fraction based on the solar neighborhood values for the UV field and the star-formation surface density. This idea follows the \citet{lagos_15} method for estimating the UV field as $\umw = \Sigma_{\rm SFR}/ \Sigma_{\rm SFR,local}$ where $\Sigma_{\rm SFR,local}$ is the the observed star-formation surface density in the solar neighborhood. This number is rather uncertain. \citet{lagos_15} suggest $10^{-3} M_{\odot}/{\rm yr}/{\rm kpc}^2$ following \citet{bonatto_11}, whose study is based on star clusters \citep[see also][]{lada_03}. This value is low compared to studies based on field stars, for example the rate of $3$--$7 \times 10^{-3} M_{\odot}/{\rm yr}/{\rm kpc}^2$ found by \citet{miller_79}. Moreover, the local UV field does not necessarily represent a galactic average because it is patchy and depends strongly on the distance to the nearest O star \citep[e.g.,][]{parravano_03, hamden_13}. 

Despite these uncertainties, we undertake a rough calibration of $f_{\rm esc}$ by comparing the $\sigmasfr$ and $\umw$ as predicted by our optically thin calculation. Although not shown in any figure, we find $\umw \propto \sigmasfr$ at high $\sigmasfr$. Such a linear relation is expected, given that star-forming cells are the source of UV in our calculation. We assume an intermediate local SFR surface density of $\Sigma_{\rm SFR,local} = 4 \times 10^{-3} M_{\odot}/{\rm yr}/{\rm kpc}^2$ \citep{robertson_08} and find that an escape fraction of $f_{\rm esc} = 10\%$ normalizes the relation to the local UV field, as observed by \citet{draine_78}. 

Based on the above estimates, we can revisit the assumption that the galaxy is optically thin. It is clear that dense, high-metallicity regions will block LW radiation effectively. Our propagation method and escape fraction represent crude averages over a complex, clumpy ISM with irregular geometry. Thus, it is not clear that the escape fraction should be constant for different types of galaxies and as a function of redshift. For example, \citet{whitaker_17} estimate the fraction of obscured UV light from star formation by comparing the UV and IR fluxes. They find that the obscured fraction strongly increases with galaxy stellar mass, indicating that the escape fraction might be lower for high-mass galaxies.

Encouragingly, we show in Appendix~\ref{sec:app:uv:convergence} that even extreme escape fractions of $0.01$ and $1$ change the molecular fraction by a factor of only about three. Nevertheless, a better understanding of the UV field would provide the most important systematic improvement in our modeling.

\subsection{Uncertainties in the Neutral Fraction}
\label{sec:discussion:neutral}

So far, we have discussed uncertainties that affect how we split neutral hydrogen into \hi and \htwo, but the underlying neutral fraction carries systematic uncertainties as well. Errors in the neutral fraction would not affect the atomic and molecular phases equally because $\fmol$ nonlinearly depends on the neutral surface density.

One potential source of error could be ionizing radiation from young stars or AGNs. \citet{rahmati_13_localradiation} found stellar radiation to significantly reduce the neutral fraction, particularly at high redshift. In their simulations, this reduction primarily affects gas with densities between $n_{\rm H} \approx 0.01$ and $1\ {\rm cm^{-3}}$ (Figure~3 in \citealt{rahmati_13_localradiation}). Much of this range lies above the star formation threshold of $n_{\rm H} = 0.106\ {\rm cm^{-3}}$, meaning that its neutral fraction is assumed to be about unity, a number that might be reduced in realistic radiative transfer simulations. 

However, the effects of local ionizing radiation would be difficult to include either in the ionization state calculations in IllustrisTNG or in the \citetalias{springel_03} ISM model because the escape fraction for ionizing radiation is highly uncertain \citep{gnedin_08}. As ionizing photons are absorbed by \hi, their mean free path is much shorter than that of LW photons, meaning that an optically thin propagation would be a very poor approximation. One could imagine a cell-by-cell correction based on the local SFR, but such a calculation is beyond the scope of this paper. 


\section{Conclusions}
\label{sec:conclusion}

We have presented a significantly improved methodological framework to estimate the molecular fraction in galaxies in cosmological simulations. We have tested five models for the \hiht transition, most of which rely on estimates of the neutral hydrogen surface density, the LW-band UV flux, and metallicity. We have applied our framework to the IllustrisTNG simulation suite, but we emphasize that it can be applied to any cosmological hydrodynamical simulation. Our main conclusions are as follows:
\begin{enumerate}

\item We have proposed a new method to model the molecular fraction based on projected maps rather than for individual volume elements. We find that the two methods agree reasonably well in their average predictions but that they diverge for individual galaxies and in the spatial distribution of molecular hydrogen. The projected modeling is slightly less sensitive to resolution effects than the volumetric modeling.

\item Projected modeling must be used when predicting the molecular fraction based on its observed correlation with the midplane pressure because the local thermal pressure in cosmological simulations is a poor approximation to the midplane pressure.

\item We have provided a rough estimate of the LW-band UV flux based on the optically thin propagation of radiation from young stars. Our calculation reproduces a scaling with star-formation surface density that matches the observed local values for an escape fraction of about $10\%$. The inferred molecular masses vary by a factor of three between extreme escape fractions of 1\% and 100\%. 

\item The average molecular fraction differs by a factor of up to three between different mass resolution levels (Appendix~\ref{sec:app:convergence}). The same difference is observed between the \htwo fractions in TNG100 and TNG300.

\item We critically investigate the Jeans length approximation to the column density of a gas cell. We find that, on average, it recovers projected surface densities to $0.4$ dex (or a factor of $2.5$), with equally large scatter.

\end{enumerate}
Modeling the molecular fraction remains a difficult enterprise. This paper represents an improvement but certainly not a conclusive answer to this challenge. We emphasize that all postprocessing calculations similar to our modeling are subject to numerous important, systematic uncertainties, namely the fraction of neutral gas, the modeling of the LW flux, and simplifications due to geometry.


\vspace{0.5cm}

We are grateful to Yannick Bah{\'e}, Shmuel Bialy, Blakesley Burkhardt, Claude-Andr{\'e} Faucher-Gigu{\`e}re, Robert Feldmann, Jonathan Freundlich, Nick Gnedin, Ben Johnson, Rahul Kannan, Andrey Kravtsov, Mark Krumholz, Joel Leja, Philip Mansfield, R{\"u}diger Pakmor, Joop Schaye, and Sandro Tacchella for useful discussions. All cosmological calculations were performed using the Python toolkit \textsc{Colossus} \citep{diemer_17_colossus}. B.D. and J.F. gratefully acknowledge the financial support of an Institute for Theory and Computation Fellowship. C.L. is funded by an Australian Research Council Discovery Early Career Researcher Award (DE150100618) and by the Australian Research Council Centre of Excellence for All Sky Astrophysics in 3 Dimensions (ASTRO 3D), through project number CE170100013. P.T. is supported through Hubble Fellowship grant \#HST-HF2-51384.001-A awarded by the Space Telescope Science Institute, which is operated by the Association of Universities for Research in Astronomy, Inc., for NASA, under contract NAS5-26555.


\appendix

In this appendix, we discuss technical aspects of our methodology, namely our method to compute the UV field, the convergence of our results with mass resolution, and mathematical details of the \hiht models. \\

\section{The UV Field}
\label{sec:app:uv}

We set the UV field to the maximum of two components, the cosmic UV background (\ref{sec:app:uv:bg}) and radiation from young stars (\ref{sec:app:uv:sf}). Our model relies on one free parameter, the escape fraction, whose value we have calibrated to $10\%$ (Section~\ref{sec:discussion:fesc}). Most of the \hiht models discussed in Section~\ref{sec:methods:models} work not with a physical value of the UV field but with $\umw$, the intensity of the UV field in units of the field at the solar neighborhood. That number was constrained by \citet{habing_68} and \citet{draine_78}, the latter finding a field $1.7$ times stronger than the former \citep[see, e.g.,][]{sternberg_14}. Most of the \hiht models assume the \citet{draine_78} field, and we follow this convention. To convert a UV field in physical units to a value of $\umw$, a number of ways have been suggested, including
\begin{itemize}
\item comparing the flux at $1000$\AA, a very good proxy for the destruction of \htwo because the absorption of photons is efficient only in a relatively narrow frequency range \citep[e.g.,][]{draine_11, gnedin_11};
\item comparing the total flux in the LW band ($912$--$1108$\AA\ or $11.2$--$13.6$ eV), that is, the spectral region relevant to photodissociation; and
\item comparing the photoionization or photoheating rates.
\end{itemize}
We choose the first option: normalize $\umw$ at $1000$\AA. At this wavelength, the \citet{draine_78} flux is $3.43 \times 10^{-8}\ \rm{photons}\ s^{-1} {\rm cm}^{-2} {\rm Hz}^{-1}$.

\subsection{UV Background}
\label{sec:app:uv:bg}

\begin{figure}
\centering
\includegraphics[trim = 3mm 6mm 4mm 4mm, clip, scale=0.74]{\figdir/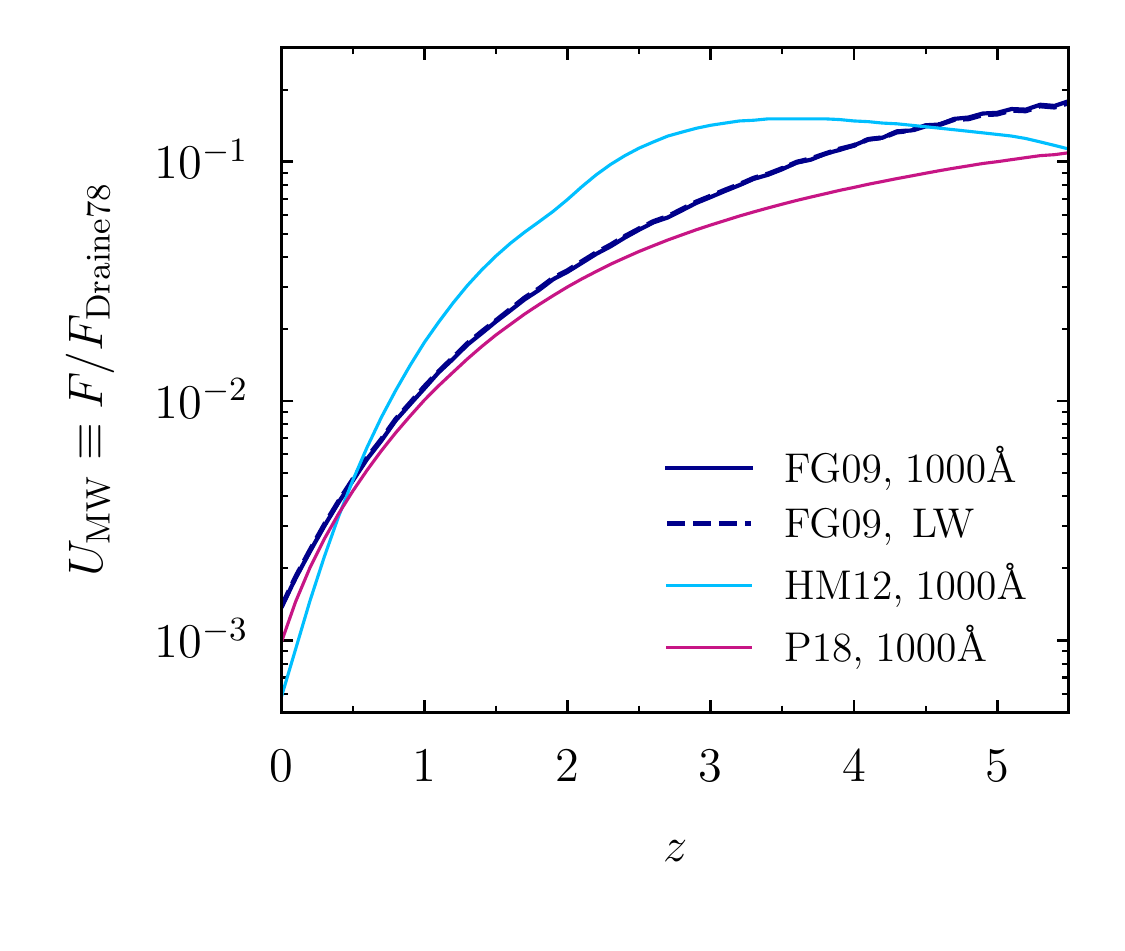}
\caption{Comparison of methods to compute $\umw$, the UV background field in units of the locally observed field according to \citet{draine_78}. Based on the \citet{fauchergiguere_09} model, we compare the flux at $1000$\AA\ (dark blue solid line) and the flux in the LW band (dark blue dashed line). The results are almost indistinguishable because the \citet{fauchergiguere_09} and \citet{draine_78} spectra resemble each other in shape in the relevant energy range. Both methods predict a low $\umw \approx 10^{-3}$ at $z = 0$. For comparison, we also show $\umw$ assuming the \citet{haardt_12} and \citet{puchwein_18} UV background models (light blue and pink lines).}
\label{fig:uvb}
\end{figure}

We adopt the time-dependent UV background of \citet[][in the updated 2011 version]{fauchergiguere_09}, the same model used in the computation of the ionization state of gas in IllustrisTNG. Figure~\ref{fig:uvb} demonstrates that the different ways to compute $\umw$ from the UV background spectrum give almost identical results because the \citet{fauchergiguere_09} and \citet{draine_78} spectra have similar shapes in the relevant region. Most importantly, $\umw$ is small at low redshift, about $10^{-3}$ at $z = 0$. We note that our computations give much lower values for $\umw$ than the computation used in \citet{lagos_15}, where the photoheating rate was used, $\umw = \dot{q}_{\rm HI} / 2.2 \times 10^{-12} {\rm eV}/{\rm s}$, where $\dot{q}_{\rm HI}$ is the \hi heating rate from the \citet{haardt_01} tables. At $z = 0$, this calculation gives $\umw = 0.072$, about $50$ times higher than our value. 

\subsection{UV from Young Stars}
\label{sec:app:uv:sf}

\begin{figure}
\centering
\includegraphics[trim = 2mm 0mm 2mm 0mm, clip, scale=0.7]{\figdir/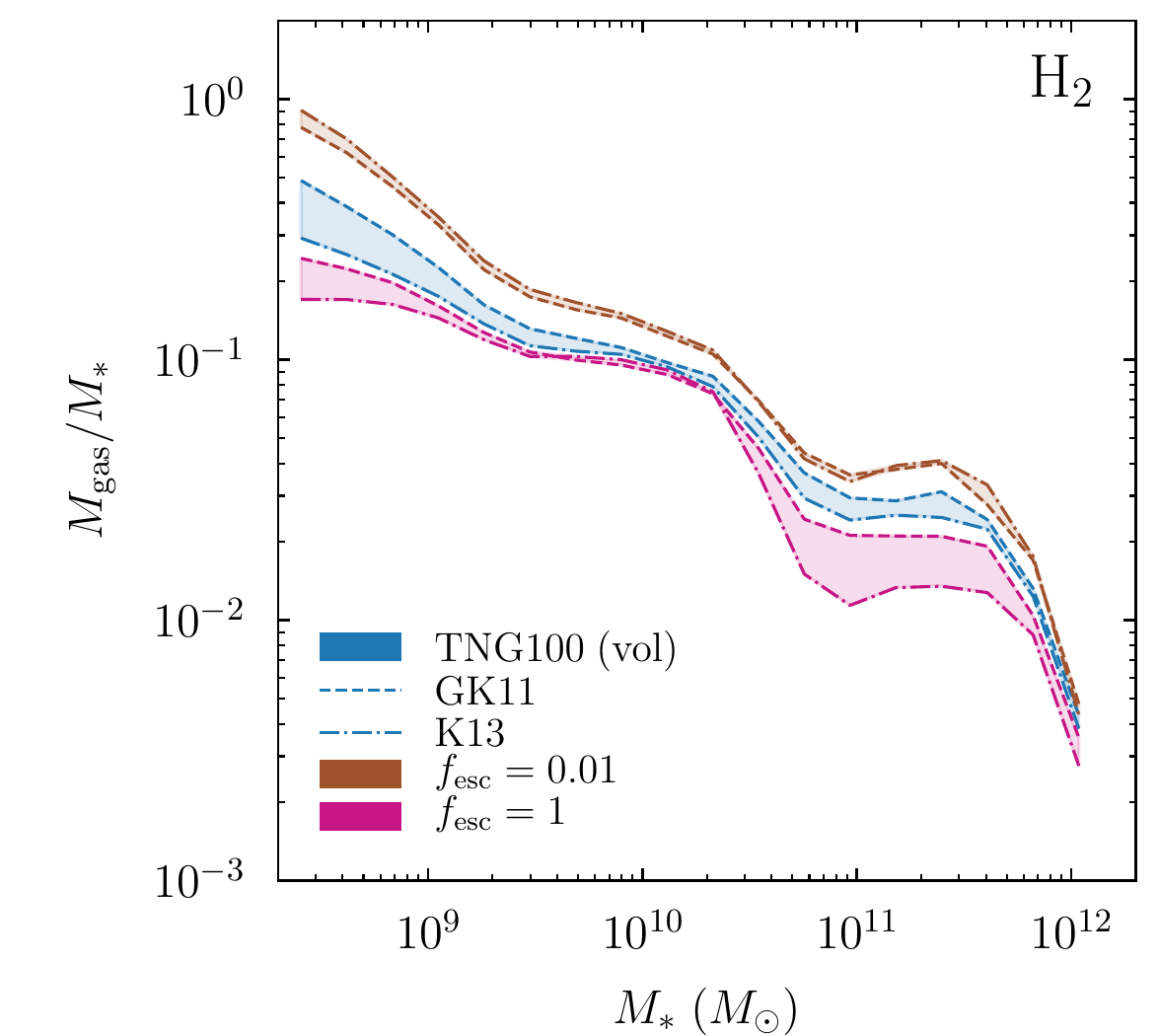}
\caption{Impact of the UV escape fraction on the \htwo mass fraction for the fiducial escape fraction of 10\% (blue), an extremely low escape fraction (1\%, brown), and no absorption (100\%, pink). The lines show the predictions of the volumetric \modelgk and \modelk models, but the same figures for other \hiht models or the projection-based versions look almost identical. Given that the probed escape fractions (and thus overall UV strengths) vary by a factor of $100$, the variation of the \htwo mass by a factor of about three is rather moderate.}
\label{fig:conv_uv}
\end{figure}

\begin{figure*}
\centering
\includegraphics[trim = 2mm 0mm 0mm 0mm, clip, scale=0.64]{\figdir/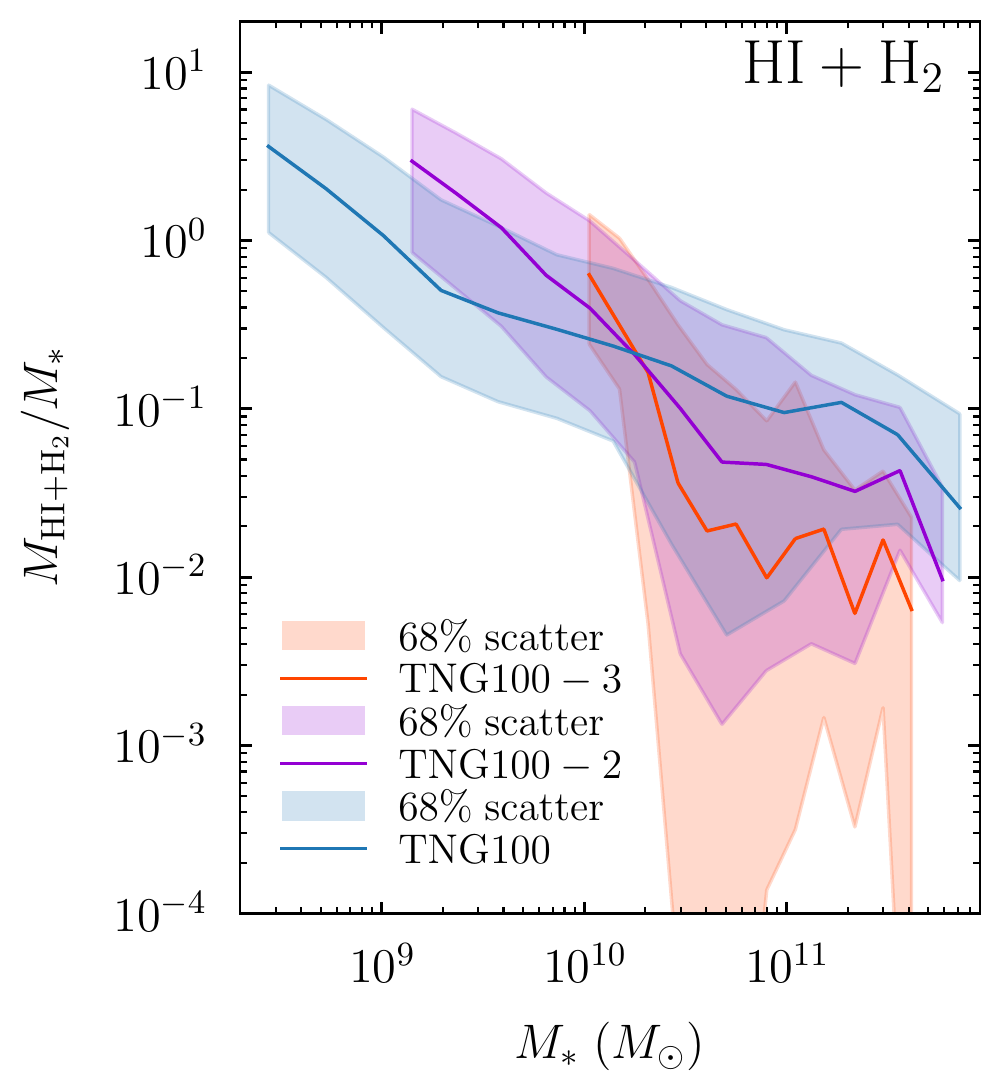}
\includegraphics[trim = 0mm 0mm 1mm 0mm, clip, scale=0.64]{\figdir/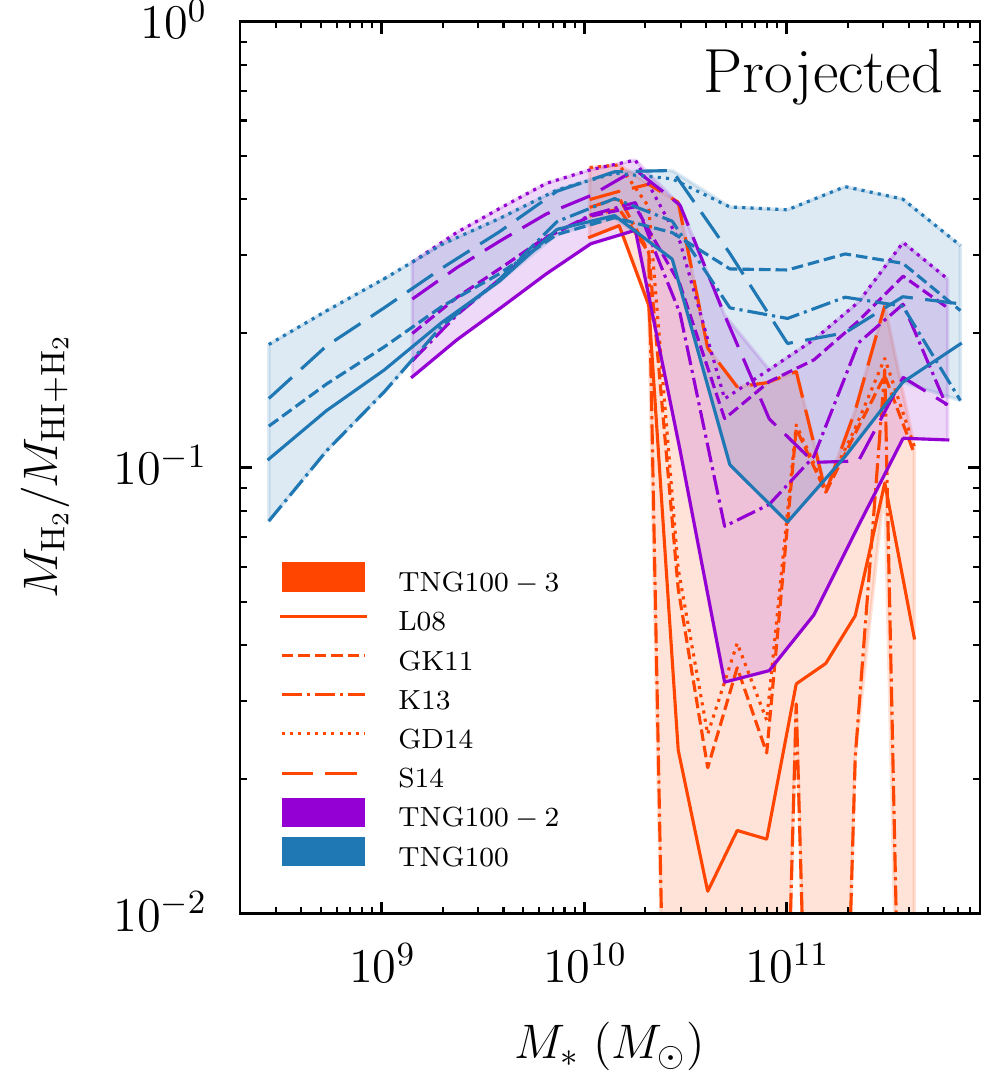}
\includegraphics[trim = 23mm 0mm 1mm 0mm, clip, scale=0.64]{\figdir/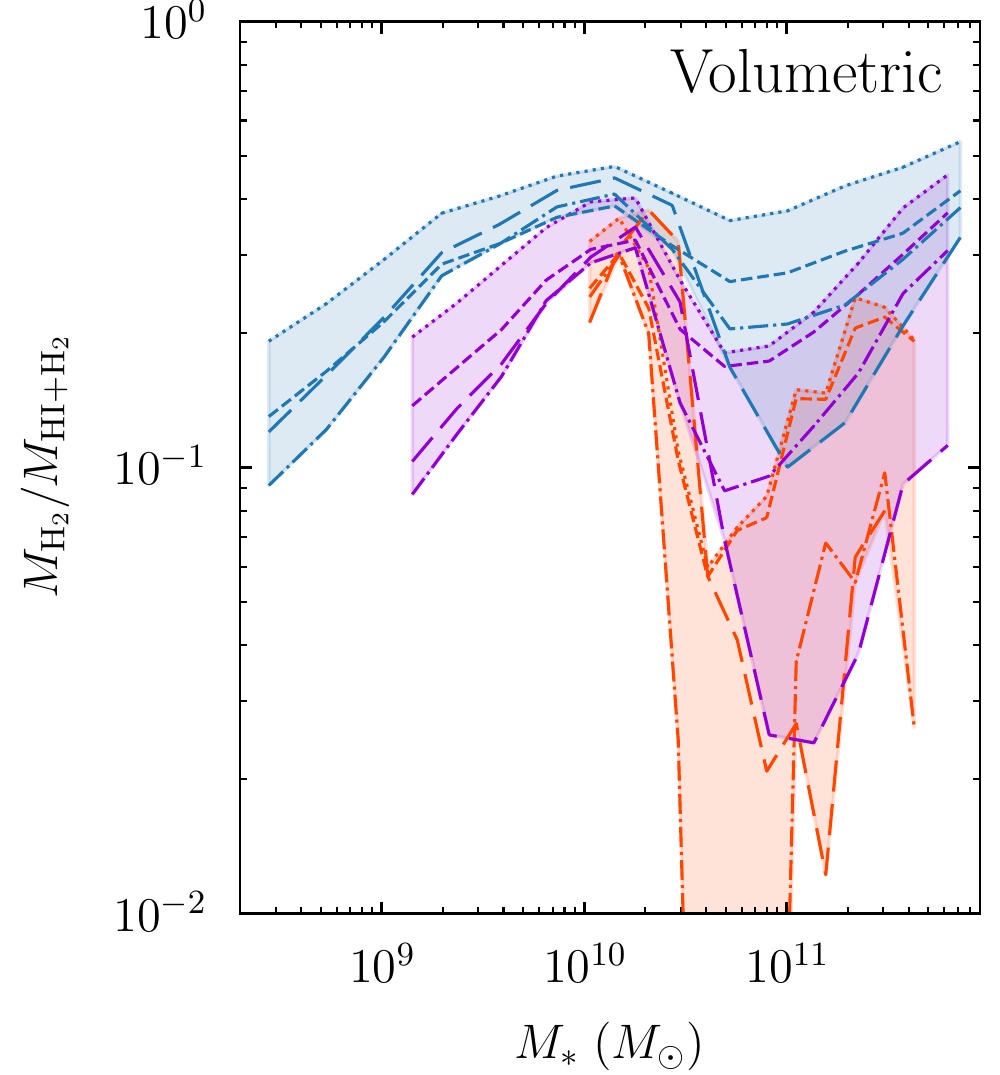}
\caption{Convergence with mass resolution of the neutral gas fraction (left) and the $\mht / \mn$ ratio (center and right) as a function of stellar mass. The blue lines and shaded areas show the TNG100 simulations, and the purple and red lines show the  lower-resolution counterparts TNG100-2 and TNG100-3, respectively. The neutral gas fraction is converged only to a factor of about three at fixed mass between the resolution levels, meaning that the atomic and molecular fractions of stellar mass will be equally affected. Thus, instead of showing the fraction with respect to stellar mass, we consider the ratio of \htwo and neutral gas, both for the projection-based (center) and volumetric (right) versions of our \hiht models. Comparing the TNG100 and TNG100-2 lines for each model, they are also converged to about a factor of two to three for most stellar masses. In TNG100-3, the convergence gets significantly worse.}
\label{fig:conv_res_fracs}
\end{figure*}

\begin{figure*}
\centering
\includegraphics[trim = 4mm 10mm 3mm 3mm, clip, scale=0.7]{\figdir/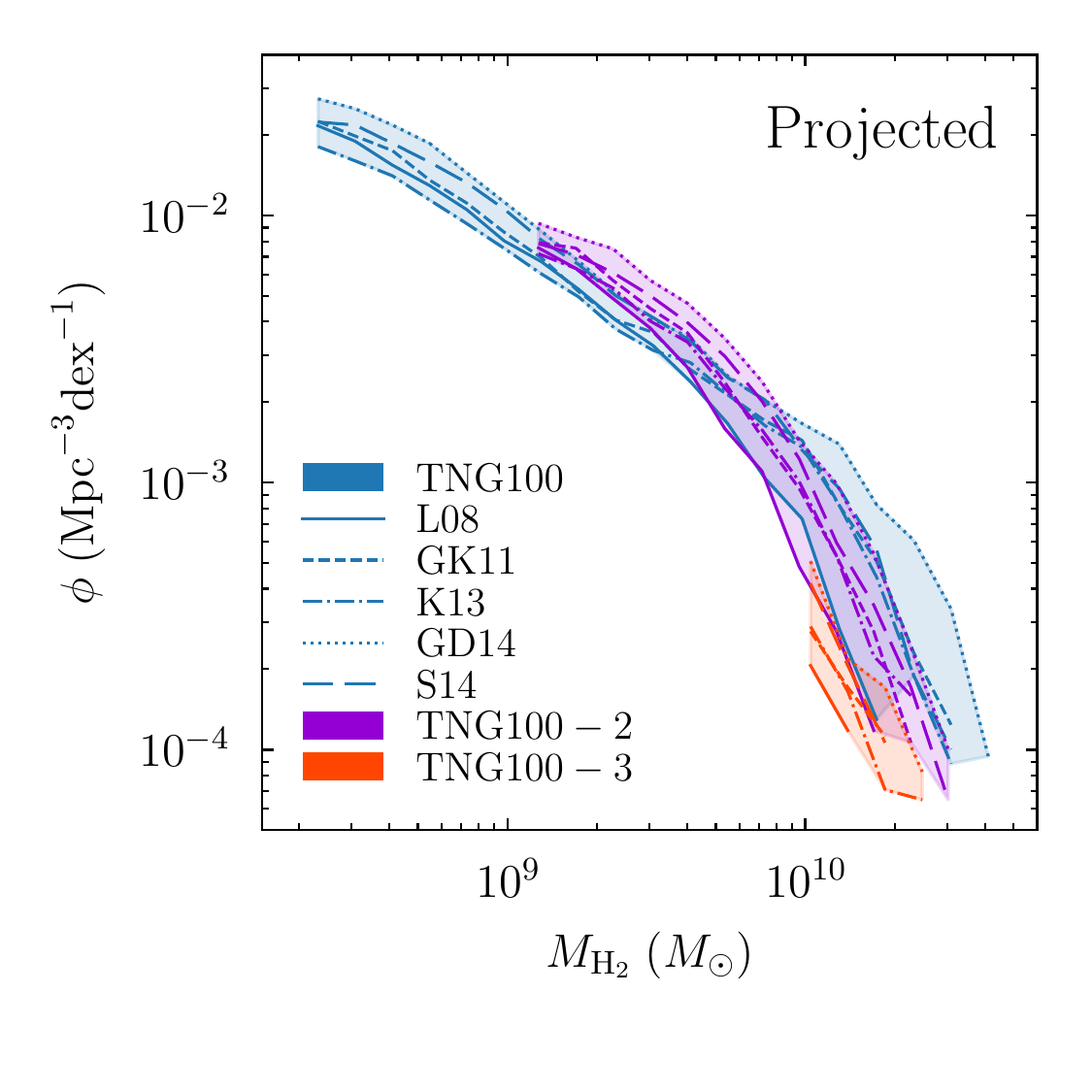}
\includegraphics[trim = 24mm 10mm 0mm 3mm, clip, scale=0.7]{\figdir/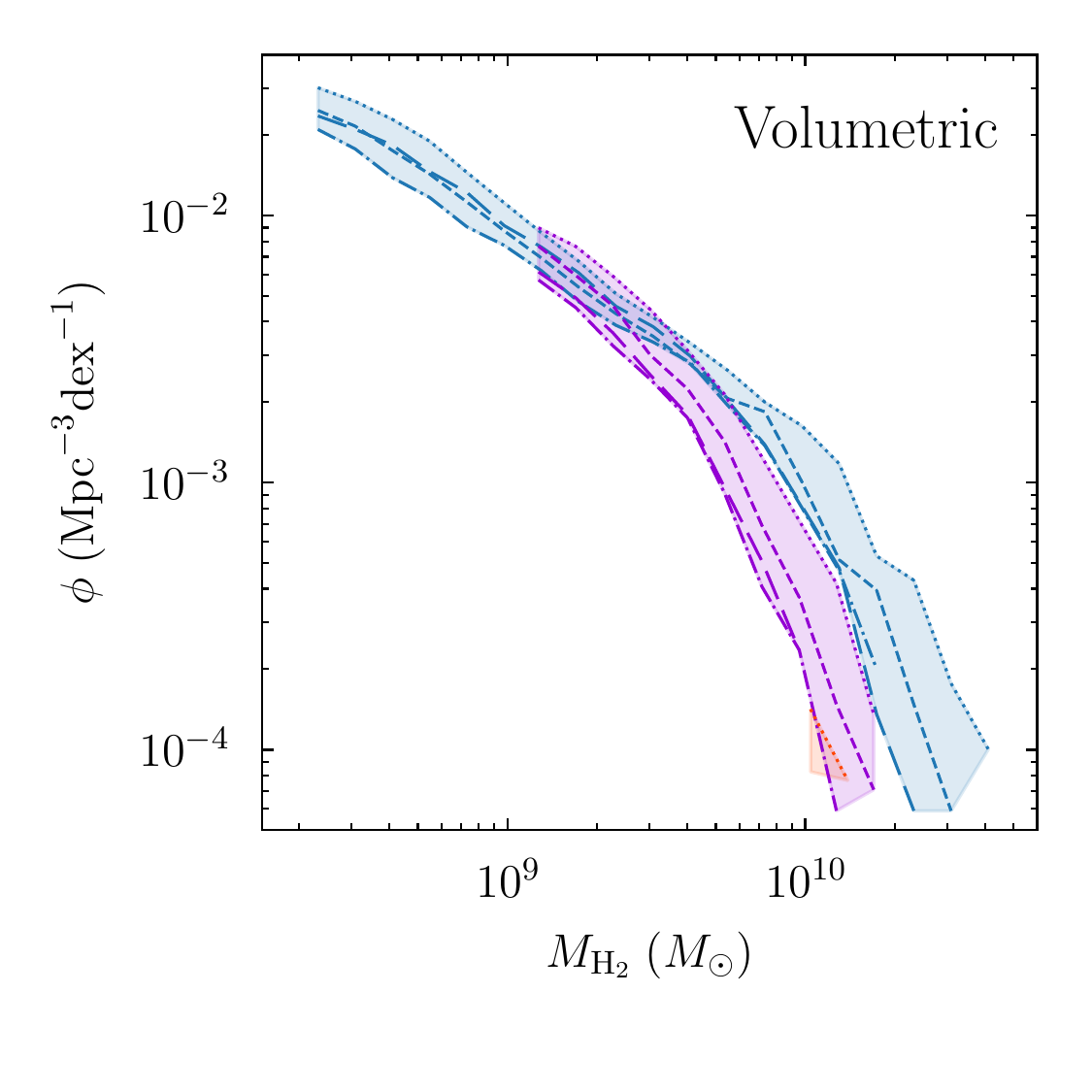}
\caption{Same as Figure~\ref{fig:conv_res_fracs} but for the \htwo mass function predicted by the projected (left) and volumetric (right) versions of our \hiht models. The projection-based models are slightly better converged in this metric than the volumetric models.}
\label{fig:conv_res_mfunc}
\end{figure*}

In most of the ISM, the UV flux is dominated by the radiation from nearby young, massive stars \citep[e.g.][]{parravano_03}. However, if we apply this UV flux only to star-forming gas cells, we create a sharp break in $\umw$ at the (somewhat arbitrary) star formation threshold (Figure~\ref{fig:ismconditions}). Moreover, cells that do not form stars will receive only the UV background, allowing significant (and presumably unrealistic) \htwo production even at very low densities (dotted lines in bottom row of Figure~\ref{fig:ismconditions}). In real galaxies, even gas that is not star-forming itself experiences UV flux due to nearby star formation. To properly estimate the corresponding flux, we would need radiative transfer calculations that are far beyond the scope of this paper.

Instead, we assume that star-forming clouds emit UV at a rate proportional to their star formation and that a fraction of this radiation is absorbed in the cloud where it originated. We assume that the resulting radiation propagates through an optically thin medium, in other words, that it is not attenuated any further and decreases with an inverse square law. We sum the contributions from all star-forming cells at the locations of the non-star forming cells.

In detail, we simulate a continuously forming population of stars using \textsc{Starburst99} \citep{leitherer_99}. We assume a \citet{kroupa_01} IMF because, out of the available IMF models, it is the most similar to the \citet{chabrier_03} IMF used in IllustrisTNG. After a few hundred megayears, the spectrum of a continuously forming population reaches equilibrium. As with the UV background, we compute $\umw$ based on the flux at $1000$\AA. At this wavelength, the spectrum depends on metallicity only weakly, with less than 50\% variation between the most extreme metallicities considered by \textsc{Starburst99}. According to the \textsc{Starburst99} calculation, the flux due to an SFR of $1 \msun / {\rm yr}$ at a distance of $1$ kpc is $3.3 \times 10^{-6}\ \rm{photons}\ s^{-1} {\rm cm}^{-2} {\rm Hz}^{-1}$. For comparison, this value corresponds to $\umw \approx 100$ and is about $500$ times stronger than the \citet{fauchergiguere_09} UV background at $z = 0$. Clearly, the contribution from young stars will dominate over the UV background as long as even a small fraction of the UV photons leave the star-forming regions and the galaxy.

We propagate the radiation using a fast Fourier transform (FFT) technique to convolve the SFR distribution with the $1/r^2$ Green's function. We perform this calculation twice on two $128^3$ grids, one grid that spans all gas cells bound to the galaxy and one higher-resolution grid that spans two gas half-mass radii. Without the high-resolution grid, the UV at the center of the galaxy can be underestimated significantly. As we cannot resolve the propagation of flux inside the grid cells, we assume a distance of $1/\sqrt{3(r_{\rm cell}/2)^3}$ for the contributions from star formation inside a given grid cell. We linearly interpolate the resulting flux grid to the positions of the gas cells. In the case of projected modeling, we simply apply the FFT to the 2D map of star formation surface density. 

Figures~\ref{fig:ismconditions} and \ref{fig:galaxy} show the results of this computation. For the cell-by-cell calculation, $\umw$ follows a more or less linear scaling with density, though with large scatter. As expected, our method spreads the UV radiation to lower-density cells, reducing their molecular fractions compared to a purely local UV field. This difference manifests itself in a spatial distribution that falls off more sharply with radius compared to a purely local UV field.

\subsection{Convergence with Escape Fraction}
\label{sec:app:uv:convergence}

The UV model described above has only one free parameter, the escape fraction, which determines the overall scaling of the field. In Section~\ref{sec:discussion:fesc}, we calibrated this number to 10\% based on the solar neighborhood values of the UV field and star-formation surface density. However, these measurements are uncertain, and our UV propagation model is extremely simplistic as it does not take dust attenuation into account (Section~\ref{sec:discussion:fesc}). Thus, we need to investigate the effect that a much lower or higher UV flux would have on our results. 

Figure~\ref{fig:conv_uv} shows the effect on the \htwo mass fraction of varying the escape fraction by a factor of $100$. We emphasize that an escape fraction of 100\% is certainly unrealistic. Nevertheless, even with these extreme values, the average \htwo masses vary by a factor of about three or less. This conclusion holds regardless of the \hiht model and whether the volumetric or projection version is used. Considering the \htwo mass functions rather than fractions leads to the same conclusions. Using a purely local UV field as in \citet{lagos_15} corresponds to gas fractions similar to the $f_{\rm esc} = 0.01$ case.

We conclude that the treatment of the UV field is likely to be our largest systematic uncertainty, but that the global properties of the galaxy population such as the \htwo mass fraction and mass function are reasonably robust to our modeling choices.

\section{Convergence with Mass Resolution}
\label{sec:app:convergence}

To assess how well converged the various gas fractions are with mass and force resolution, we have run our models on the lower resolutions of TNG100, TNG100-2, and TNG100-3 (see Table~\ref{table:sims}). Figure~\ref{fig:conv_res_fracs} shows the median ratios of neutral gas to stellar mass and \htwo mass to neutral gas mass. One of the main issues is readily apparent from the left panel: the median $\mn / \mstar$ ratio (which is independent of the \hiht models) differs by a factor of about three between resolution levels. This finding is not surprising given that TNG300 (which has a mass resolution comparable to TNG100-2) shows similar deviations in the neutral gas fraction (Figure~\ref{fig:fracs}). This difference will directly affect the $\mhi/M_*$ and $\mht/M_*$ ratios, making them less valuable as a convergence check. Instead, we compare the $\mhi/\mn$ and $\mht/\mn$ ratios in the center and right panels of Figure~\ref{fig:conv_res_fracs}. 

The model predictions for the lowest resolution level, TNG100-3, differ by orders of magnitude and tend toward negligible \htwo fractions. Clearly, our modeling breaks down at such low resolutions, and we thus focus on the comparison between TNG100 and TNG100-2. Here, the convergence is strongly dependent on mass: at stellar masses between $10^9$ and $10^{10} \msun$, TNG100-2 and TNG100 are almost perfectly converged, while they diverge by a factor of up to four at higher stellar masses. We have repeated the same experiment while binning in halo mass instead of stellar mass and find the same results. The convergence of the projected models is slightly better, particularly when ignoring the \modell model, which is not plotted for the volumetric models.

Figure~\ref{fig:conv_res_mfunc} shows the convergence of the \htwo mass function. The volumetric models again converge well at masses around $\mht \approx 10^9 \msun$ and differ by a factor of up to four in the mass function at the highest masses. The projected models perform better in this metric, with differences within a factor of about two at all masses.

In Section~\ref{sec:results:models}, we noted that the projected versions of the models exhibited slightly smaller differences between TNG100 and TNG300. Combined with the convergence tests in Figures \ref{fig:conv_res_fracs} and \ref{fig:conv_res_mfunc}, we conclude that the projected versions are slightly less sensitive to resolution effects. However, this statement depends on the \hiht model used and is difficult to generalize.

\section{Details of \hi/\htwo Models}
\label{sec:app:models}

\begin{figure}
\centering
\includegraphics[trim = 2mm 2mm 1mm 0mm, clip, scale=0.75]{\figdir/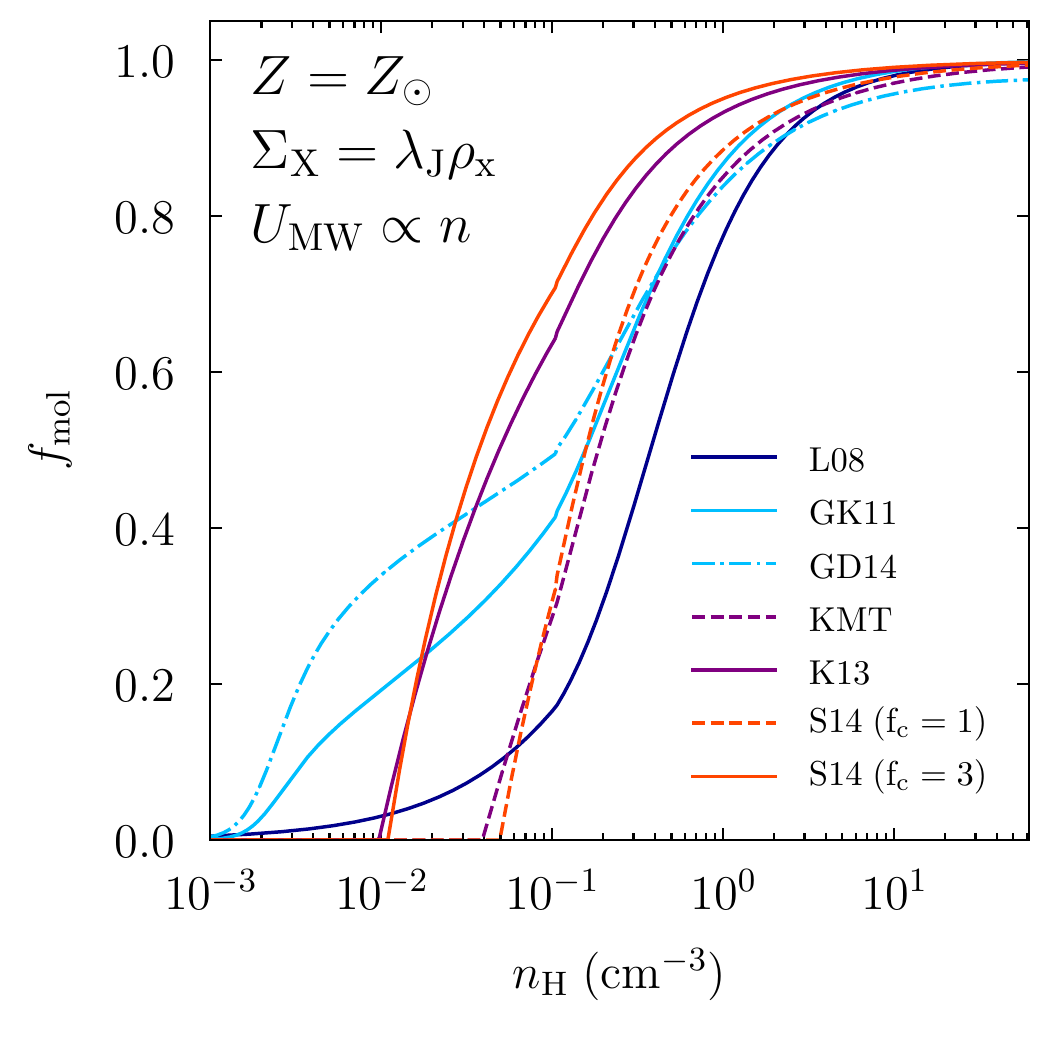}
\caption{Molecular fraction as a function of density as predicted by the \hiht models used in this paper. This calculation depends on a number of variables that are computed similarly to the blue lines in Figure~\ref{fig:ismconditions}: we choose solar metallicity, $T = \num{10000}$ K below the star formation threshold, the \citetalias{springel_03} model's equation of state above the star formation threshold, and the $\umw \approx 10\ n_{\rm H} / (1{\rm cm}^{-3})$ relation shown in Figure~\ref{fig:ismconditions}. In reality, these conditions vary across our galaxy sample. This figure corresponds to the volumetric implementation of the models in that we estimate surface densities from the density on the $x$ axis and the Jeans length. The dark blue line shows the prediction of the volumetric \modell model, which we have shown to be unphysical (Section~\ref{sec:methods:models:l08}). The light blue lines show that the \modelgk and \modelgd models agree well and that they predict nonzero molecular fractions out to lower densities than other models. The dashed lines show the original \citetalias{krumholz_09_kmt1} and \models models, which agree well. The \modelk model predicts higher molecular fractions at low density due to the addition of the hydrostatic pressure floor (Appendix~\ref{sec:app:models:k13}). Applying a clumping factor of $f_{\rm c} = 3$ to the \models model brings it into close agreement with \modelk (Appendix~\ref{sec:app:models:s14}).}
\label{fig:model_predictions}
\end{figure}

In this appendix, we list the mathematical expressions for the \hiht models used in this paper and describe how their predictions are computed in detail. If symbols appear in multiple models but with different meanings, they are understood to be defined only within the respective section. Figure~\ref{fig:model_predictions} compares the $\fmol$ predictions of the models as a function of density.

\subsection{Leroy et al. 2008 (L08)}
\label{sec:app:models:l08}

As discussed in Section~\ref{sec:methods:models:l08}, we ignore the volumetric version of the \modell model because the assumption that the midplane pressure is approximated by the thermal pressure does not hold in our simulations. To compute the projected midplane pressure in projection as given in Equation~(\ref{eq:peff}), we need maps of stellar and gas surface density and velocity dispersion in the $z$ direction. For the surface density, we follow \modell in setting $\sigmag$ to all neutral gas, that is, $\sigman + \Sigma_{\rm He}$. Including ionized gas results in molecular masses about $0.1$ dex higher. 

In computing the velocity dispersion maps for gas and stars, we need to make certain choices because their exact meaning depends on the observations we wish to compare to. Here, we assume that the velocity dispersion is measured from a line width in each map pixel (e.g., using an integral field unit spectrograph). We compute a map of the mean $z$-velocity, find the pixel at the $x$--$y$ position of each stellar particle or gas cell $i$, and subtract its value from the $z$-velocity of the particles, giving their relative velocity $v_{\rm z,i,rel}$. For the gas velocity dispersion map, we weight particles by their neutral hydrogen mass. We also include the velocity dispersion due to the thermal energy of the gas cells,
\begin{equation}
\sigma_{\rm gas, pixel} = \sqrt{\frac{\sum_i m_{\rm HI+H_2,i} \left(v_{\rm z,i,rel}^2 + \sigma_{\rm p,i}^2 \right)}{\sum_i m_{\rm HI+H_2,i}}}
\end{equation}
where $\sigma_{\rm p}^2 = P/\rho = u (\gamma - 1)$. We set a floor of $1\ {\rm km}/{\rm s}$ on our velocity dispersion maps to avoid spurious results in regions with few particles or gas cells. We note that \modell assume a fixed $\sigma_{\rm gas} = 11\ {\rm km}/{\rm s}$; we have verified that using this constant has a negligible impact on our results.

Finally, we note that the stellar and gas velocity dispersions are likely subject to resolution effects, meaning that the second term in Equation~(\ref{eq:peff}) may be inaccurate. For the majority of map pixels, the first term strongly dominates, meaning that $\peff$ depends only on the gas surface density. Moreover, Figures~\ref{fig:conv_res_fracs} and \ref{fig:conv_res_mfunc} show that the projected \modell model converges at least as well as the other models with resolution. Nevertheless, a detailed study of the reliability of stellar and gas velocity dispersions would be desirable.

\subsection{Gnedin \& Kravtsov 2011 (GK11)}
\label{sec:app:models:gk11}

\modelgk find that their simulation results for the molecular fraction are well described by
\begin{equation}
\fmol = \left( 1 + \frac{\Sigma_{\rm c}}{\sigman} \right)^{-2} \,,
\end{equation}
where $\Sigma_{\rm c}$ is a critical threshold surface density that depends only on $\umw$ and $\dmw$,
\begin{equation}
\Sigma_{\rm c} = 2 \times 10^7 \frac{\msun}{{\rm kpc}^2} \left( \frac{ \left[ \ln \left(1 + g \dmw^{3/7} (\umw/15)^{4/7} \right) \right]^{4/7}}{D_{\rm MW} \sqrt{1 + U_{\rm MW} D_{\rm MW}^2}} \right) \,,
\end{equation}
where
\begin{equation}
g = \frac{1 + \alpha s + s^2}{1 + s} \,, \alpha = 5 \frac{\umw / 2}{1 + (\umw / 2)^2} \,,
\end{equation}
and
\begin{equation}
s = \frac{0.04}{\dmw + 1.5 \times 10^{-3} \times \ln \left(1 + (3 \umw)^{1.7}\right)} \,.
\end{equation}
\modelgk point out that this formula becomes inaccurate at very low $\dmw \lsim 0.01$.

\subsection{Gnedin \& Draine 2014 (GD14)}
\label{sec:app:models:gd14}

The model is mathematically somewhat different from the \modelgk model:
\begin{equation}
\rmol = \left( \frac{\sigman}{\Sigma_{\rm c}} \right)^\alpha \,,
\end{equation}
with
\begin{equation}
\alpha = 0.5 + \frac{1}{1 + \sqrt{U_{\rm MW} D_{\rm MW}^2 / 600}} \,.
\end{equation}
As in the \modelgk model, $\Sigma_{\rm c}$ is a critical threshold surface density:
\begin{equation}
\Sigma_{\rm c} = 5 \times 10^7 \frac{\msun}{{\rm kpc}^2} \left( \frac{\sqrt{0.001 + 0.1 U_{\rm MW}}}{g \left(1 + 1.69 \sqrt{0.001 + 0.1 U_{\rm MW}}\right)} \right)  \,,
\end{equation}
where the factor $g$ depends on the dust-to-gas ratio,
\begin{equation}
g = \sqrt{D_{\rm MW}^2 + D_*^2}
\end{equation}
and
\begin{equation}
D_* = 0.17 \frac{2 + S^5}{1 + S^5} \,.
\end{equation}
The last equation includes a factor that depends on the resolution, $S \equiv L_{\rm cell} / 100\ {\rm pc}$, where $L_{\rm cell}$ is the size of the computational cells. Cells in a moving-mesh code have no well-defined side length; we estimate it as $L_{\rm cell} = V_{\rm cell}^{1/3} = (m_{\rm cell}/\rho_{\rm cell})^{1/3}$. For the projection-based version of this model, we use the side length of the map pixels to compute $S$. As expected, we find that the \modelgk and \modelgd models agree well. Figure~\ref{fig:model_predictions} shows that they predict slightly higher molecular fractions at low density compared to the \modelk and \models models.

\subsection{Krumholz 2013 (K13)}
\label{sec:app:models:k13}

The \modelk model is based on considerations of the interplay between \htwo recombination and UV dissociation in molecular clouds. \citet{mckee_10} approximate the molecular fraction as
\begin{equation}
\fmol=  \left\{
\begin{array}{ll}
      1 - 3s / (4 + s) & {\rm if} \, s < 2 \\
      0 & {\rm if} \, s \geq 2 \\
\end{array} 
\right. \,,
\end{equation}
where 
\begin{equation}
\label{equ:k13_1}
s \equiv \frac{\ln(1 + 0.6 \chi + 0.01 \chi^2)}{0.6 \tau_{\rm c}} \,.
\end{equation}
The two critical variables in this model are $\chi$ which quantifies the balance between UV dissociation and recombination, and $\tau_{\rm c}$, the optical depth of a cloud. In the \citetalias{krumholz_09_kmt1} and \modelk models, the UV field can be determined self-consistently from the predicted SFR. In this context, the simulation has already computed an SFR, which underlies our estimate of the UV field. Given these input variables, \modelk provides an approximation for $\chi$:
\begin{equation}
\chi \equiv 7.2 \umw \left( \frac{n_{\rm CNM}}{10\ {\rm cm}^{-3}} \right)^{-1}
\end{equation}
where the density appears in the denominator because a higher density leads to a higher recombination rate. However, this density is not given by the density in our simulated gas cells because their density represents an average over a large region, whereas $n_{\rm CNM}$ above refers to the typical density of the cold neutral medium (CNM), about $100$ K. Thus, we use the \modelk estimate of $n_{\rm CNM}$, which is based on the key insight that there is a relatively narrow range of pressures where a two-phase ISM can exist in equilibrium \citep{wolfire_03}. The pressure corresponding to this density range is modified by the UV field, which sets the overall cooling rate. \citet{wolfire_03} give an analytical expression for the minimum density in two-phase equilibrium. \citetalias{krumholz_09_kmt1} postulate that the equilibrium density of the CNM, $n_{\rm CNM,2p}$, should be three times that minimum density and, after certain simplifying assumptions, write
\begin{equation}
\label{equ:k13_ncnm2p}
n_{\rm CNM,2p} = 3 n_{\rm CNM,min} = 23 \umw \frac{4.1}{1 + 3.1 \dmw^{0.365}} {\rm cm}^{-3} \,.
\end{equation}
The problem with this expression is that it predicts a vanishing density and pressure in regions with a vanishing UV field, which is unrealistic because hydrostatic equilibrium demands a finite pressure. Thus, a minimum pressure is set at low densities according to the hydrostatic equilibrium,
\begin{equation}
n_{\rm CNM} = {\rm max}(n_{\rm CNM,2p}, n_{\rm CNM,hydro}) \,,
\end{equation}
with
\begin{equation}
n_{\rm CNM,hydro} = \frac{P_{\rm th}}{1.1 k_{\rm B} T_{\rm CNM,max}} \,,
\end{equation}
where the factor of $1.1$ accounts for helium. The temperature of $T_{\rm CNM,max} = 243$ K represents the maximum temperature of the CNM according to \citet{wolfire_03}, meaning that $n_{\rm CNM,hydro}$ represents a lower limit. As with the \modell model, we might be tempted to interpret $P_{\rm th}$ as the thermal pressure in the simulated gas cells, but the high-pressure region at the midplane of the disk is not resolved, meaning that the cells' thermal pressure would underestimate the hydrostatic pressure. Instead, we use \modelk's prescription, which is based on the hydrostatic equilibrium model of \citet{ostriker_10}:
\begin{equation}
\label{eq:k13_pth}
P_{\rm th} = \frac{\pi G \sigmahi^2}{4 \alpha} \left( 1 + 2 \rmol + \sqrt{(1 + 2 \rmol)^2 + \frac{32 \zeta_{\rm d} \alpha \tilde{f}_{\rm w} c_{\rm w}^2 \rho_{\rm sd}}{\pi G \sigmahi^2}} \right) \,,
\end{equation}
where $\alpha = 5$ reduces the thermal pressure to account for turbulent, magnetic, and cosmic-ray pressure, $\zeta_{\rm d} = 0.33$ is a geometric factor, $c_{\rm w} = 8\ {\rm km}/{\rm s}$ is the sound speed in the warm neutral medium, and $\tilde{f}_{\rm w} = 0.5$ is the ratio of the thermal velocity dispersion to $c_{\rm w}$.

Up to this point, we have not used the density or surface density in our simulation. These quantities enter via the $\rho_{\rm sd}$ parameter, which represents the midplane density of stars and dark matter. We could set $\rho_{\rm sd}$ to a constant \citep[e.g.,][]{lagos_15}, but that would mean neglecting the spatial dependence of the hydrostatic floor: we expect this term to be larger at the dense centers of galaxies and to become insignificant in the outskirts. Moreover, the values of $\rho_{\rm sd}$ quoted in the observational literature vary between $10^{-5}$ and $0.1 \msun/{\rm pc}^3$, presumably due to the different types of galaxies observed and the radial variation of the density \citep{holmberg_00, bruzzese_15, watts_18}. The value of $\rho_{\rm sd}$ significantly influences the predicted molecular fractions, with order-of-magnitude changes over the possible range of $\rho_{\rm sd}$. Thus, we attempt to approximate $\rho_{\rm sd}$ from the simulation data. For the cell-by-cell modeling, we subtract the density of each gas cell from the total density of matter surrounding the cell (computed using an SPH kernel over the 64 nearest neighbors). In some cells, this number becomes negative and we set it to zero. In the case of projection-based modeling, we compute the stellar and dark matter densities separately: $\rho_{\rm sd} = \rho_* + \rho_{\rm dm}$. We base the stellar density on a simple disk model, $\rho_* \approx \sigmastar / (2 h_*)$, with a scale height of
\begin{equation}
h_* = \frac{\sigma_*^2}{\pi G (\sigmastar + \sigman)} \,,
\end{equation}
where $\sigma_*$ is computed as discussed in Appendix~\ref{sec:app:models:l08} (though without a contribution from thermal pressure). We approximate the dark matter density in the disk as a \citet{navarro_97} profile with a concentration given by the relation of \citet{diemer_15}. For host halos (as defined by the halo finder), we take $M_{\rm 200c}$ from the halo catalog. For subhalos, we estimate it as the mass in dark matter that is bound to the galaxy. This approximation is far from exact, but varying it makes no appreciable difference to the predicted $\fmol$ because the dark matter tends to be a subdominant contribution to $\rho_{\rm sd}$ in regions of appreciable molecular density. 

Based on this modeling, we find values of $\rho_{\rm sd}$ that vary between zero and $\sim 1 \msun/{\rm pc}^3$. The exact range depends strongly on the properties of a given galaxy. While setting $\rho_{\rm sd}$ to a constant leads to similar average molecular masses, we find that modeling a spatially dependent $\rho_{\rm sd}$ changes the spatial distribution of \htwo significantly.

The second important factor in Equation~(\ref{equ:k13_1}) is the optical depth of dust:
\begin{equation}
\tau_{\rm c} \equiv 0.066 f_{\rm c} \dmw \frac{\sigman}{M_{\odot} {\rm pc}^{-2}} \,.
\end{equation}
Here, $f_{\rm c}$ is a clumping factor that accounts for the scale on which the surface density is measured; the actual surface density of star-forming clouds is much higher than the surface density that occurs in the large simulation cells. On scales of $100$ pc or less, we expect $f_{\rm c} \approx 1$. For kiloparsec scales, \modelk suggests $f_{\rm c} \approx 5$ which is the value we adopt. We note that \citetalias{krumholz_09_kmt1} give an alternative formula for $\chi$, which does not explicitly depend on the UV field. This expression contains another clumping factor of $30$ that accounts for higher \htwo formation rates in dense clouds that are not resolved in the simulation. Conversely, the surface density measured in simulations may represent the column density from multiple clouds and would thus overestimate the column density of individual clouds. Thus, the clumping factor accounts for multiple mismatches between theory and simulation and should be seen as a free parameter.

The expression for $P_{\rm th}$ (Equation~\ref{eq:k13_pth}) depends on $\rmol$ which is the quantity we wish to compute, meaning that we need to iterate to find the solution. If implemented naively, the $\rmol$ can take more than $50$ iterations to converge to an accuracy of $10^{-3}$ because the solution tends to oscillate for certain values. We find a great speed-up to fewer than $10$ iterations when initializing the equation with $\fmol = 0.5$ and averaging the new and old solution with a weight of $0.7$ toward the new solution. Compared to the solution where the $\rmol$ terms are dropped from Equation~\ref{eq:k13_pth}, iterating makes a difference of more than $20\%$ for some values of the input parameters.

For a comparison between the \citetalias{krumholz_09_kmt1} and \modelk models, Figure~\ref{fig:model_predictions} shows their predictions for a specific set of physical parameters. The addition of the hydrostatic pressure floor and the different clumping factors lead to higher molecular fractions at low density in the \modelk model.

\subsection{Sternberg et al. 2014 (S14)}
\label{sec:app:models:s14}

The \models model is based on a physical picture similar to that in the \citetalias{krumholz_09_kmt1} model. It predicts the total column density of atomic hydrogen in a slab irradiated by UV on both sides:
\begin{equation}
\label{equ:s14_1}
N_{\rm HI} = 5.3 \times 10^{20} {\rm cm}^{-2} \left( \frac{1}{Z/Z_{\odot}} \ln \left[ \frac{\alpha G / f_{\rm c}} {2} + 1\right] \right) \,.
\end{equation}
Like $\chi$ in the \citetalias{krumholz_09_kmt1} model, the $\alpha G$ parameter quantifies the ratio of \htwo photodissociation and recombination \citep{sternberg_88, bialy_16, bialy_17}:
\begin{equation}
\alpha G = \frac{D_0 G}{R\ \nneutral} \,,
\end{equation}
where $D_0 = 5.8 \times 10^{-11} U_{\rm MW} {\rm s}^{-1}$ is the photodissociation rate, $R = 3 \times 10^{-17} {\rm cm}^3/{\rm s}$ is the recombination rate, and the self-shielding factor is 
\begin{equation}
G = 3 \times 10^{-5} Z/Z_{\odot} \left( \frac{9.9}{1 + 8.9 Z/Z_{\odot}} \right) \,.
\end{equation}
The difference between $\alpha G$ and $\chi$ lies in the treatment of the LW absorption by \htwo molecules as opposed to dust (Section 4.1 in \models). For $Z = Z_{\odot}$, we reproduce Equation (3) in \citet{bialy_17},
\begin{equation}
\alpha G \approx 60 U_{\rm MW} \left(\frac{\nneutral}{{\rm cm}^{-3}}\right)^{-1} \,,
\end{equation}
highlighting that $\alpha G$ depends only on the volume density of neutral hydrogen, the radiation field, and metallicity. We translate the atomic column density from Equation~(\ref{equ:s14_1}) into a molecular fraction as
\begin{equation}
\fmol = \left\{
\begin{array}{ll}
0 & {\rm if} \, N_{\rm HI} > \Nneutral \\
1 - N_{\rm HI} / \Nneutral & {\rm if} \, N_{\rm HI} \leq \Nneutral \\
\end{array} 
\right. \,.
\end{equation}
For the volumetric version of the model, we once again use the column density of neutral gas derived from the Jeans approximation. However, as discussed in Appendix~\ref{sec:app:models:k13}, we cannot set $\nneutral$ to the volume density in the simulation cell because it refers to the density of molecular clouds, orders of magnitude higher than typical densities in the simulation. We revert to the same strategy as the \modelk model and set $\nneutral = n_{\rm CNM,2p}$  (Equation~\ref{equ:k13_ncnm2p}). With this modification, the \models model agrees almost exactly with the original \citetalias{krumholz_09_kmt1} model (Figure~\ref{fig:model_predictions}). Without the clumping factor $f_{\rm c}$, however, this model predicts molecular fractions an order of magnitude lower than all other models due to the mismatch between the simulated surface densities and those of actual molecular clouds. Thus, we have introduced a clumping factor of $f_{\rm c} = 3$ into the \models model in Equation~(\ref{equ:s14_1}). This factor brings the model into good overall agreement with \modelk, as shown in Section~\ref{sec:results}.


\bibliographystyle{aasjournal}
\iflocal
\bibliography{../../../Docs/_LatexInclude/gf.bib}
\else
\bibliography{gf.bib}
\fi

\end{document}

%% file: commands.tex
\newcommand{\msun}{\>{M_{\odot}}}

\def\gcm3{\mathrm{g} / \mathrm{cm}^3}

\def\mstar{M_{\ast}}

\def\gtsima{$\; \buildrel > \over \sim \;$}
\def\ltsima{$\; \buildrel < \over \sim \;$}
\def\prosima{$\; \buildrel \propto \over \sim \;$}
\def\gsim{\lower.7ex\hbox{\gtsima}}
\def\lsim{\lower.7ex\hbox{\ltsima}}
\def\simgt{\lower.7ex\hbox{\gtsima}}
\def\simlt{\lower.7ex\hbox{\ltsima}}
\def\simpr{\lower.7ex\hbox{\prosima}}


%% file: macros.tex
\def\hi{H\,{\sc i}\xspace}
\def\htwo{${\rm H}_2$\xspace}
\def\hiht{\hi/${\rm H}_2$\xspace}

\def\rmol{R_{\rm mol}}
\def\fmol{f_{\rm mol}}
\def\peff{P_{\rm eff}}

\def\sigmag{\Sigma_{\rm gas}}
\def\sigmah{\Sigma_{\rm H}}
\def\sigman{\Sigma_{\rm HI+H_2}}
\def\sigmaht{\Sigma_{\rm H_2}}
\def\sigmahi{\Sigma_{\rm HI}}
\def\sigmasfr{\Sigma_{\rm SFR}}
\def\sigmastar{\Sigma_{*}}

\def\rhog{\rho_{\rm gas}}
\def\rhoh{\rho_{\rm H}}
\def\rhon{\rho_{\rm HI+H_2}}
\def\rhoht{\rho_{\rm H_2}}
\def\rhohi{\rho_{\rm HI}}

\def\nneutral{n_{\rm HI+H_2}}
\def\Nneutral{N_{\rm HI+H_2}}
\def\fneutral{f_{\rm HI+H_2}}

\def\mn{M_{\rm HI+H_2}}
\def\mht{M_{\rm H_2}}
\def\mhi{M_{\rm HI}}

\def\umw{U_{\rm MW}}
\def\dmw{D_{\rm MW}}
\def\zmw{Z_{\rm MW}}
\def\ljeans{\lambda_{\rm J}}

\def\modell{\citetalias{leroy_08}\xspace}
\def\modelgk{\citetalias{gnedin_11}\xspace}
\def\modelk{\citetalias{krumholz_13}\xspace}
\def\modelgd{\citetalias{gnedin_14}\xspace}
\def\models{\citetalias{sternberg_14}\xspace}


%% file: citation_fix.tex
\usepackage{etoolbox}
\makeatletter
\patchcmd{\NAT@citex}
  {\@citea\NAT@hyper@{\NAT@nmfmt{\NAT@nm}\NAT@date}}
  {\@citea\NAT@nmfmt{\NAT@nm}\NAT@hyper@{\NAT@date}}
  {}
  {}
\patchcmd{\NAT@citex}
  {\@citea\NAT@hyper@{%
     \NAT@nmfmt{\NAT@nm}%
     \hyper@natlinkbreak{\NAT@aysep\NAT@spacechar}{\@citeb\@extra@b@citeb}%
     \NAT@date}}
  {\@citea\NAT@nmfmt{\NAT@nm}%
   \NAT@aysep\NAT@spacechar%
   \NAT@hyper@{\NAT@date}}
  {}
  {}
\patchcmd{\NAT@citex}
  {\@citea\NAT@hyper@{%
     \NAT@nmfmt{\NAT@nm}%
     \hyper@natlinkbreak{\NAT@spacechar\NAT@@open\if*#1*\else#1\NAT@spacechar\fi}%
       {\@citeb\@extra@b@citeb}%
     \NAT@date}}
  {\@citea\NAT@nmfmt{\NAT@nm}%
   \NAT@spacechar\NAT@@open\if*#1*\else#1\NAT@spacechar\fi%
   \NAT@hyper@{\NAT@date}}
  {}
  {}
\makeatother

%% file: paper1.bbl
\begin{thebibliography}{}
\expandafter\ifx\csname natexlab\endcsname\relax\def\natexlab#1{#1}\fi
\providecommand{\url}[1]{\href{#1}{#1}}

\bibitem[{{Asplund} {et~al.}(2009){Asplund}, {Grevesse}, {Sauval}, \&
  {Scott}}]{asplund_09}
{Asplund}, M., {Grevesse}, N., {Sauval}, A.~J., \& {Scott}, P. 2009, Annual
  Review of Astronomy and Astrophysics, 47, 481

\bibitem[{{Bah{\'e}} {et~al.}(2016){Bah{\'e}}, {Crain}, {Kauffmann}, {Bower},
  {Schaye}, {Furlong}, {Lagos}, {Schaller}, {Trayford}, {Dalla Vecchia}, \&
  {Theuns}}]{bahe_16}
{Bah{\'e}}, Y.~M., {Crain}, R.~A., {Kauffmann}, G., {et~al.} 2016, \mnras, 456,
  1115

\bibitem[{{Bekki}(2015)}]{bekki_15}
{Bekki}, K. 2015, \mnras, 449, 1625

\bibitem[{{Bett} {et~al.}(2010){Bett}, {Eke}, {Frenk}, {Jenkins}, \&
  {Okamoto}}]{bett_10}
{Bett}, P., {Eke}, V., {Frenk}, C.~S., {Jenkins}, A., \& {Okamoto}, T. 2010,
  \mnras, 404, 1137

\bibitem[{{Bialy} {et~al.}(2017){Bialy}, {Burkhart}, \& {Sternberg}}]{bialy_17}
{Bialy}, S., {Burkhart}, B., \& {Sternberg}, A. 2017, \apj, 843, 92

\bibitem[{{Bialy} \& {Sternberg}(2016)}]{bialy_16}
{Bialy}, S., \& {Sternberg}, A. 2016, \apj, 822, 83

\bibitem[{{Bigiel} {et~al.}(2008){Bigiel}, {Leroy}, {Walter}, {Brinks}, {de
  Blok}, {Madore}, \& {Thornley}}]{bigiel_08}
{Bigiel}, F., {Leroy}, A., {Walter}, F., {et~al.} 2008, \aj, 136, 2846

\bibitem[{{Bigiel} {et~al.}(2011){Bigiel}, {Leroy}, {Walter}, {Brinks}, {de
  Blok}, {Kramer}, {Rix}, {Schruba}, {Schuster}, {Usero}, \&
  {Wiesemeyer}}]{bigiel_11}
{Bigiel}, F., {Leroy}, A.~K., {Walter}, F., {et~al.} 2011, \apjl, 730, L13

\bibitem[{{Black} \& {Dalgarno}(1976)}]{black_76}
{Black}, J.~H., \& {Dalgarno}, A. 1976, \apj, 203, 132

\bibitem[{{Blitz} \& {Rosolowsky}(2004)}]{blitz_04}
{Blitz}, L., \& {Rosolowsky}, E. 2004, \apjl, 612, L29

\bibitem[{{Blitz} \& {Rosolowsky}(2006)}]{blitz_06}
---. 2006, \apj, 650, 933

\bibitem[{{Bolatto} {et~al.}(2013){Bolatto}, {Wolfire}, \&
  {Leroy}}]{bolatto_13}
{Bolatto}, A.~D., {Wolfire}, M., \& {Leroy}, A.~K. 2013, \araa, 51, 207

\bibitem[{{Bolatto} {et~al.}(2017){Bolatto}, {Wong}, {Utomo}, {Blitz}, {Vogel},
  {S{\'a}nchez}, {Barrera-Ballesteros}, {Cao}, {Colombo}, {Dannerbauer},
  {Garc{\'{\i}}a-Benito}, {Herrera-Camus}, {Husemann}, {Kalinova}, {Leroy},
  {Leung}, {Levy}, {Mast}, {Ostriker}, {Rosolowsky}, {Sandstrom}, {Teuben},
  {van de Ven}, \& {Walter}}]{bolatto_17}
{Bolatto}, A.~D., {Wong}, T., {Utomo}, D., {et~al.} 2017, \apj, 846, 159

\bibitem[{{Bonatto} \& {Bica}(2011)}]{bonatto_11}
{Bonatto}, C., \& {Bica}, E. 2011, \mnras, 415, 2827

\bibitem[{{Browning} {et~al.}(2003){Browning}, {Tumlinson}, \&
  {Shull}}]{browning_03}
{Browning}, M.~K., {Tumlinson}, J., \& {Shull}, J.~M. 2003, \apj, 582, 810

\bibitem[{{Bruzzese} {et~al.}(2015){Bruzzese}, {Meurer}, {Lagos}, {Elson},
  {Werk}, {Blakeslee}, \& {Ford}}]{bruzzese_15}
{Bruzzese}, S.~M., {Meurer}, G.~R., {Lagos}, C.~D.~P., {et~al.} 2015, \mnras,
  447, 618

\bibitem[{{Chabrier}(2003)}]{chabrier_03}
{Chabrier}, G. 2003, \pasp, 115, 763

\bibitem[{{Christensen} {et~al.}(2012){Christensen}, {Quinn}, {Governato},
  {Stilp}, {Shen}, \& {Wadsley}}]{christensen_12}
{Christensen}, C., {Quinn}, T., {Governato}, F., {et~al.} 2012, \mnras, 425,
  3058

\bibitem[{{Crain} {et~al.}(2017){Crain}, {Bah{\'e}}, {Lagos}, {Rahmati},
  {Schaye}, {McCarthy}, {Marasco}, {Bower}, {Schaller}, {Theuns}, \& {van der
  Hulst}}]{crain_17}
{Crain}, R.~A., {Bah{\'e}}, Y.~M., {Lagos}, C.~d.~P., {et~al.} 2017, \mnras,
  464, 4204

\bibitem[{{Dale} {et~al.}(2012){Dale}, {Ercolano}, \& {Bonnell}}]{dale_12}
{Dale}, J.~E., {Ercolano}, B., \& {Bonnell}, I.~A. 2012, \mnras, 424, 377

\bibitem[{{Dav{\'e}} {et~al.}(2016){Dav{\'e}}, {Thompson}, \&
  {Hopkins}}]{dave_16}
{Dav{\'e}}, R., {Thompson}, R., \& {Hopkins}, P.~F. 2016, \mnras, 462, 3265

\bibitem[{{Davis} {et~al.}(1985){Davis}, {Efstathiou}, {Frenk}, \&
  {White}}]{davis_85}
{Davis}, M., {Efstathiou}, G., {Frenk}, C.~S., \& {White}, S.~D.~M. 1985, \apj,
  292, 371

\bibitem[{{Diemer}(2017)}]{diemer_17_colossus}
{Diemer}, B. 2017, ArXiv e-prints, arXiv:1712.04512

\bibitem[{{Diemer} \& {Kravtsov}(2015)}]{diemer_15}
{Diemer}, B., \& {Kravtsov}, A.~V. 2015, \apj, 799, 108

\bibitem[{{Dolag} {et~al.}(2009){Dolag}, {Borgani}, {Murante}, \&
  {Springel}}]{dolag_09}
{Dolag}, K., {Borgani}, S., {Murante}, G., \& {Springel}, V. 2009, \mnras, 399,
  497

\bibitem[{{Draine}(1978)}]{draine_78}
{Draine}, B.~T. 1978, \apjs, 36, 595

\bibitem[{{Draine}(2003{\natexlab{a}})}]{draine_03a}
---. 2003{\natexlab{a}}, \araa, 41, 241

\bibitem[{{Draine}(2003{\natexlab{b}})}]{draine_03b}
---. 2003{\natexlab{b}}, \apj, 598, 1017

\bibitem[{{Draine}(2011)}]{draine_11}
---. 2011, {Physics of the Interstellar and Intergalactic Medium} (Princeton
  University Press)

\bibitem[{{Draine} \& {Bertoldi}(1996)}]{draine_96}
{Draine}, B.~T., \& {Bertoldi}, F. 1996, \apj, 468, 269

\bibitem[{{Dubois} {et~al.}(2014){Dubois}, {Pichon}, {Welker}, {Le Borgne},
  {Devriendt}, {Laigle}, {Codis}, {Pogosyan}, {Arnouts}, {Benabed}, {Bertin},
  {Blaizot}, {Bouchet}, {Cardoso}, {Colombi}, {de Lapparent}, {Desjacques},
  {Gavazzi}, {Kassin}, {Kimm}, {McCracken}, {Milliard}, {Peirani}, {Prunet},
  {Rouberol}, {Silk}, {Slyz}, {Sousbie}, {Teyssier}, {Tresse}, {Treyer},
  {Vibert}, \& {Volonteri}}]{dubois_14}
{Dubois}, Y., {Pichon}, C., {Welker}, C., {et~al.} 2014, \mnras, 444, 1453

\bibitem[{{Duffy} {et~al.}(2012){Duffy}, {Kay}, {Battye}, {Booth}, {Dalla
  Vecchia}, \& {Schaye}}]{duffy_12}
{Duffy}, A.~R., {Kay}, S.~T., {Battye}, R.~A., {et~al.} 2012, \mnras, 420, 2799

\bibitem[{{Elmegreen}(1989)}]{elmegreen_89}
{Elmegreen}, B.~G. 1989, \apj, 338, 178

\bibitem[{{Elmegreen}(1993)}]{elmegreen_93}
---. 1993, \apj, 411, 170

\bibitem[{{Faucher-Gigu{\`e}re} {et~al.}(2009){Faucher-Gigu{\`e}re}, {Lidz},
  {Zaldarriaga}, \& {Hernquist}}]{fauchergiguere_09}
{Faucher-Gigu{\`e}re}, C.-A., {Lidz}, A., {Zaldarriaga}, M., \& {Hernquist}, L.
  2009, \apj, 703, 1416

\bibitem[{{Ferland} {et~al.}(1998){Ferland}, {Korista}, {Verner}, {Ferguson},
  {Kingdon}, \& {Verner}}]{ferland_98}
{Ferland}, G.~J., {Korista}, K.~T., {Verner}, D.~A., {et~al.} 1998, \pasp, 110,
  761

\bibitem[{{Forbes} {et~al.}(2012){Forbes}, {Krumholz}, \&
  {Burkert}}]{forbes_12}
{Forbes}, J., {Krumholz}, M., \& {Burkert}, A. 2012, \apj, 754, 48

\bibitem[{{Forbes} {et~al.}(2014){Forbes}, {Krumholz}, {Burkert}, \&
  {Dekel}}]{forbes_14}
{Forbes}, J.~C., {Krumholz}, M.~R., {Burkert}, A., \& {Dekel}, A. 2014, \mnras,
  438, 1552

\bibitem[{{Fu} {et~al.}(2010){Fu}, {Guo}, {Kauffmann}, \& {Krumholz}}]{fu_10}
{Fu}, J., {Guo}, Q., {Kauffmann}, G., \& {Krumholz}, M.~R. 2010, \mnras, 409,
  515

\bibitem[{{Genel} {et~al.}(2014){Genel}, {Vogelsberger}, {Springel}, {Sijacki},
  {Nelson}, {Snyder}, {Rodriguez-Gomez}, {Torrey}, \& {Hernquist}}]{genel_14}
{Genel}, S., {Vogelsberger}, M., {Springel}, V., {et~al.} 2014, \mnras, 445,
  175

\bibitem[{{Glover} \& {Clark}(2012)}]{glover_12}
{Glover}, S.~C.~O., \& {Clark}, P.~C. 2012, \mnras, 421, 9

\bibitem[{{Gnedin} \& {Draine}(2014)}]{gnedin_14}
{Gnedin}, N.~Y., \& {Draine}, B.~T. 2014, \apj, 795, 37

\bibitem[{{Gnedin} \& {Kravtsov}(2011)}]{gnedin_11}
{Gnedin}, N.~Y., \& {Kravtsov}, A.~V. 2011, \apj, 728, 88

\bibitem[{{Gnedin} {et~al.}(2008){Gnedin}, {Kravtsov}, \& {Chen}}]{gnedin_08}
{Gnedin}, N.~Y., {Kravtsov}, A.~V., \& {Chen}, H.-W. 2008, \apj, 672, 765

\bibitem[{{Gnedin} {et~al.}(2009){Gnedin}, {Tassis}, \& {Kravtsov}}]{gnedin_09}
{Gnedin}, N.~Y., {Tassis}, K., \& {Kravtsov}, A.~V. 2009, \apj, 697, 55

\bibitem[{{Grand} {et~al.}(2017){Grand}, {G{\'o}mez}, {Marinacci}, {Pakmor},
  {Springel}, {Campbell}, {Frenk}, {Jenkins}, \& {White}}]{grand_17}
{Grand}, R.~J.~J., {G{\'o}mez}, F.~A., {Marinacci}, F., {et~al.} 2017, \mnras,
  467, 179

\bibitem[{{Haardt} \& {Madau}(2001)}]{haardt_01}
{Haardt}, F., \& {Madau}, P. 2001, in Clusters of Galaxies and the High
  Redshift Universe Observed in X-rays, ed. D.~M. {Neumann} \& J.~T.~V. {Tran}

\bibitem[{{Haardt} \& {Madau}(2012)}]{haardt_12}
{Haardt}, F., \& {Madau}, P. 2012, \apj, 746, 125

\bibitem[{{Habing}(1968)}]{habing_68}
{Habing}, H.~J. 1968, \bain, 19, 421

\bibitem[{{Hamden} {et~al.}(2013){Hamden}, {Schiminovich}, \&
  {Seibert}}]{hamden_13}
{Hamden}, E.~T., {Schiminovich}, D., \& {Seibert}, M. 2013, \apj, 779, 180

\bibitem[{{Hodge} {et~al.}(2015){Hodge}, {Riechers}, {Decarli}, {Walter},
  {Carilli}, {Daddi}, \& {Dannerbauer}}]{hodge_15}
{Hodge}, J.~A., {Riechers}, D., {Decarli}, R., {et~al.} 2015, \apjl, 798, L18

\bibitem[{{Holmberg} \& {Flynn}(2000)}]{holmberg_00}
{Holmberg}, J., \& {Flynn}, C. 2000, \mnras, 313, 209

\bibitem[{{Hopkins} \& {Lee}(2016)}]{hopkins_16_dust}
{Hopkins}, P.~F., \& {Lee}, H. 2016, \mnras, 456, 4174

\bibitem[{{Hu} {et~al.}(2017){Hu}, {Naab}, {Glover}, {Walch}, \&
  {Clark}}]{hu_17}
{Hu}, C.-Y., {Naab}, T., {Glover}, S.~C.~O., {Walch}, S., \& {Clark}, P.~C.
  2017, \mnras, 471, 2151

\bibitem[{{Hu} {et~al.}(2016){Hu}, {Naab}, {Walch}, {Glover}, \&
  {Clark}}]{hu_16}
{Hu}, C.-Y., {Naab}, T., {Walch}, S., {Glover}, S.~C.~O., \& {Clark}, P.~C.
  2016, \mnras, 458, 3528

\bibitem[{{Jura}(1975{\natexlab{a}})}]{jura_75_thick}
{Jura}, M. 1975{\natexlab{a}}, \apj, 197, 581

\bibitem[{{Jura}(1975{\natexlab{b}})}]{jura_75_thin}
---. 1975{\natexlab{b}}, \apj, 197, 575

\bibitem[{{Katz} {et~al.}(2017){Katz}, {Kimm}, {Sijacki}, \&
  {Haehnelt}}]{katz_17}
{Katz}, H., {Kimm}, T., {Sijacki}, D., \& {Haehnelt}, M.~G. 2017, \mnras, 468,
  4831

\bibitem[{{Kennicutt}(1998)}]{kennicutt_98}
{Kennicutt}, Jr., R.~C. 1998, \apj, 498, 541

\bibitem[{{Kroupa}(2001)}]{kroupa_01}
{Kroupa}, P. 2001, \mnras, 322, 231

\bibitem[{{Krumholz}(2013)}]{krumholz_13}
{Krumholz}, M.~R. 2013, \mnras, 436, 2747

\bibitem[{{Krumholz} \& {Dekel}(2012)}]{krumholz_12}
{Krumholz}, M.~R., \& {Dekel}, A. 2012, \apj, 753, 16

\bibitem[{{Krumholz} \& {Gnedin}(2011)}]{krumholz_11_comparison}
{Krumholz}, M.~R., \& {Gnedin}, N.~Y. 2011, \apj, 729, 36

\bibitem[{{Krumholz} {et~al.}(2011){Krumholz}, {Leroy}, \&
  {McKee}}]{krumholz_11_h2sfr}
{Krumholz}, M.~R., {Leroy}, A.~K., \& {McKee}, C.~F. 2011, \apj, 731, 25

\bibitem[{{Krumholz} {et~al.}(2009{\natexlab{a}}){Krumholz}, {McKee}, \&
  {Tumlinson}}]{krumholz_09_kmt1}
{Krumholz}, M.~R., {McKee}, C.~F., \& {Tumlinson}, J. 2009{\natexlab{a}}, \apj,
  699, 850

\bibitem[{{Krumholz} {et~al.}(2009{\natexlab{b}}){Krumholz}, {McKee}, \&
  {Tumlinson}}]{krumholz_09_kmt2}
---. 2009{\natexlab{b}}, \apj, 693, 216

\bibitem[{{Kuhlen} {et~al.}(2012){Kuhlen}, {Krumholz}, {Madau}, {Smith}, \&
  {Wise}}]{kuhlen_12}
{Kuhlen}, M., {Krumholz}, M.~R., {Madau}, P., {Smith}, B.~D., \& {Wise}, J.
  2012, \apj, 749, 36

\bibitem[{{Lada} \& {Lada}(2003)}]{lada_03}
{Lada}, C.~J., \& {Lada}, E.~A. 2003, \araa, 41, 57

\bibitem[{{Lagos} {et~al.}(2011{\natexlab{a}}){Lagos}, {Baugh}, {Lacey},
  {Benson}, {Kim}, \& {Power}}]{lagos_11_hih2}
{Lagos}, C.~D.~P., {Baugh}, C.~M., {Lacey}, C.~G., {et~al.} 2011{\natexlab{a}},
  \mnras, 418, 1649

\bibitem[{{Lagos} {et~al.}(2011{\natexlab{b}}){Lagos}, {Lacey}, {Baugh},
  {Bower}, \& {Benson}}]{lagos_11_sflaw}
{Lagos}, C.~D.~P., {Lacey}, C.~G., {Baugh}, C.~M., {Bower}, R.~G., \& {Benson},
  A.~J. 2011{\natexlab{b}}, \mnras, 416, 1566

\bibitem[{{Lagos} {et~al.}(2015){Lagos}, {Crain}, {Schaye}, {Furlong}, {Frenk},
  {Bower}, {Schaller}, {Theuns}, {Trayford}, {Bah{\'e}}, \& {Dalla
  Vecchia}}]{lagos_15}
{Lagos}, C.~d.~P., {Crain}, R.~A., {Schaye}, J., {et~al.} 2015, \mnras, 452,
  3815

\bibitem[{{Lagos} {et~al.}(2016){Lagos}, {Theuns}, {Schaye}, {Furlong},
  {Bower}, {Schaller}, {Crain}, {Trayford}, \& {Matthee}}]{lagos_16}
{Lagos}, C.~d.~P., {Theuns}, T., {Schaye}, J., {et~al.} 2016, \mnras, 459, 2632

\bibitem[{{Larson}(1981)}]{larson_81}
{Larson}, R.~B. 1981, \mnras, 194, 809

\bibitem[{{Leitherer} {et~al.}(1999){Leitherer}, {Schaerer}, {Goldader},
  {Delgado}, {Robert}, {Kune}, {de Mello}, {Devost}, \&
  {Heckman}}]{leitherer_99}
{Leitherer}, C., {Schaerer}, D., {Goldader}, J.~D., {et~al.} 1999, \apjs, 123,
  3

\bibitem[{{Leroy} {et~al.}(2008){Leroy}, {Walter}, {Brinks}, {Bigiel}, {de
  Blok}, {Madore}, \& {Thornley}}]{leroy_08}
{Leroy}, A.~K., {Walter}, F., {Brinks}, E., {et~al.} 2008, \aj, 136, 2782

\bibitem[{{Leroy} {et~al.}(2013){Leroy}, {Walter}, {Sandstrom}, {Schruba},
  {Munoz-Mateos}, {Bigiel}, {Bolatto}, {Brinks}, {de Blok}, {Meidt}, {Rix},
  {Rosolowsky}, {Schinnerer}, {Schuster}, \& {Usero}}]{leroy_13}
{Leroy}, A.~K., {Walter}, F., {Sandstrom}, K., {et~al.} 2013, \aj, 146, 19

\bibitem[{{Leroy} {et~al.}(2017){Leroy}, {Schinnerer}, {Hughes}, {Kruijssen},
  {Meidt}, {Schruba}, {Sun}, {Bigiel}, {Aniano}, {Blanc}, {Bolatto},
  {Chevance}, {Colombo}, {Gallagher}, {Garcia-Burillo}, {Kramer}, {Querejeta},
  {Pety}, {Thompson}, \& {Usero}}]{leroy_17}
{Leroy}, A.~K., {Schinnerer}, E., {Hughes}, A., {et~al.} 2017, \apj, 846, 71

\bibitem[{{Marasco} {et~al.}(2016){Marasco}, {Crain}, {Schaye}, {Bah{\'e}},
  {van der Hulst}, {Theuns}, \& {Bower}}]{marasco_16}
{Marasco}, A., {Crain}, R.~A., {Schaye}, J., {et~al.} 2016, \mnras, 461, 2630

\bibitem[{{Marinacci} {et~al.}(2017{\natexlab{a}}){Marinacci}, {Grand},
  {Pakmor}, {Springel}, {G{\'o}mez}, {Frenk}, \& {White}}]{marinacci_17}
{Marinacci}, F., {Grand}, R.~J.~J., {Pakmor}, R., {et~al.} 2017{\natexlab{a}},
  \mnras, 466, 3859

\bibitem[{{Marinacci} {et~al.}(2017{\natexlab{b}}){Marinacci}, {Vogelsberger},
  {Pakmor}, {Torrey}, {Springel}, {Hernquist}, {Nelson}, {Weinberger},
  {Pillepich}, {Naiman}, \& {Genel}}]{marinacci_18}
{Marinacci}, F., {Vogelsberger}, M., {Pakmor}, R., {et~al.} 2017{\natexlab{b}},
  ArXiv e-prints, arXiv:1707.03396

\bibitem[{{McKee} \& {Krumholz}(2010)}]{mckee_10}
{McKee}, C.~F., \& {Krumholz}, M.~R. 2010, \apj, 709, 308

\bibitem[{{McKinnon} {et~al.}(2016){McKinnon}, {Torrey}, \&
  {Vogelsberger}}]{mckinnon_16}
{McKinnon}, R., {Torrey}, P., \& {Vogelsberger}, M. 2016, \mnras, 457, 3775

\bibitem[{{McKinnon} {et~al.}(2017){McKinnon}, {Torrey}, {Vogelsberger},
  {Hayward}, \& {Marinacci}}]{mckinnon_17}
{McKinnon}, R., {Torrey}, P., {Vogelsberger}, M., {Hayward}, C.~C., \&
  {Marinacci}, F. 2017, \mnras, 468, 1505

\bibitem[{{McKinnon} {et~al.}(2018){McKinnon}, {Vogelsberger}, {Torrey},
  {Marinacci}, \& {Kannan}}]{mckinnon_18}
{McKinnon}, R., {Vogelsberger}, M., {Torrey}, P., {Marinacci}, F., \& {Kannan},
  R. 2018, \mnras, arXiv:1805.04521

\bibitem[{{Miller} \& {Scalo}(1979)}]{miller_79}
{Miller}, G.~E., \& {Scalo}, J.~M. 1979, \apjs, 41, 513

\bibitem[{{Naiman} {et~al.}(2018){Naiman}, {Pillepich}, {Springel},
  {Ramirez-Ruiz}, {Torrey}, {Vogelsberger}, {Pakmor}, {Nelson}, {Marinacci},
  {Hernquist}, {Weinberger}, \& {Genel}}]{naiman_18}
{Naiman}, J.~P., {Pillepich}, A., {Springel}, V., {et~al.} 2018, \mnras,
  arXiv:1707.03401

\bibitem[{{Narayanan} {et~al.}(2018){Narayanan}, {Conroy}, {Dave}, {Johnson},
  \& {Popping}}]{narayanan_18}
{Narayanan}, D., {Conroy}, C., {Dave}, R., {Johnson}, B., \& {Popping}, G.
  2018, ArXiv e-prints, arXiv:1805.06905

\bibitem[{{Navarro} {et~al.}(1997){Navarro}, {Frenk}, \& {White}}]{navarro_97}
{Navarro}, J.~F., {Frenk}, C.~S., \& {White}, S.~D.~M. 1997, \apj, 490, 493

\bibitem[{{Nelson} {et~al.}(2018){Nelson}, {Pillepich}, {Springel},
  {Weinberger}, {Hernquist}, {Pakmor}, {Genel}, {Torrey}, {Vogelsberger},
  {Kauffmann}, {Marinacci}, \& {Naiman}}]{nelson_18}
{Nelson}, D., {Pillepich}, A., {Springel}, V., {et~al.} 2018, \mnras, 475, 624

\bibitem[{{Nickerson} {et~al.}(2018){Nickerson}, {Teyssier}, \&
  {Rosdahl}}]{nickerson_18}
{Nickerson}, S., {Teyssier}, R., \& {Rosdahl}, J. 2018, ArXiv e-prints,
  arXiv:1802.00445

\bibitem[{{Obreschkow} {et~al.}(2009){Obreschkow}, {Kl{\"o}ckner}, {Heywood},
  {Levrier}, \& {Rawlings}}]{obreschkow_09b}
{Obreschkow}, D., {Kl{\"o}ckner}, H.-R., {Heywood}, I., {Levrier}, F., \&
  {Rawlings}, S. 2009, \apj, 703, 1890

\bibitem[{{Ostriker} {et~al.}(2010){Ostriker}, {McKee}, \&
  {Leroy}}]{ostriker_10}
{Ostriker}, E.~C., {McKee}, C.~F., \& {Leroy}, A.~K. 2010, \apj, 721, 975

\bibitem[{{Parravano} {et~al.}(2003){Parravano}, {Hollenbach}, \&
  {McKee}}]{parravano_03}
{Parravano}, A., {Hollenbach}, D.~J., \& {McKee}, C.~F. 2003, \apj, 584, 797

\bibitem[{{Pelupessy} {et~al.}(2006){Pelupessy}, {Papadopoulos}, \& {van der
  Werf}}]{pelupessy_06}
{Pelupessy}, F.~I., {Papadopoulos}, P.~P., \& {van der Werf}, P. 2006, \apj,
  645, 1024

\bibitem[{{Pillepich} {et~al.}(2018{\natexlab{a}}){Pillepich}, {Springel},
  {Nelson}, {Genel}, {Naiman}, {Pakmor}, {Hernquist}, {Torrey}, {Vogelsberger},
  {Weinberger}, \& {Marinacci}}]{pillepich_18_tng}
{Pillepich}, A., {Springel}, V., {Nelson}, D., {et~al.} 2018{\natexlab{a}},
  \mnras, 473, 4077

\bibitem[{{Pillepich} {et~al.}(2018{\natexlab{b}}){Pillepich}, {Nelson},
  {Hernquist}, {Springel}, {Pakmor}, {Torrey}, {Weinberger}, {Genel}, {Naiman},
  {Marinacci}, \& {Vogelsberger}}]{pillepich_18}
{Pillepich}, A., {Nelson}, D., {Hernquist}, L., {et~al.} 2018{\natexlab{b}},
  \mnras, 475, 648

\bibitem[{{Planck Collaboration} {et~al.}(2016){Planck Collaboration}, {Ade},
  {Aghanim}, {Arnaud}, {Ashdown}, {Aumont}, {Baccigalupi}, {Banday},
  {Barreiro}, {Bartlett}, \& et~al.}]{planck_16}
{Planck Collaboration}, {Ade}, P.~A.~R., {Aghanim}, N., {et~al.} 2016, \aap,
  594, A13

\bibitem[{{Popping} {et~al.}(2015){Popping}, {Behroozi}, \&
  {Peeples}}]{popping_15}
{Popping}, G., {Behroozi}, P.~S., \& {Peeples}, M.~S. 2015, \mnras, 449, 477

\bibitem[{{Popping} {et~al.}(2017){Popping}, {Somerville}, \&
  {Galametz}}]{popping_17}
{Popping}, G., {Somerville}, R.~S., \& {Galametz}, M. 2017, \mnras, 471, 3152

\bibitem[{{Popping} {et~al.}(2014){Popping}, {Somerville}, \&
  {Trager}}]{popping_14}
{Popping}, G., {Somerville}, R.~S., \& {Trager}, S.~C. 2014, \mnras, 442, 2398

\bibitem[{{Puchwein} {et~al.}(2018){Puchwein}, {Haardt}, {Haehnelt}, \&
  {Madau}}]{puchwein_18}
{Puchwein}, E., {Haardt}, F., {Haehnelt}, M.~G., \& {Madau}, P. 2018, ArXiv
  e-prints, arXiv:1801.04931

\bibitem[{{Rahmati} {et~al.}(2013{\natexlab{a}}){Rahmati}, {Pawlik}, {Rai{\v
  c}evic}, \& {Schaye}}]{rahmati_13}
{Rahmati}, A., {Pawlik}, A.~H., {Rai{\v c}evic}, M., \& {Schaye}, J.
  2013{\natexlab{a}}, \mnras, 430, 2427

\bibitem[{{Rahmati} {et~al.}(2015){Rahmati}, {Schaye}, {Bower}, {Crain},
  {Furlong}, {Schaller}, \& {Theuns}}]{rahmati_15}
{Rahmati}, A., {Schaye}, J., {Bower}, R.~G., {et~al.} 2015, \mnras, 452, 2034

\bibitem[{{Rahmati} {et~al.}(2013{\natexlab{b}}){Rahmati}, {Schaye}, {Pawlik},
  \& {Rai{\v c}evi{\'c}}}]{rahmati_13_localradiation}
{Rahmati}, A., {Schaye}, J., {Pawlik}, A.~H., \& {Rai{\v c}evi{\'c}}, M.
  2013{\natexlab{b}}, \mnras, 431, 2261

\bibitem[{{R{\'e}my-Ruyer} {et~al.}(2014){R{\'e}my-Ruyer}, {Madden},
  {Galliano}, {Galametz}, {Takeuchi}, {Asano}, {Zhukovska}, {Lebouteiller},
  {Cormier}, {Jones}, {Bocchio}, {Baes}, {Bendo}, {Boquien}, {Boselli},
  {DeLooze}, {Doublier-Pritchard}, {Hughes}, {Karczewski}, \&
  {Spinoglio}}]{remyruyer_14}
{R{\'e}my-Ruyer}, A., {Madden}, S.~C., {Galliano}, F., {et~al.} 2014, \aap,
  563, A31

\bibitem[{{Robertson} \& {Kravtsov}(2008)}]{robertson_08}
{Robertson}, B.~E., \& {Kravtsov}, A.~V. 2008, \apj, 680, 1083

\bibitem[{{Rosdahl} {et~al.}(2015){Rosdahl}, {Schaye}, {Teyssier}, \&
  {Agertz}}]{rosdahl_15}
{Rosdahl}, J., {Schaye}, J., {Teyssier}, R., \& {Agertz}, O. 2015, \mnras, 451,
  34

\bibitem[{{Saintonge} {et~al.}(2011){Saintonge}, {Kauffmann}, {Wang}, {Kramer},
  {Tacconi}, {Buchbender}, {Catinella}, {Graci{\'a}-Carpio}, {Cortese},
  {Fabello}, {Fu}, {Genzel}, {Giovanelli}, {Guo}, {Haynes}, {Heckman},
  {Krumholz}, {Lemonias}, {Li}, {Moran}, {Rodriguez-Fernandez}, {Schiminovich},
  {Schuster}, \& {Sievers}}]{saintonge_11_depletion}
{Saintonge}, A., {Kauffmann}, G., {Wang}, J., {et~al.} 2011, \mnras, 415, 61

\bibitem[{{Salpeter}(1955)}]{salpeter_55}
{Salpeter}, E.~E. 1955, \apj, 121, 161

\bibitem[{{Schaye}(2001)}]{schaye_01}
{Schaye}, J. 2001, \apj, 559, 507

\bibitem[{{Schaye} \& {Dalla Vecchia}(2008)}]{schaye_08}
{Schaye}, J., \& {Dalla Vecchia}, C. 2008, \mnras, 383, 1210

\bibitem[{{Schaye} {et~al.}(2010){Schaye}, {Dalla Vecchia}, {Booth}, {Wiersma},
  {Theuns}, {Haas}, {Bertone}, {Duffy}, {McCarthy}, \& {van de
  Voort}}]{schaye_10}
{Schaye}, J., {Dalla Vecchia}, C., {Booth}, C.~M., {et~al.} 2010, \mnras, 402,
  1536

\bibitem[{{Schaye} {et~al.}(2015){Schaye}, {Crain}, {Bower}, {Furlong},
  {Schaller}, {Theuns}, {Dalla Vecchia}, {Frenk}, {McCarthy}, {Helly},
  {Jenkins}, {Rosas-Guevara}, {White}, {Baes}, {Booth}, {Camps}, {Navarro},
  {Qu}, {Rahmati}, {Sawala}, {Thomas}, \& {Trayford}}]{schaye_15}
{Schaye}, J., {Crain}, R.~A., {Bower}, R.~G., {et~al.} 2015, \mnras, 446, 521

\bibitem[{{Schmidt}(1959)}]{schmidt_59}
{Schmidt}, M. 1959, \apj, 129, 243

\bibitem[{{Schruba} {et~al.}(2018){Schruba}, {Bialy}, \&
  {Sternberg}}]{schruba_18}
{Schruba}, A., {Bialy}, S., \& {Sternberg}, A. 2018, ArXiv e-prints,
  arXiv:1805.05353

\bibitem[{{Schruba} {et~al.}(2011){Schruba}, {Leroy}, {Walter}, {Bigiel},
  {Brinks}, {de Blok}, {Dumas}, {Kramer}, {Rosolowsky}, {Sandstrom},
  {Schuster}, {Usero}, {Weiss}, \& {Wiesemeyer}}]{schruba_11}
{Schruba}, A., {Leroy}, A.~K., {Walter}, F., {et~al.} 2011, \aj, 142, 37

\bibitem[{{Shull}(1978)}]{shull_78}
{Shull}, J.~M. 1978, \apj, 219, 877

\bibitem[{{Sijacki} {et~al.}(2015){Sijacki}, {Vogelsberger}, {Genel},
  {Springel}, {Torrey}, {Snyder}, {Nelson}, \& {Hernquist}}]{sijacki_15}
{Sijacki}, D., {Vogelsberger}, M., {Genel}, S., {et~al.} 2015, \mnras, 452, 575

\bibitem[{{Somerville} \& {Dav{\'e}}(2015)}]{somerville_15}
{Somerville}, R.~S., \& {Dav{\'e}}, R. 2015, \araa, 53, 51

\bibitem[{{Spitzer} \& {Zweibel}(1974)}]{spitzer_74}
{Spitzer}, Lyman, J., \& {Zweibel}, E.~G. 1974, \apjl, 191, L127

\bibitem[{{Springel}(2010)}]{springel_10}
{Springel}, V. 2010, \mnras, 401, 791

\bibitem[{{Springel} \& {Hernquist}(2003)}]{springel_03}
{Springel}, V., \& {Hernquist}, L. 2003, \mnras, 339, 289

\bibitem[{{Springel} {et~al.}(2001){Springel}, {White}, {Tormen}, \&
  {Kauffmann}}]{springel_01_subfind}
{Springel}, V., {White}, S.~D.~M., {Tormen}, G., \& {Kauffmann}, G. 2001,
  \mnras, 328, 726

\bibitem[{{Springel} {et~al.}(2018){Springel}, {Pakmor}, {Pillepich},
  {Weinberger}, {Nelson}, {Hernquist}, {Vogelsberger}, {Genel}, {Torrey},
  {Marinacci}, \& {Naiman}}]{springel_18}
{Springel}, V., {Pakmor}, R., {Pillepich}, A., {et~al.} 2018, \mnras, 475, 676

\bibitem[{{Sternberg}(1988)}]{sternberg_88}
{Sternberg}, A. 1988, \apj, 332, 400

\bibitem[{{Sternberg} \& {Dalgarno}(1989)}]{sternberg_89}
{Sternberg}, A., \& {Dalgarno}, A. 1989, \apj, 338, 197

\bibitem[{{Sternberg} {et~al.}(2014){Sternberg}, {Le Petit}, {Roueff}, \& {Le
  Bourlot}}]{sternberg_14}
{Sternberg}, A., {Le Petit}, F., {Roueff}, E., \& {Le Bourlot}, J. 2014, \apj,
  790, 10

\bibitem[{{Tepper-Garc{\'{\i}}a} {et~al.}(2012){Tepper-Garc{\'{\i}}a},
  {Richter}, {Schaye}, {Booth}, {Dalla Vecchia}, \& {Theuns}}]{teppergarcia_12}
{Tepper-Garc{\'{\i}}a}, T., {Richter}, P., {Schaye}, J., {et~al.} 2012, \mnras,
  425, 1640

\bibitem[{{Thompson} {et~al.}(2014){Thompson}, {Nagamine}, {Jaacks}, \&
  {Choi}}]{thompson_14}
{Thompson}, R., {Nagamine}, K., {Jaacks}, J., \& {Choi}, J.-H. 2014, \apj, 780,
  145

\bibitem[{{Torrey} {et~al.}(2014){Torrey}, {Vogelsberger}, {Genel}, {Sijacki},
  {Springel}, \& {Hernquist}}]{torrey_14}
{Torrey}, P., {Vogelsberger}, M., {Genel}, S., {et~al.} 2014, \mnras, 438, 1985

\bibitem[{{Torrey} {et~al.}(2018){Torrey}, {Vogelsberger}, {Hernquist},
  {McKinnon}, {Marinacci}, {Simcoe}, {Springel}, {Pillepich}, {Naiman},
  {Pakmor}, {Weinberger}, {Nelson}, \& {Genel}}]{torrey_18}
{Torrey}, P., {Vogelsberger}, M., {Hernquist}, L., {et~al.} 2018, \mnras, 477,
  L16

\bibitem[{{van Dishoeck} \& {Black}(1986)}]{vandishoeck_86}
{van Dishoeck}, E.~F., \& {Black}, J.~H. 1986, The Astrophysical Journal
  Supplement Series, 62, 109

\bibitem[{{Villaescusa-Navarro} {et~al.}(2018){Villaescusa-Navarro}, {Genel},
  {Castorina}, {Obuljen}, {Spergel}, {Hernquist}, {Nelson}, {Carucci},
  {Pillepich}, {Marinacci}, {Diemer}, {Vogelsberger}, {Weinberger}, \&
  {Pakmor}}]{villaescusanavarro_18}
{Villaescusa-Navarro}, F., {Genel}, S., {Castorina}, E., {et~al.} 2018, ArXiv
  e-prints, arXiv:1804.09180

\bibitem[{{Vogelsberger} {et~al.}(2013){Vogelsberger}, {Genel}, {Sijacki},
  {Torrey}, {Springel}, \& {Hernquist}}]{vogelsberger_13}
{Vogelsberger}, M., {Genel}, S., {Sijacki}, D., {et~al.} 2013, \mnras, 436,
  3031

\bibitem[{{Vogelsberger} {et~al.}(2014{\natexlab{a}}){Vogelsberger}, {Genel},
  {Springel}, {Torrey}, {Sijacki}, {Xu}, {Snyder}, {Nelson}, \&
  {Hernquist}}]{vogelsberger_14_illustris}
{Vogelsberger}, M., {Genel}, S., {Springel}, V., {et~al.} 2014{\natexlab{a}},
  \mnras, 444, 1518

\bibitem[{{Vogelsberger} {et~al.}(2014{\natexlab{b}}){Vogelsberger}, {Genel},
  {Springel}, {Torrey}, {Sijacki}, {Xu}, {Snyder}, {Bird}, {Nelson}, \&
  {Hernquist}}]{vogelsberger_14_nature}
---. 2014{\natexlab{b}}, \nat, 509, 177

\bibitem[{{Watts} {et~al.}(2018){Watts}, {Meurer}, {Lagos}, {Bruzzese},
  {Kroupa}, \& {Jerabkova}}]{watts_18}
{Watts}, A.~B., {Meurer}, G.~R., {Lagos}, C.~D.~P., {et~al.} 2018, \mnras,
  arXiv:1804.07072

\bibitem[{{Weinberger} {et~al.}(2017){Weinberger}, {Springel}, {Hernquist},
  {Pillepich}, {Marinacci}, {Pakmor}, {Nelson}, {Genel}, {Vogelsberger},
  {Naiman}, \& {Torrey}}]{weinberger_17}
{Weinberger}, R., {Springel}, V., {Hernquist}, L., {et~al.} 2017, \mnras, 465,
  3291

\bibitem[{{Whitaker} {et~al.}(2017){Whitaker}, {Pope}, {Cybulski}, {Casey},
  {Popping}, \& {Yun}}]{whitaker_17}
{Whitaker}, K.~E., {Pope}, A., {Cybulski}, R., {et~al.} 2017, \apj, 850, 208

\bibitem[{{Wolfire} {et~al.}(2003){Wolfire}, {McKee}, {Hollenbach}, \&
  {Tielens}}]{wolfire_03}
{Wolfire}, M.~G., {McKee}, C.~F., {Hollenbach}, D., \& {Tielens}, A.~G.~G.~M.
  2003, \apj, 587, 278

\bibitem[{{Wong} \& {Blitz}(2002)}]{wong_02}
{Wong}, T., \& {Blitz}, L. 2002, \apj, 569, 157

\end{thebibliography}
